\documentclass[useAMS,usenatbib]{mn2e}	
\usepackage{graphicx}
\usepackage{subfigure}
\usepackage{verbatim}

\title[Triggering type II quasars]{The importance of galaxy interactions in triggering type II quasar activity}
\author[P.~S.~ Bessiere et al.]
{\parbox{\textwidth}{P.~S.~ Bessiere$^{1}$\thanks{E-mail:p.bessiere@sheffield.ac.uk},
C.~N.~Tadhunter$^{1}$,	
C.~Ramos Almeida$^{2,3}$,
\&
M.~Villar Mart\'{i}n$^{4,5}$}\vspace{0.4cm}\\
\parbox{\textwidth}{$^{1}$ Department of Physics and Astronomy, University of Sheffield, Sheffield, S3 7RH, UK\\
${^2}$ Instituto de Astrof\'\i sica de Canarias (IAC), C/V\'ia L\'actea, s/n, E-38205, La Laguna, Tenerife, Spain\\
${^3}$ Departamento de Astrof\'\i sica, Universidad de La Laguna, E-38205, La Laguna, Tenerife, Spain\\
${^4}$ Instituto de Astrof\'{i}sica de Andaluc\'{i}a (CSIC), Glorieta de la Astronom\'{i}a s/n, 18088 Granada, Spain\\
${^5}$ Centro de Astrobiolog\'{i}a (INTA-CSIS), Carretera de Ajalvier, km 4, 28850 Torrej\'{o}n de Ardoz, Madrid, Spain
}}

\begin{document}
\date{}
\pagerange{\pageref{firstpage}--\pageref{lastpage}} \pubyear{2012}
\maketitle
\label{firstpage}

\begin{abstract}
We present deep Gemini GMOS-S optical broad-band images for a complete sample of 20 SDSS selected type II quasars taken from \citet{zakamska03}, with redshifts in the range $0.3 < z < 0.41$ and [OIII]$\lambda5007$ emission line luminosities $L_{[OIII]} > 10^{8.5} L_\odot$. The images were taken with the aim of investigating the interaction status of the quasar host galaxies, in order to determine the significance of galaxy interactions in triggering nuclear activity. We find that 15 of our sample of 20 ($75\pm20\%$) show evidence for interaction in the form of tails, shells, fans, irregular features, amorphous halos and double nuclei. The median surface brightness of the features is $\tilde{\mu}_r^{corr} =  23.4~mag~arcsec^{-2}$ and the range is $\Delta \mu_r^{corr} \simeq [20.9, 24.7]~mag~arcsec^{-2}$.

We find a similar rate of interaction signatures in the type II quasars as in a comparison sample of quiescent early-type galaxies at similar redshift ($67\pm14\%$) taken from \citet{ramos12}. However the surface brightness of the detected features is up to 2 magnitudes brighter for the type II quasars than for the quiescent early-types, which have surface brightnesses in the range $\Delta \mu_r^{corr} \simeq [22.1, 26.1]~mag~arcsec^{-2}$  and a median surface brightness $\tilde{\mu}_r^{corr} = 24.3~mag~arcsec^{-2}$. Despite the relatively small sample size, this may indicate that the mergers witnessed in the comparison sample galaxies could have different progenitors, or we may be viewing the interactions at different stages. We also compare our results with those of \citet{ramos11} who made a similar analysis using a complete sample of radio-loud AGN. They find a higher rate of interaction signatures in the radio loud AGN ($95^{~+5}_{-21}\%$) than the type II quasars, but a very similar range of surface brightnesses for the morphological features $\Delta \mu_r^{corr} \simeq [20.9, 24.8]~mag~arcsec^{-2}$, possibly indicating a similarity in the types of triggering interactions.

The wide range of features detected in the type II quasar sample suggests that AGN activity can be triggered  before, during or after the coalescence of the black holes, with 6 of the 20 objects ($30\pm12\%$) having double nuclei. Overall, the results presented here are consistent with the idea that galaxy interaction plays an important role in the triggering of quasar activity. We also use time scale arguments to show that it is unlikely that most radio-quiet quasars cycle through a radio-loud phase as part of a single quasar triggering event.
\end{abstract}


\section{Introduction}
\label{intro}

The role played by Active Galactic Nuclei (AGN) in the evolution of galaxies has come into sharp focus in recent years. Rather than being viewed as exotic objects, it is now thought that they may play a central role in the evolution of galaxies. This shift in perception is due to the discovery of a tight correlation between the masses of the super-massive black holes (SMBH) that are thought to lay at the heart of all massive galaxies \citep{kormendy95}, and many of the properties of the galaxy bulges (e.g. \citealt{magorrian88,ferrarese00, gebhardt00}). It has also been shown that there is a strong correlation between the evolution of AGN activity with redshift and the star formation history of the Universe (\citealt{madau96, ueda03}). It was first speculated, and later demonstrated through simulations, that AGN activity is a strong candidate for mediating these relationships (e.g. \citealt*{dimatteo05,springel05}), with feedback from the central engine blowing out the gas from the inner region, thereby quenching star formation, switching off the AGN and consequently, halting the growth of the SMBH.

If it is the case that all massive galaxies go through an AGN phase, then to accurately incorporate nuclear activity into galaxy evolution models, it is important to determine how and when AGN are triggered in the course of galaxy evolution. Mergers of gas-rich galaxies are commonly suggested as the trigger for luminous AGN (e.g. \citealt{toomre72, heckman86, hopkins06}). Such mergers have the potential to funnel material into the central regions of galaxies where it can be accreted onto the SMBH \citep{jogee06}, triggering the AGN. 

If galaxy mergers are indeed the trigger for quasar activity, then the interaction will be accompanied by morphological disturbance of the host galaxy in the form of tidal tails, fans, shells etc., as well as close pairs which have yet to coalesce, manifesting as double nuclei or linked by bridges. In a gas-rich merger, we would expect such features to be visible on a time-scale of $\sim$1 Gyr \citep{lotz08}, while we would expect the quasar activity to last $\sim$10 -- 100 Myr \citep{martini01}. This means that we would expect the quasar activity to be concurrent with the presence of tidal features and morphological disturbance of the host galaxy. If this is the case, then by attempting to detect these features in the host galaxies of quasars, we will be able to determine whether they have been involved in interactions.

This approach has been taken in the past with varying results. A study of radio galaxies, radio-loud and radio-quiet quasars, based on single orbit HST imaging by \citet{mclure99} and \citet{dunlop03} found that the majority of host galaxies appear to be undisturbed ellipticals. \citet{bahcall97} also use HST imaging to analyse a sample of type I quasars (both radio-loud and radio-quiet) in which they find 3/20 show signs of gravitational interactions. Other morphological studies which utilise shallow HST images, such as those of \citet{cisternas11} and \citet{grogin03}, based on 140 and 37 X-ray selected AGN respectively, produce similar results, showing little direct evidence for galaxy interactions being a significant factor in the triggering of the nuclear activity. \citet{grogin03} also find no increase in the number of close pairs associated with AGN over that found in the field population. 

In contrast to this, ground based studies (e.g. \citealt{heckman86, smith89, ramos11}) and deeper HST imaging studies (e.g. \citealt{bennert08}) do suggest that large proportions of the host galaxies of powerful AGN show signs of morphological disturbance. \citet{ellison11} also find an increase in the proportion of galaxy pairs associated with Seyfert galaxies.

At least part of this apparent discrepancy may be due to the differences in the depth of the images used to make the classifications. Diffuse, low surface brightness features will remain undetected in shallow HST images, and only galaxies with bright and obvious features will be classified as being morphologically disturbed. This is clearly demonstrated by \citet{bennert08}, who find that deeper HST images of five of the galaxies classified by \citet{dunlop03} as undisturbed show signs of morphological disturbance. Factors such as this, and differences in the way samples are selected, can make it difficult to compare the results of different studies.

Another important factor may be the luminosity of the AGN: it has been suggested on the basis of galaxy evolution simulations \citep*{hopkins08}, and the details of the $M_{BH} - \sigma$ relationship \citep*{kormendy11}, that there may be a dichotomy in AGN triggering mechanisms, with low/moderate luminosity AGN ($L_{BOL}<10^{38}~W$) triggered by secular processes (e.g. \citet{combes01}) and more luminous quasar-like AGN triggered in gas-rich mergers. In this context, it is notable that the studies of \citet{cisternas11} and \citet{grogin03} concern low/moderate luminosity AGN.

Attempting to detect morphological features in the host galaxies of luminous, quasars-like AGN ($L_{BOL}>10^{38}~W$) also comes with its own inherent problems, not least of which is that the extremely bright point source of the quasar can wash out fainter features associated with a merger, leading to misclassification. Therefore, it is important to undertake studies of luminous AGN in which the quasar nucleus is hidden from our direct view at optical wavelengths. \citet{ramos11} (hereafter RA11), have recently taken this approach, using deep Gemini GMOS-S images of a complete sample of powerful radio galaxies (PRG; \citealt{tadhunter93})\footnote{Full details of the 2Jy sample and available data can be found at http://2jy.extragalactic.info/Home.html} where, in the majority of cases, the broad-line AGN is obscured or does not dominate the light of the galaxy. They find that 94\% of the strong line radio galaxies (SLRG) in their sample show evidence of morphological peculiarities at relatively high levels of surface brightness ($\tilde{\mu}_V^{corr} =  23.6~mag~arcsec^{-2}$).

The study of RA11 concerned a sample of luminous, radio-loud AGN. However, it is also important to consider the triggering mechanism for radio-quiet quasars which comprise the overwhelming majority of luminous AGN. Currently the relationship between luminous radio-loud and radio-quiet AGN is uncertain, and it is not clear whether the two types are triggered in the same manner. For example, it has been suggested that each episode of quasar activity might cycle through radio-loud and radio-quiet phases \citep*{nipoti05}, in which case we would expect the triggering mechanism and galaxy morphologies to be similar for the two types. Fortunately, large samples of optically selected type II quasars (e.g. \citealt{zakamska03}), of which $\sim90\%$ are found to be radio-quiet \citep{zakamska04} are now available, providing the opportunity to compare the morphologies of luminous radio-quiet and radio-loud AGN.

Type II quasars have been the focus of a number of previous studies (e.g. \citealt{zakamska06,greene09,villar10,villar11}) in which the interaction status of the host galaxies has been analysed. In the case of \citet{villar11}, 5/13 of the type II quasars from their imaging study show evidence of interaction. However, as pointed out by the authors, due to the shallowness of the continuum images, this is a lower limit. \citet{zakamska06} find 4 out of 9 objects in their sample show evidence of tidal interaction. However, once again, there is an issue with the sensitivity of HST to diffuse, low surface brightness features. When trying to assess the importance of interactions and mergers in the triggering of quasar activity, these previous studies, though useful, suffer from small sample sizes, incompleteness and the fact that detailed studies of the interaction status of the host galaxies was not the prime focus of the work, so they were not optimized for this purpose. 

To overcome the issue of the comparability of the images used in previous studies, as well as those of sample size and completeness, we have obtained deep images of a moderate size, complete sample of type II quasars, allowing us to draw meaningful statistics from our findings. The images were obtained in an identical fashion to those used in RA11, using the GMOS instrument on the Gemini South telescope, with the same instrument configuration and average observing conditions. In this way, we are sensitive to the same features at the same levels of surface brightness. The purpose of doing this is to perform an identical morphological analysis of the host galaxies to that carried out in RA11, in order to compare the rates of interaction, the interaction stages at which the AGN are triggered (i.e. pre- or post-coalescence), and the surface brightnesses of the detected features. By making a comparison of the interaction/merger status of the host galaxies, and an analysis of any detected features, we expect to shed light on whether both types of powerful AGN share the same origins, or whether there is some fundamental underlying difference between the two.

In order to truly quantify the significance of mergers in the triggering of AGN, it is also important to compare the rate of interaction with a suitable sample of quiescent galaxies. \citet{ramos12} (hereafter RA12) provide just such a comparison, with a control sample selected to match the PRG in morphology, luminosity and redshift. They find that, considering only early-type galaxies which have tidal features indicative of mergers above the surface brightness limit of the features detected in the 2Jy sample, between 48\% and 53\% show evidence of morphological disturbance. The fact that this rate of interaction is lower than that found for the PRG would appear to support the idea that mergers play a significant role in the triggering of AGN. We will also utilise this well defined control sample: comparing our results for type II quasars with those for the quiescent population of `red sequence' galaxies to determine whether there are any significant differences in the rates of interactions and the types of features detected.

In Section \ref{sample_obs}, we describe the sample selection, observations and data reduction. In Section \ref{da}, we discuss the techniques used to analyse the images which are presented in Section \ref{images} along with detailed notes on the individual objects. In Section \ref{results}, we present the results of our analysis which we then discuss in detail in Section \ref{discussion}, including a comparison of our results with those of RA11 and RA12. Section \ref{conclusions} summarizes the work we have done here and its implications. Throughout, we assume a cosmology with $H_0=70~km~s^{-1}~Mpc^{-1}$, $\Omega_m=0.27$ and $\Omega_{\Lambda}=0.73$.

\section{Sample selection, Observations and Data Reduction}
\label{sample_obs}

The sample of 20 type II quasars used in our study is derived from the catalogue of candidate objects presented by \citet{zakamska03}, who selected objects from the Sloan Digital Sky Survey (SDSS \citealt{york00}) with high ionisation, narrow emission lines, but no indication of broad permitted lines. To characterise the nature of the ionising source, they used diagnostic emission line ratios to demonstrate the energetic dominance of the AGN. \citet{zakamska03} provide full information on the criteria used to select type II objects based on emission line ratios.

Because the AGN is itself obscured in type II objects, the luminosity of the [OIII]$\lambda5007$ emission line ($L_{ [OIII]}$) is used as a proxy for its bolometric luminosity. It is assumed that a type II object will have the same intrinsic AGN luminosity for a given $L_{[OIII]}$ as a type I object. \citet{zakamska03} make their cut between objects that are considered to be quasars, and those with Seyfert-like luminosities at $L_{[OIII]} >10^{8.5} L_\odot$, which is roughly equivalent to an AGN absolute magnitude $M_B < -23$ mag.

\begin{table*}
\centering
	\begin{minipage}{140mm}
	
     \caption{Full classification of the sample objects: Columns 1 and 2 give the SDSS identifier and abbreviated identifier that will be used throughout this paper. Columns 3 and 4 list the spectroscopic redshift and cosmology corrected scale taken from the NED database. Column 5 corresponds to the value of log$(L_{[OIII]}/L_\odot)$ taken from \citet{zakamska03} upon which the original sample selection was based, while the values in brackets are updated values taken from \citet{reyes08}. Column 6 corresponds to log $L_{5Ghz}$ W/Hz/Sr calculated using the integrated flux at 1.4 GHz from either the FIRST or NVSS surveys, assuming a spectral index of $\alpha = -0.75$ where it is unknown. If the object was undetected by either survey, then an upper limit is given corresponding to the detection limit of the survey. }
    \label{sample}
	\begin{tabular}{ l l c c c c  }
	\hline
	Name & Abbreviated name & z & Scale & log$(L_{[OIII]}/L_\odot)$& log $L_{5 GHz}$\\ 
		 &					&	&(kpc arcsec$^{-1}$) &			   &(W Hz$^{-1}$ Sr$^{-1}$)\\
	\hline 
	J002531-104022 & J0025-10 & 0.303 & 4.48 & 8.73(8.65) &	22.08 \\ 
	J011429+000037 & J0114+00 & 0.389 & 5.28 & 8.66(8.46) &	24.70\\
	J012341+004435 & J0123+00 & 0.399 & 5.36 & 9.13(9.14) &	23.27 \\
	J014237+144117 & J0142+14 & 0.389 & 5.28 & 8.76(8.87) &	$<22.58$\\
	J015911+143922 & J0159+14 & 0.319 & 4.64 & 8.56(8.39) &	$22.38$\\
	J021757-011324 & J0217-01 & 0.375 & 5.16 & 8.55(8.37) &	22.35\\
	J021758-001302 & J0217-00 & 0.344 & 4.88 & 8.75(8.81) &	22.15 \\
	J021834-004610 & J0218-00 & 0.372 & 5.13 & 8.85(8.62) &	$<22.13$\\
	J022701+010712 & J0227+01 & 0.363 & 5.05 & 8.90(8.70) &	$<22.11$\\
	J023411-074538 & J0234-07 & 0.310 & 4.55 & 8.77(8.78) &	22.35 \\
	J024946+001003 & J0249+00 & 0.408 & 5.44 & 8.63(8.75) &	22.14\\
	J032029+003153 & J0320+00 & 0.384 & 5.23 & 8.52(8.40) &	$<22.17$\\	
	J033248-001012 & J0332-00 & 0.310 & 4.55 & 8.50(8.53) &	23.36\\
	J033435+003724 & J0334+00 & 0.407 & 5.43 & 8.61(8.75) &	$<22.23$\\ 
	J084856+013647 & J0848-01 & 0.350 & 4.95 & 8.56(8.46) &	24.00  \\
	J090414-002144 & J0904-00 & 0.353 & 4.98 & 8.93(8.89) &	23.52\\
	J092318+010144 & J0923+01 & 0.386 & 5.27 & 8.94(8.78) &	22.09 \\
	J092356+012002 & J0924+01 & 0.380 & 5.22 & 8.59(8.46) &	23.33\\
	J094836+002104 & J0948+00 & 0.324 & 4.71 & 8.52(8.57) &	22.35\\	
	J235818-000919 & J2358-00 & 0.402 & 5.39 & 9.32(9.29) &	$<22.21$\\  
	\hline 	  
	\end{tabular}
	\end{minipage}       
\end{table*}

Our sample comprises all objects from \citet{zakamska03} with  RAs in the range $23 < RA < 10$ hr, declinations  $<+20$ degrees, redshifts in the range $0.3<z<0.41$ and [OIII] emission line luminosities $L_{[OIII]} > 10^{8.5}L_\odot$ (see Table \ref{sample}).  The redshift limits ensure that the objects are sufficiently close and bright for the detection of fainter, extended features to be possible. The $L_{[OIII]}$ limit was chosen to ensure that the objects were indeed true quasars. Subsequent updates to the emission line luminosities, made by \citet{reyes08}, suggest that six objects from the original sample selection fall marginally below this $L_{ [OIII]}$ cut. However, because the original sample selection was based on \citet{zakamska03}, we choose to retain the full sample of 20 objects, although the updated $L_{ [OIII]}$ values are included in Table \ref{sample} (bracketed values) as a matter of interest. The full sample of 20 objects is complete and unbiased in terms of host galaxy properties. 

Due to the fact that the prime focus of this study is the evidence for morphological disturbance in the host galaxies of type II quasars, we have not imposed any selection criteria regarding the radio properties of the sample. However, given that we do make a comparison in the rates of morphological disturbance between this sample and the PRG of RA11, we will discuss the radio properties of the type II quasars further in Section \ref{comp_2jy_samp}.

Deep optical imaging data for all 20 objects from the sample was obtained using the Gemini Multi-Object Spectrograph (GMOS-S) mounted on the 8.1m Gemini South telescope at Cerro Pach\'{o}n, Chile. The observations were carried out in queue mode between August 2009 and September 2011 in good seeing conditions, with a median value of FWHM = 0.8 arcsec (ranging between 0.53 arcsec and 1.08 arcsec, see Table \ref{obs}). To determine the seeing in each of the 20 images, measurements of Gaussian fits to four foreground stars were averaged. Full details of the observations can be found in Table \ref{obs}.

The GMOS-S detector \citep{hook04} consists of three adjacent $2048\times 4096$ pixel CCDs separated by two gaps of approximately 2.8 arcsec. This gives a field of view of $5.5\times 5.5$ arcmin$^2$ and has a pixel scale of 0.146 arcsec pixel$^{-1}$. 

For this study, all the objects were imaged in the $r^{\prime}$-band filter (r$_{G0326}, \lambda_{eff}=6300$ \AA, $\Delta \lambda = 1360$ \AA). One of the issues that may arise when using the $r^\prime$ broad-band filter  in $0.3<z<0.41$ redshift range is that of emission line contamination. In all of the objects the [OIII]$\lambda\lambda5007,4959$ lines are shifted into the band. This may result in the detected features being dominated by strong emission lines rather than continuum emission, potentially affecting the comparison with the quiescent early-type sample\footnote{Extended, AGN-photoionized regions are unlikely to be present in the quiescent comparison sample objects.}. However, for 10 (50\%) of the objects in our sample we have long-slit spectroscopy observations for which the slit passes through at least one of the morphological features identified in Table \ref{surf_bright} (\citealt{villar11,bessiere12}: in preparation). We find that in 9 out of 10 of these objects (90\%) the extended features are dominated by continuum rather than line emission (see individual notes in Section \ref{images}). This is consistent with the results of RA11 who found that, for the 22 PRG in the 2Jy sample with the requisite data, 91\% have at least some tidal features which are continuum rather than emission-line dominated. Therefore, it is unlikely that emission line contamination is a serious issue for the tidal features detected in our full sample of type II quasars.

\begin{table*}
\centering
  \begin{minipage}{140mm}
  \caption{Details of the observations carried out at Gemini South using the GMOS instrument: Column 1 gives the object name, column 2 the exposure times, columns 3 and 4 give the date of the observation and the seeing conditions (in arcsec) as measured from the final reduced images. Column 5 gives the Galactic extinction and column 6 gives the apparent magnitude of the object within a 30 kpc aperture, after the removal of any other bright objects that fall within the radius. Column 7 is the magnitude in the $r^\prime$ band corrected for Galactic extinction and k-corrected.}
  \label{obs}
    \begin{tabular}{l c c c c c r}
      \hline
      Name & Exptime(s) & obs. date & Seeing & A$_\lambda$ (mag) & Mag(AB) & Mag(corr)\\
      \hline
      J0025-10 & $250\times 4$ & 2011-09-19 & 0.86 & 0.083 & 18.32 & 17.84\\
      J0114+00 & $250\times 4$ & 2010-01-12 & 1.03 & 0.073 & 19.22 & 18.56\\
      J0123+00 & $250\times 8$ & 2009-12-09 & 0.73 & 0.089 & 19.73 & 19.03\\
      J0142+14 & $250\times 4$ & 2010-01-13 & 1.06 & 0.133 & 18.85 & 18.13\\
      J0159+14 & $250\times 9$ & 2011-08-04 & 0.93 & 0.150 & 19.98 & 19.40\\
      J0217-01 & $250\times 4$ & 2009-12-09 & 0.84 & 0.087 & 19.19 & 18.55\\
      J0217-00 & $250\times 4$ & 2009-10-20 & 0.77 & 0.090 & 18.87 & 18.30\\
      J0218-00 & $250\times 4$ & 2009-12-23 & 0.84 & 0.097 & 18.91 & 18.27\\
      J0227+01 & $250\times 4$ & 2010-01-07 & 0.66 & 0.078 & 19.19 & 18.59\\
      J0234-07 & $250\times 4$ & 2010-01-12 & 0.98 & 0.106 & 19.67 & 19.16\\
      J0249+00 & $250\times 8$ & 2009-12-14 & 0.73 & 0.149 & 20.08 & 19.30\\ 
      J0320+00 & $250\times 4$ & 2009-11-12 & 0.76 & 0.269 & 19.13 & 18.29\\     
      J0332-00 & $250\times 4$ & 2009-10-23 & 0.51 & 0.287 & 18.53 & 17.83\\
      J0334+00 & $250\times 8$ & 2009-12-24 & 0.63 & 0.318 & 20.10 & 19.15\\ 
      J0848+01 & $250\times 4$ & 2009-12-22 & 0.65 & 0.098 & 19.14 & 18.55\\
      J0904-00 & $250\times 4$ & 2009-12-24 & 0.53 & 0.099 & 18.74 & 18.14\\
      J0923+01 & $250\times 4$ & 2009-12-13 & 0.66 & 0.073 & 19.18 & 18.52\\ 
      J0924+01 & $250\times 7$ & 2009-12-24 & 0.90 & 0.111 & 20.08 & 19.40\\
      J0948+00 & $250\times 4$ & 2009-12-26 & 0.83 & 0.205 & 19.44 & 18.80\\           
      J2358-00 & $250\times 4$ & 2009-08-23 & 1.08 & 0.096 & 19.08 & 18.37\\
      \hline              
      \end{tabular}
  \end{minipage}       
\end{table*}

Other possible sources of AGN contamination include scattered quasar light \citep[e.g.][]{tadhunter92} and nebular continuum \citep{dickson95}. However, both of these components would be expected to contribute more strongly at blue/UV rest-frame wavelengths, rather than the longer wavelengths encompassed by our r$^\prime$ filter images. Moreover, both of these components are strongly correlated with the emission line flux \citep[e.g.][]{tadhunter02}. Therefore, since most of the extended regions for which we have adequate spectroscopic information do not show strong emission lines, it is unlikely that scattered quasar light and nebular continuum make a major contribution to the extended features detected in the type II quasars studied here.

Between four and nine 250 second exposures were taken for each object, allowing for the detection of the low surface brightness features that are indicative of merger events. The images were taken in a four point square dither pattern with a 10 arcsec step size. This technique allows for the removal of the gaps between the CCDs when the images are combined, improved removal of other image imperfections, and better flat fielding.

The data were reduced using the dedicated Gemini GMOS package within the IRAF\footnote{IRAF is distributed by the National Optical Astronomy Observatory, which is operated by the Association of Universities for Research in Astronomy (AURA) under cooperative agreement with the National Science Foundation $(http://iraf.noao.edu/)$.}  environment. The reduction was carried out it four steps. The separate images from the three CCDs that comprised each exposure were first combined into a single image which was then flat fielded and bias subtracted. These images were then mosaicked and finally co-added to produce the final image.

In order to calibrate the magnitude scale, we used calibration images of Landolt standard star fields taken with the $r^{\prime}$-band filter at both the beginning and end of the 2009-2010 observing period to determine the photometric zero point. Using these, we determined that the photometric accuracy of our observations is $\pm 0.08$ mag. This calibration is more uncertain for our image of Q0159+14, as this was taken with the GMOS-S instrument mounted on the telescope in a different configuration. However, the magnitude derived from the Gemini image agrees with the SDSS Model $r_{AB}$ magnitude for this object to within 0.2 magnitudes.

We then measured the apparent $r^\prime_{AB}$ magnitude of each object within a 30 kpc diameter aperture, removing contamination from other sources that fell within the aperture (e.g. foreground galaxies or stars). These values are given in column 6 of Table \ref{obs}. We then corrected for Galactic extinction using the E(B-V) values available at the NASA/IPAC Infrared science archive\footnote{http://irsa.ipac.caltech.edu/applications/DUST/} based on the extinction law of \citet*{cardelli89}. To make the magnitudes consistent within this sample, and also to make them directly comparable to other samples, k-corrections using the values reported by \citet{frei94} and \citet*{fukugita95} were also applied. The apparent magnitude of each object, corrected for Galactic extinction and k-corrected,  are given in column 7 of Table \ref{obs}.

\section{Data analysis}
\label{da}
\begin{figure*}
\centering
\subfigure[J0025-10, {U[L](O[D])}]{\includegraphics[width=7.5cm]{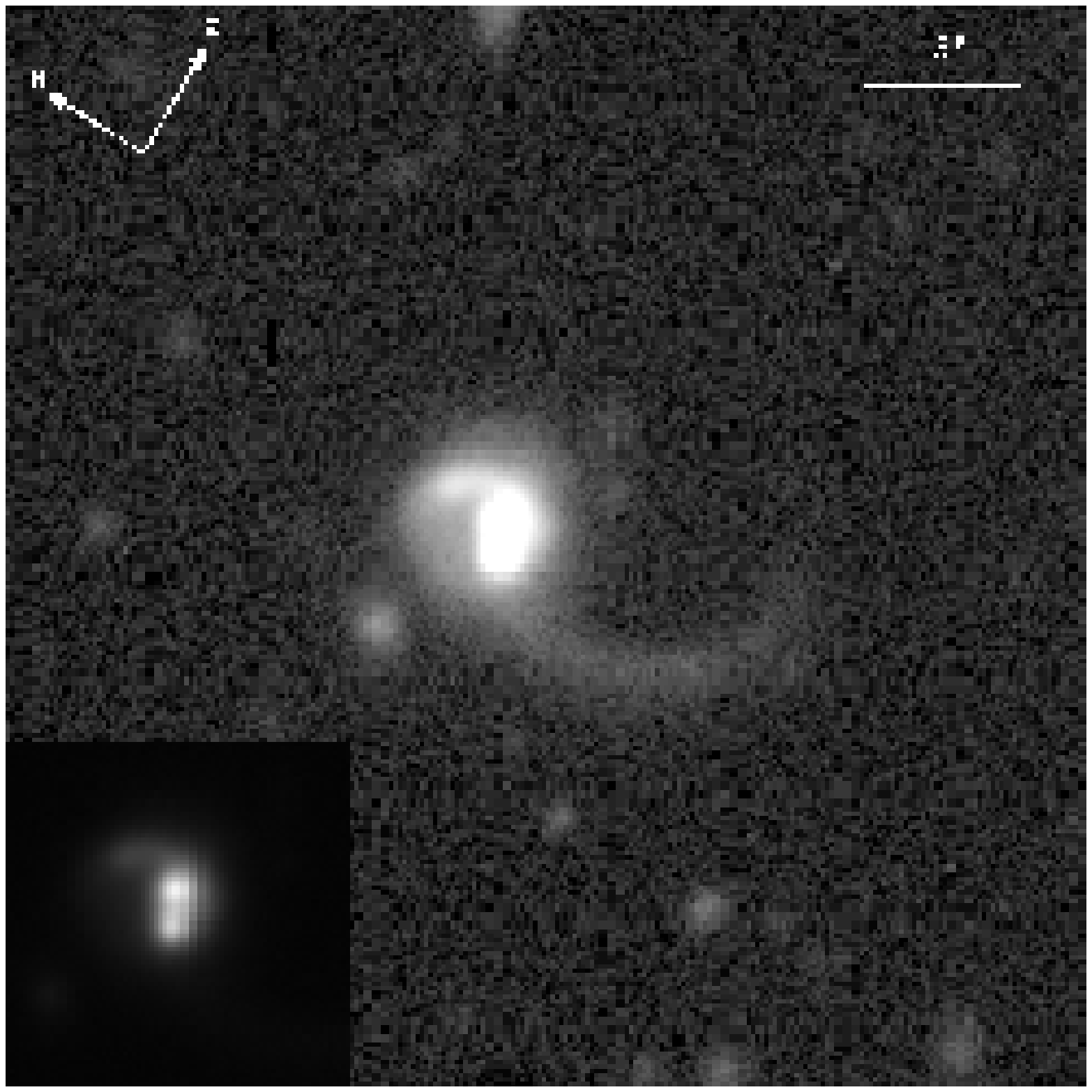}
\label{q0025}}
\subfigure[J0114+00, {U[D](S[D])}]{\includegraphics[width=7.5cm]{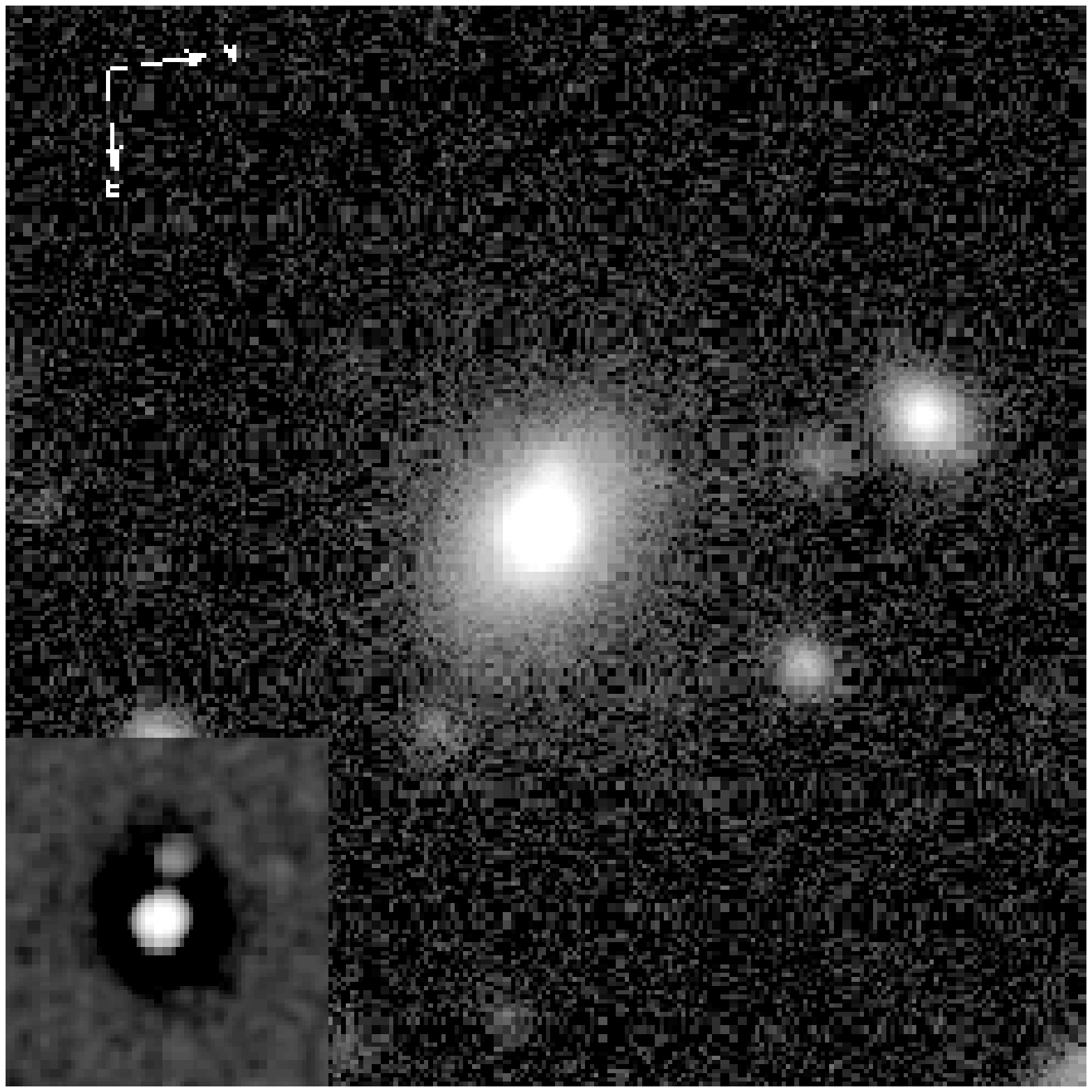}
\label{q0114}}
\subfigure[J0123+00, {M[L](S[D])}]{\includegraphics[width=7.5cm]{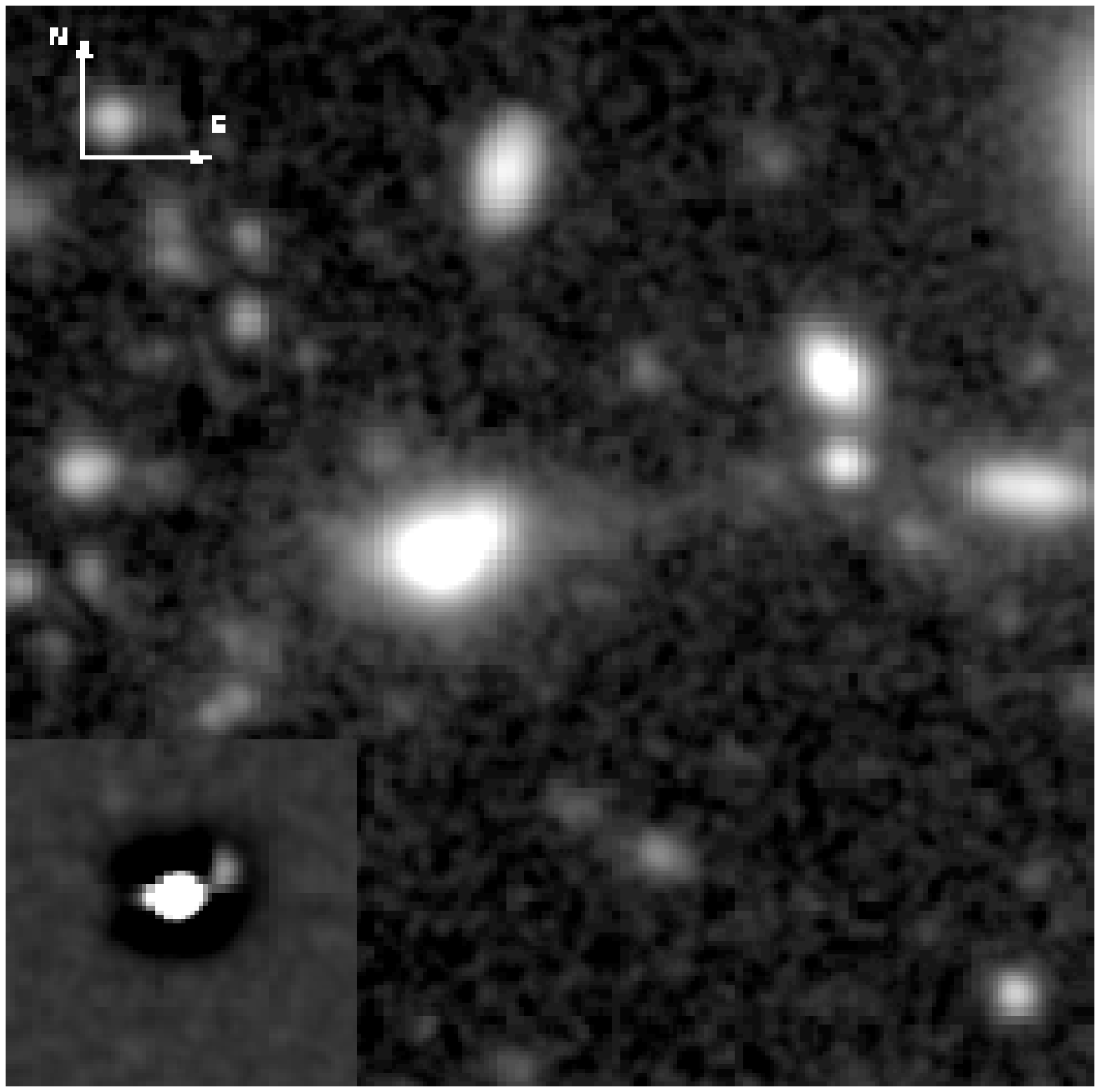}
\label{q0123}}
\subfigure[J0142+14, {O[L]}]{\includegraphics[width=7.5cm]{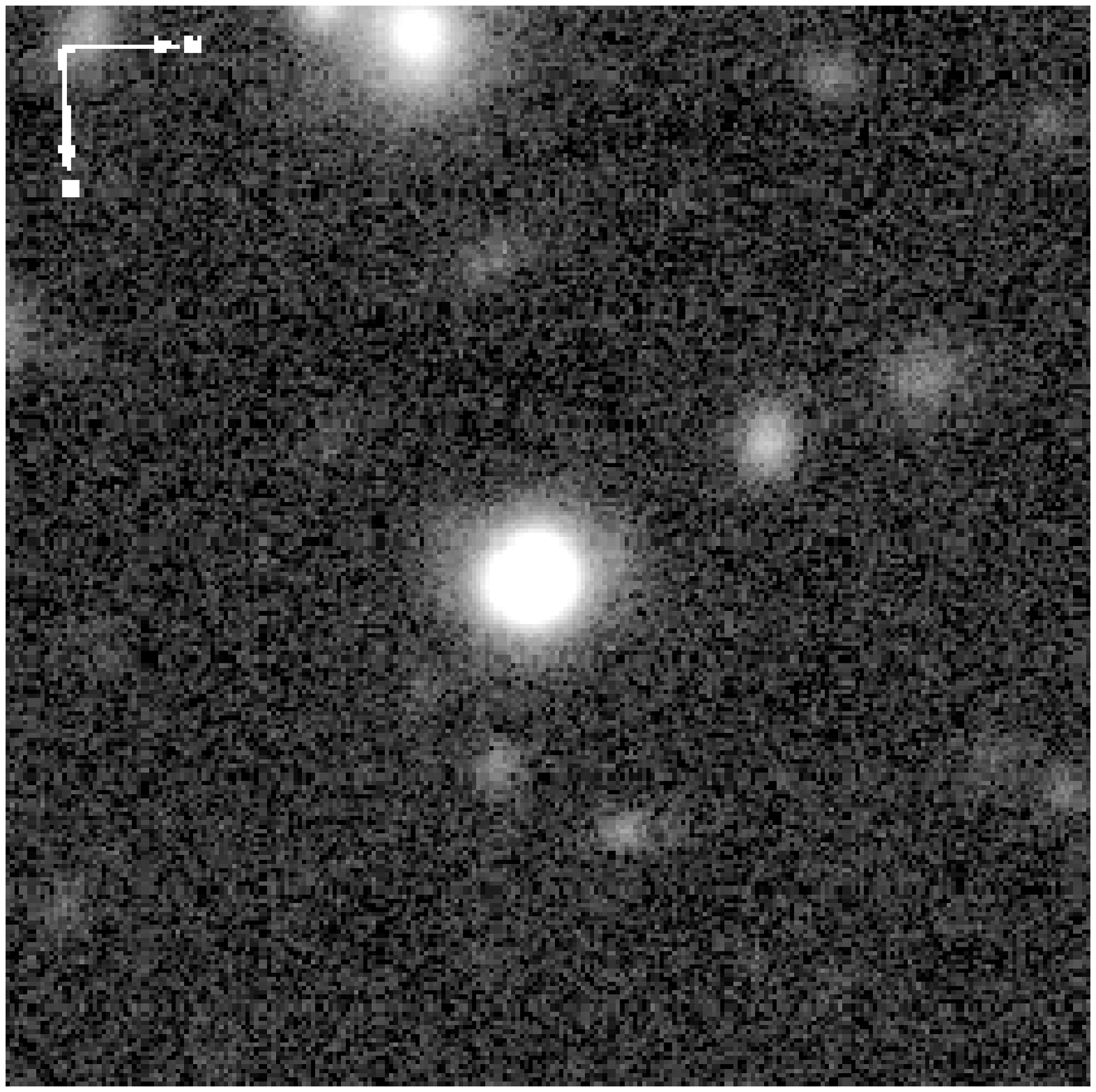}
\label{q0142}}
\subfigure[J0159+14, {O[L]}]{\includegraphics[width=7.5cm]{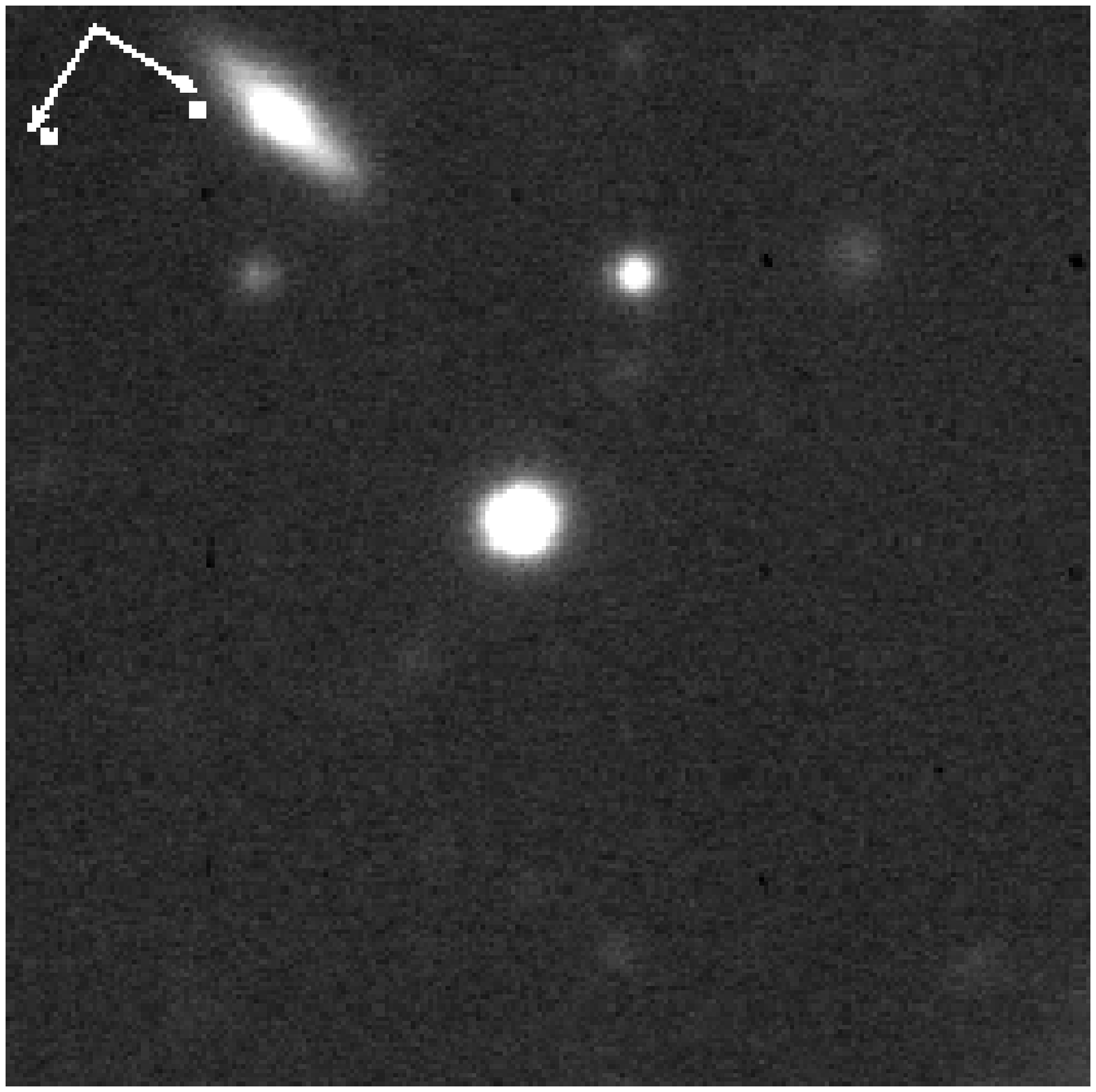}
\label{q0159}}
\subfigure[J0217-00, {O[L](U[L])}]{\includegraphics[width=7.5cm]{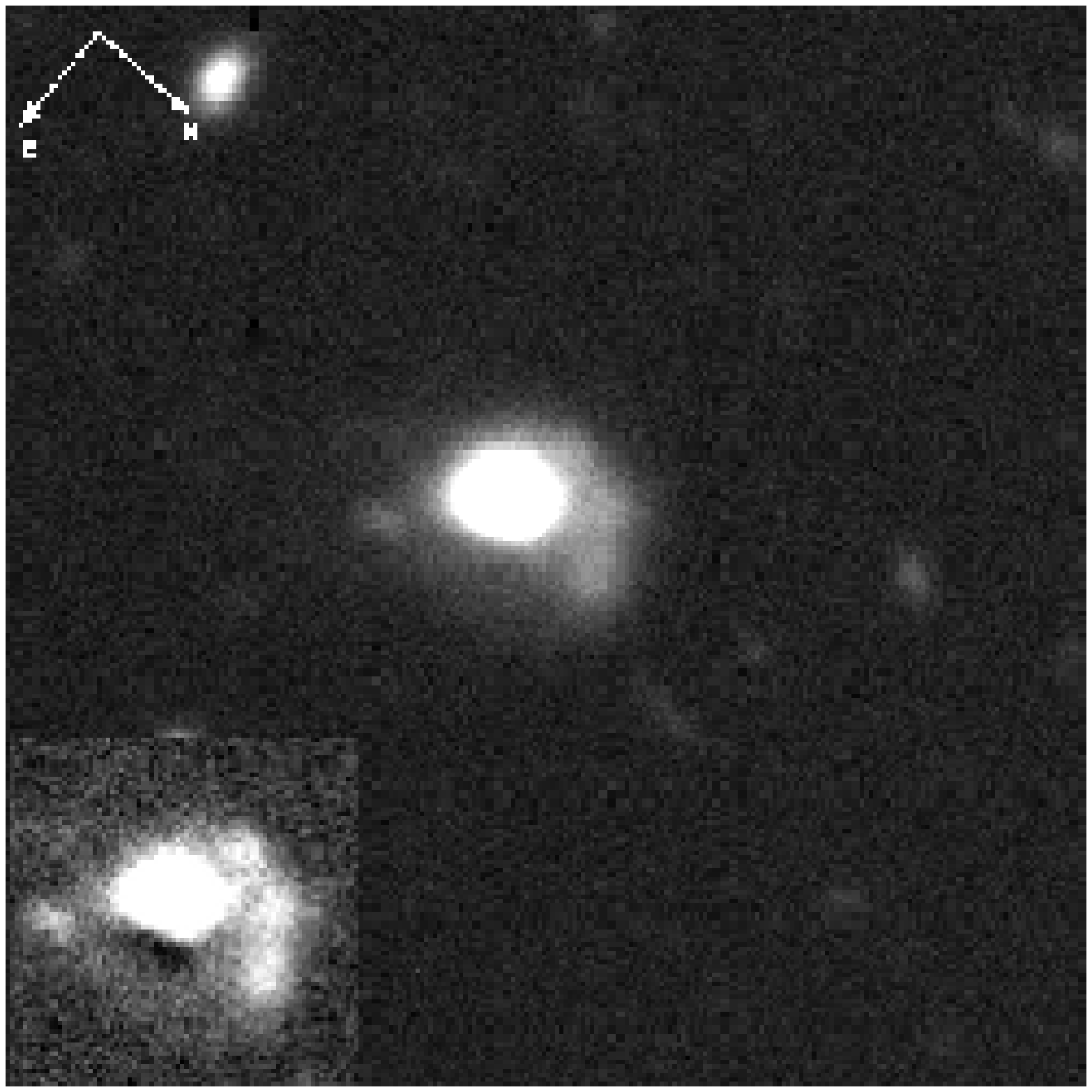}
\label{q21700}}
\end{figure*}

\begin{figure*}
\centering
\subfigure[J0217-01, {O[D]}]{\includegraphics[width=7.5cm]{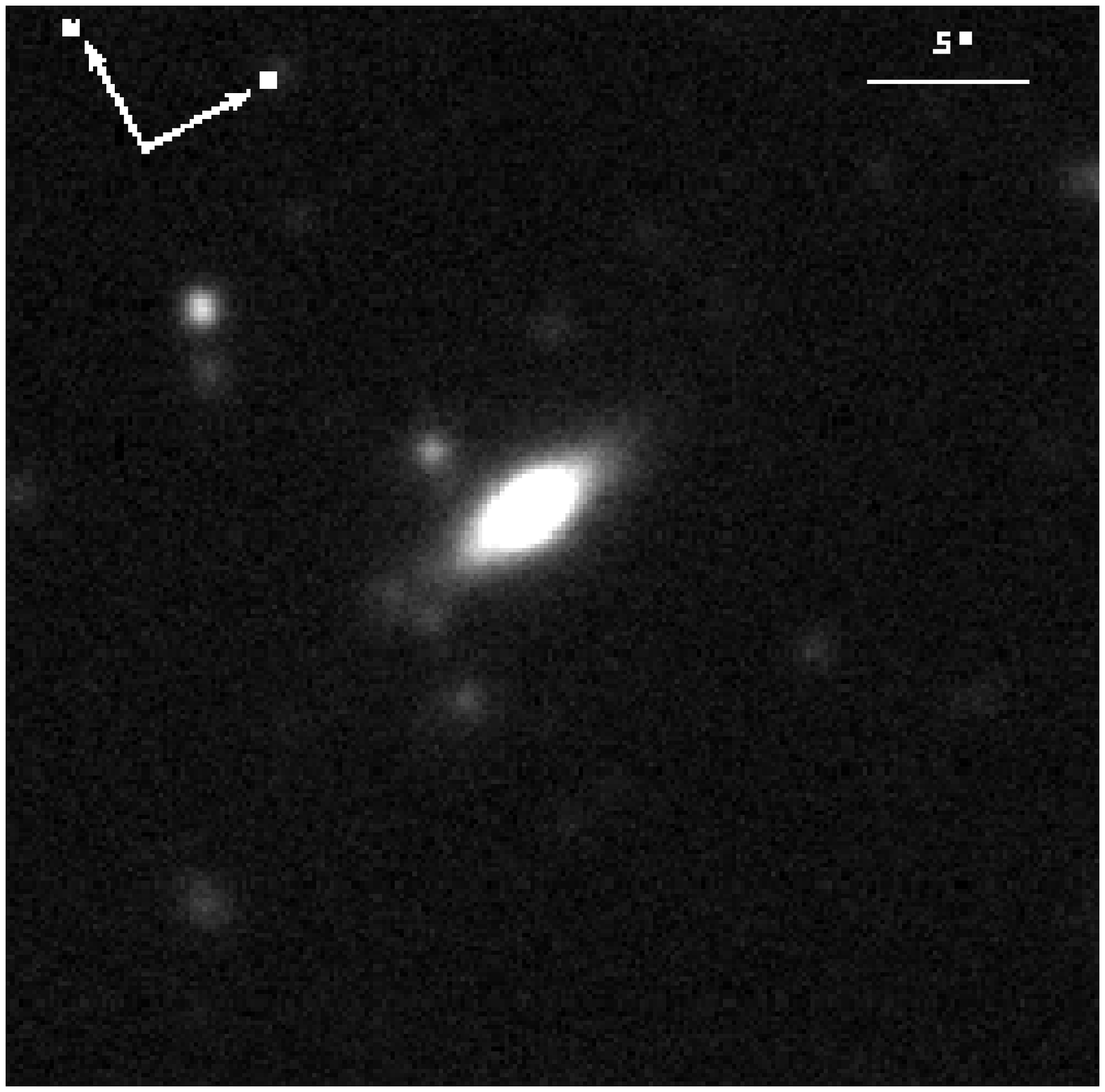}
\label{q021701}}
\subfigure[J0218-00, {O[D](M[D])}]{\includegraphics[width=7.5cm]{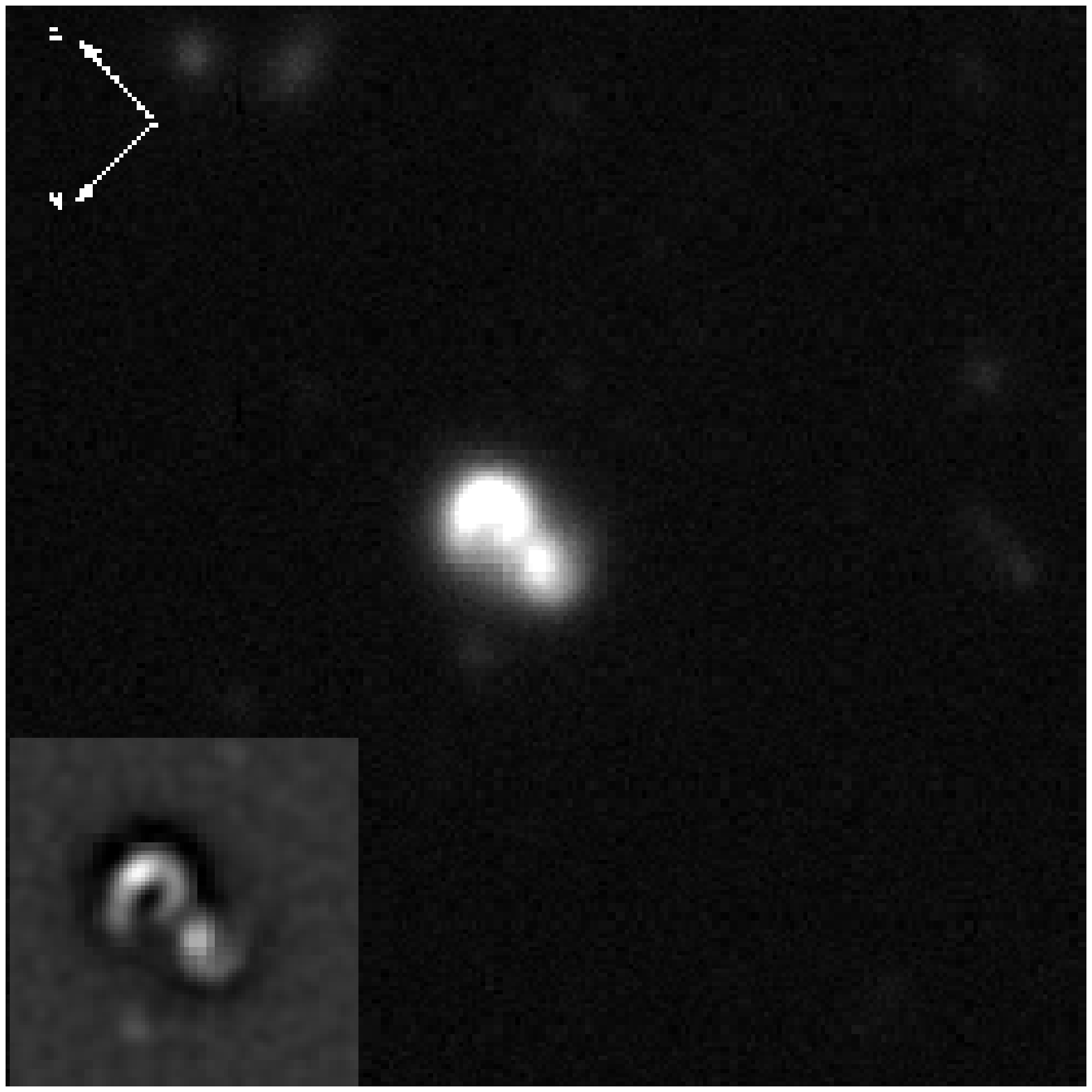}
\label{q0218}}
\subfigure[J0227+01, {M[D](U[D])}]{\includegraphics[width=7.5cm]{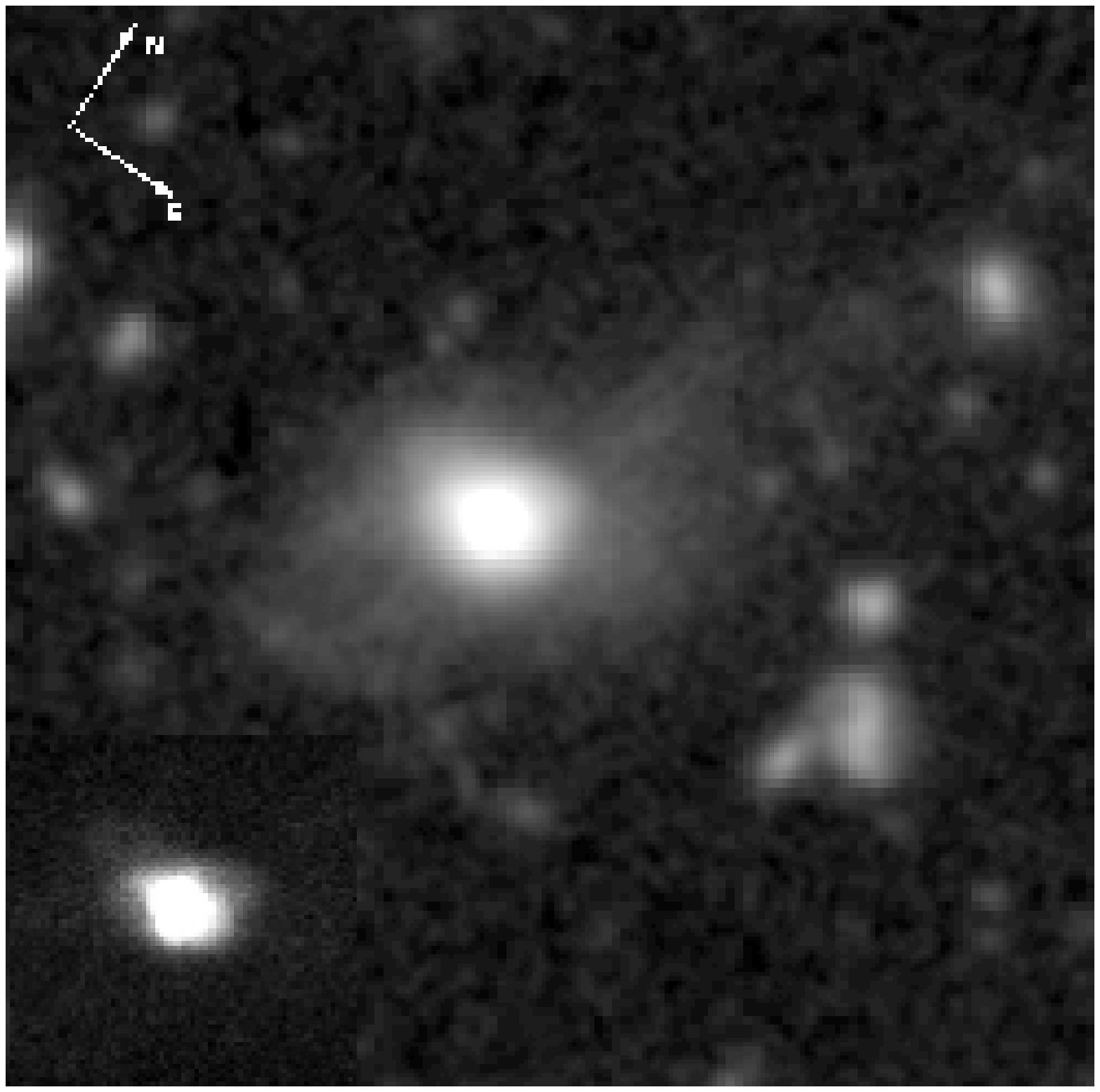}
\label{q0227}}
\subfigure[J0234-07, {O[L]}]{\includegraphics[width=7.5cm]{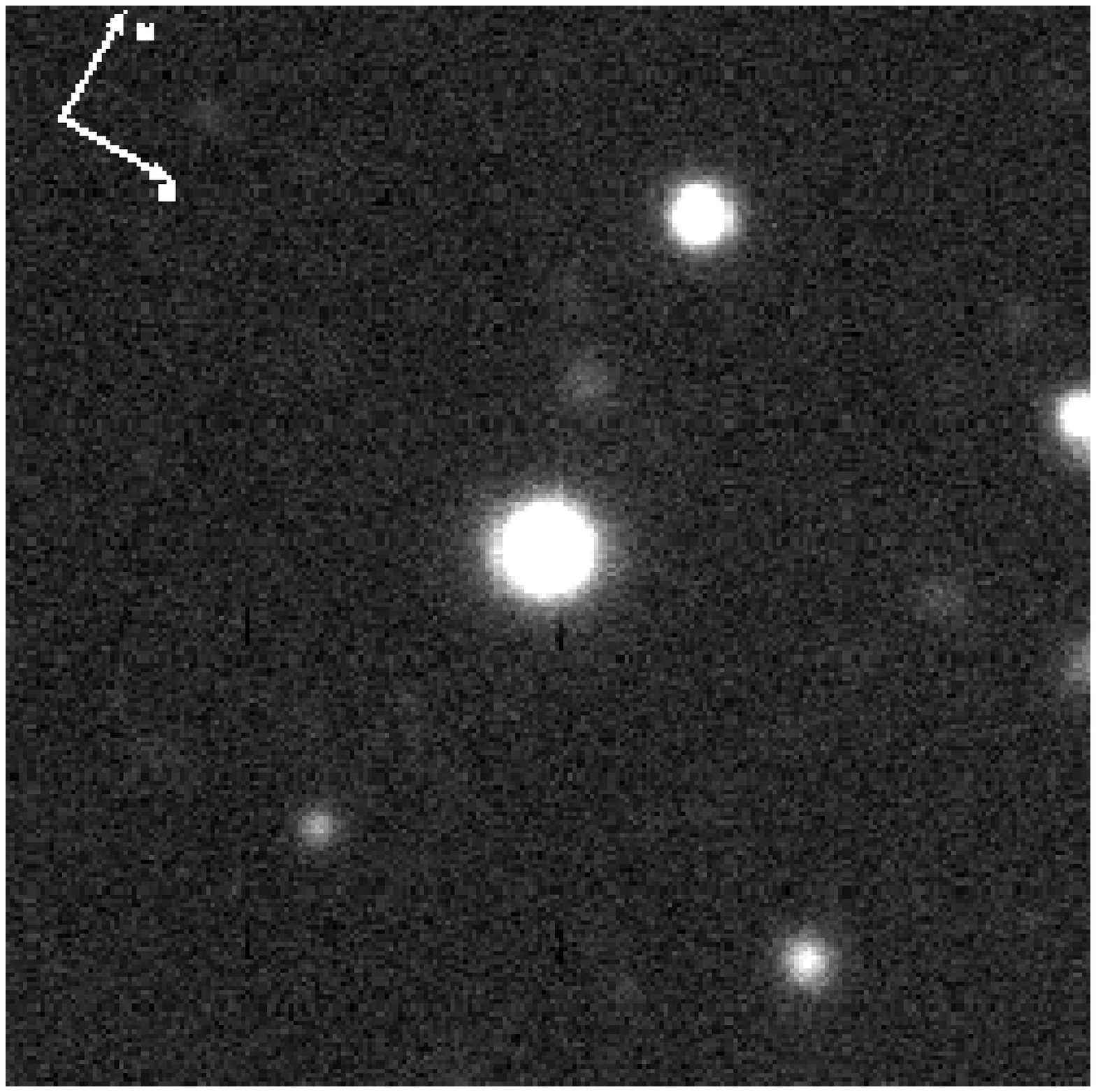}
\label{q0234}}
\subfigure[J0249+00, {M[L]}]{\includegraphics[width=7.5cm]{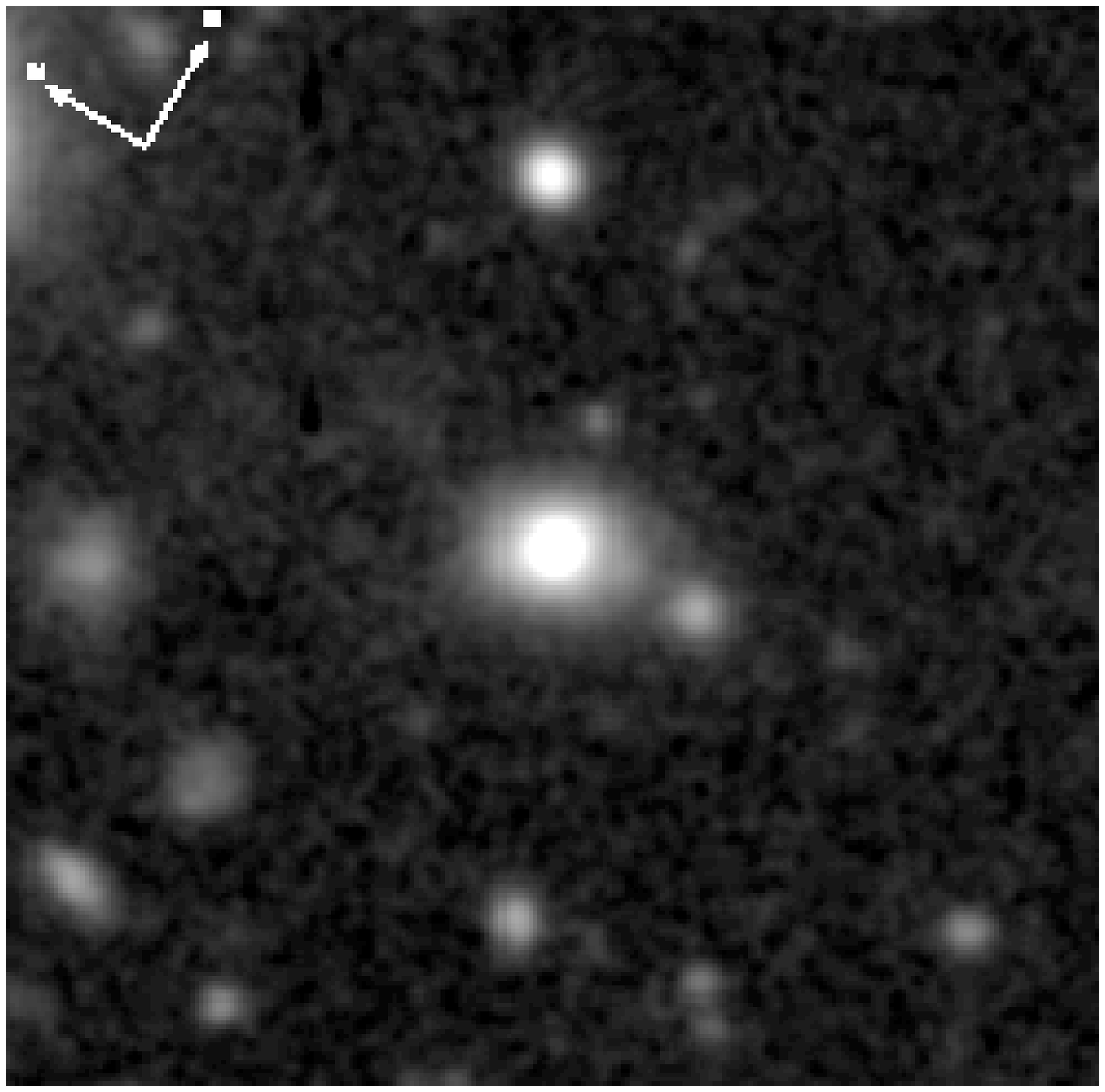}
\label{q0249}}
\subfigure[J0320+00, {O[L]}]{\includegraphics[width=7.5cm]{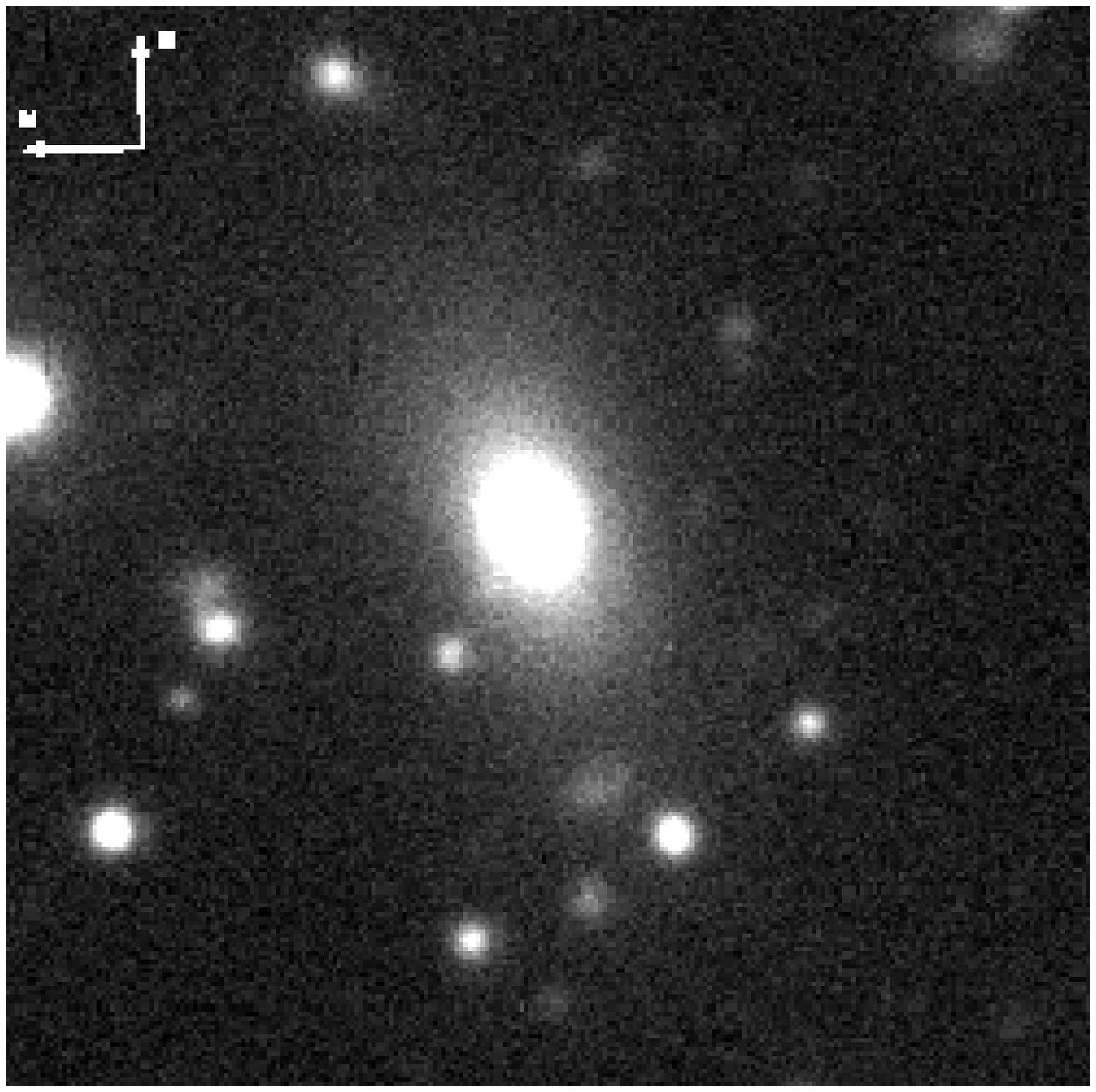}
\label{q0320}}
\end{figure*}

\begin{figure*}
\centering
\subfigure[J0332-00, {M[L](S[D])}]{\includegraphics[width=7.5cm]{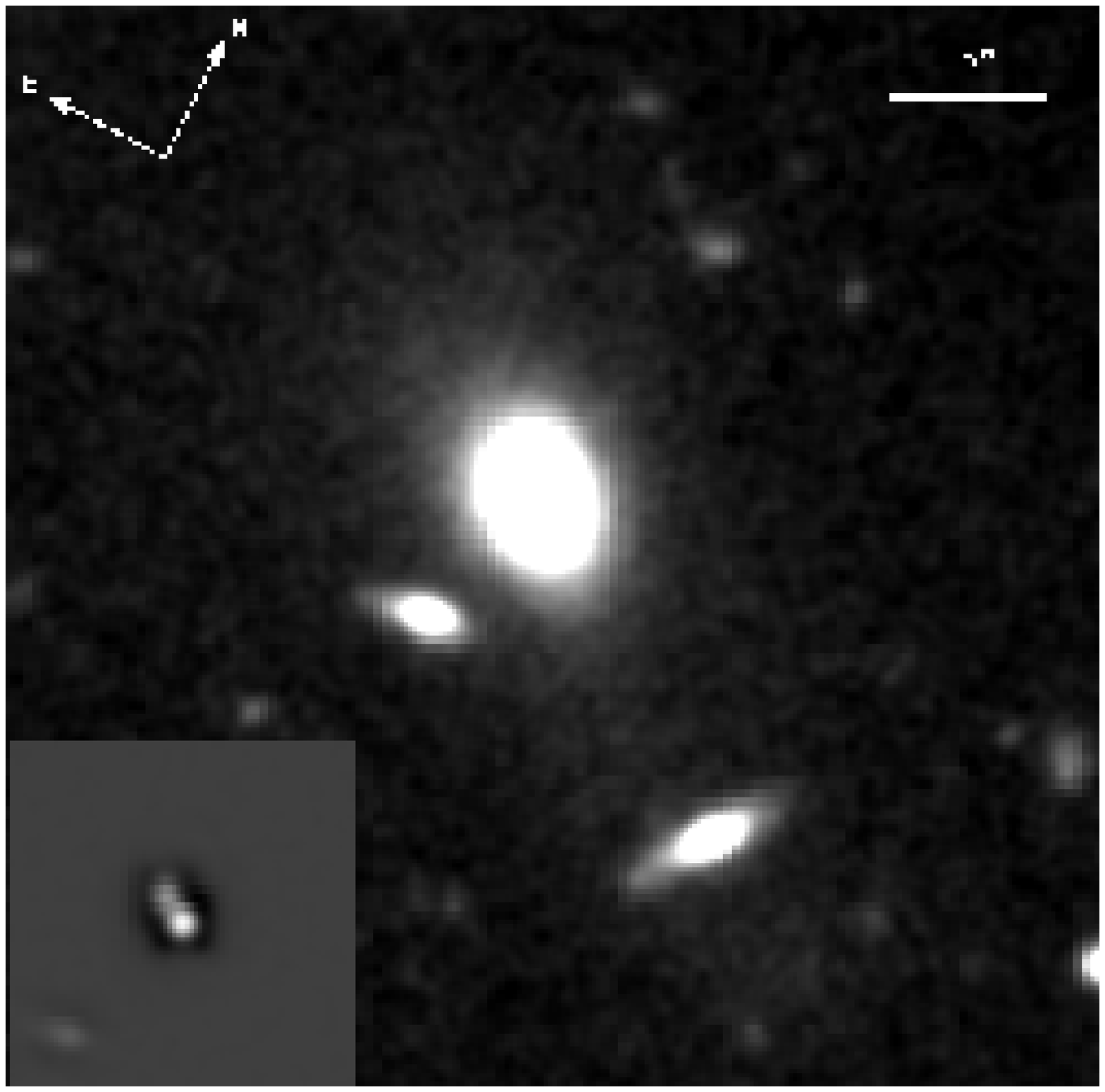}
\label{q0332}}
\subfigure[J0334+00, {M[L]}]{\includegraphics[width=7.5cm]{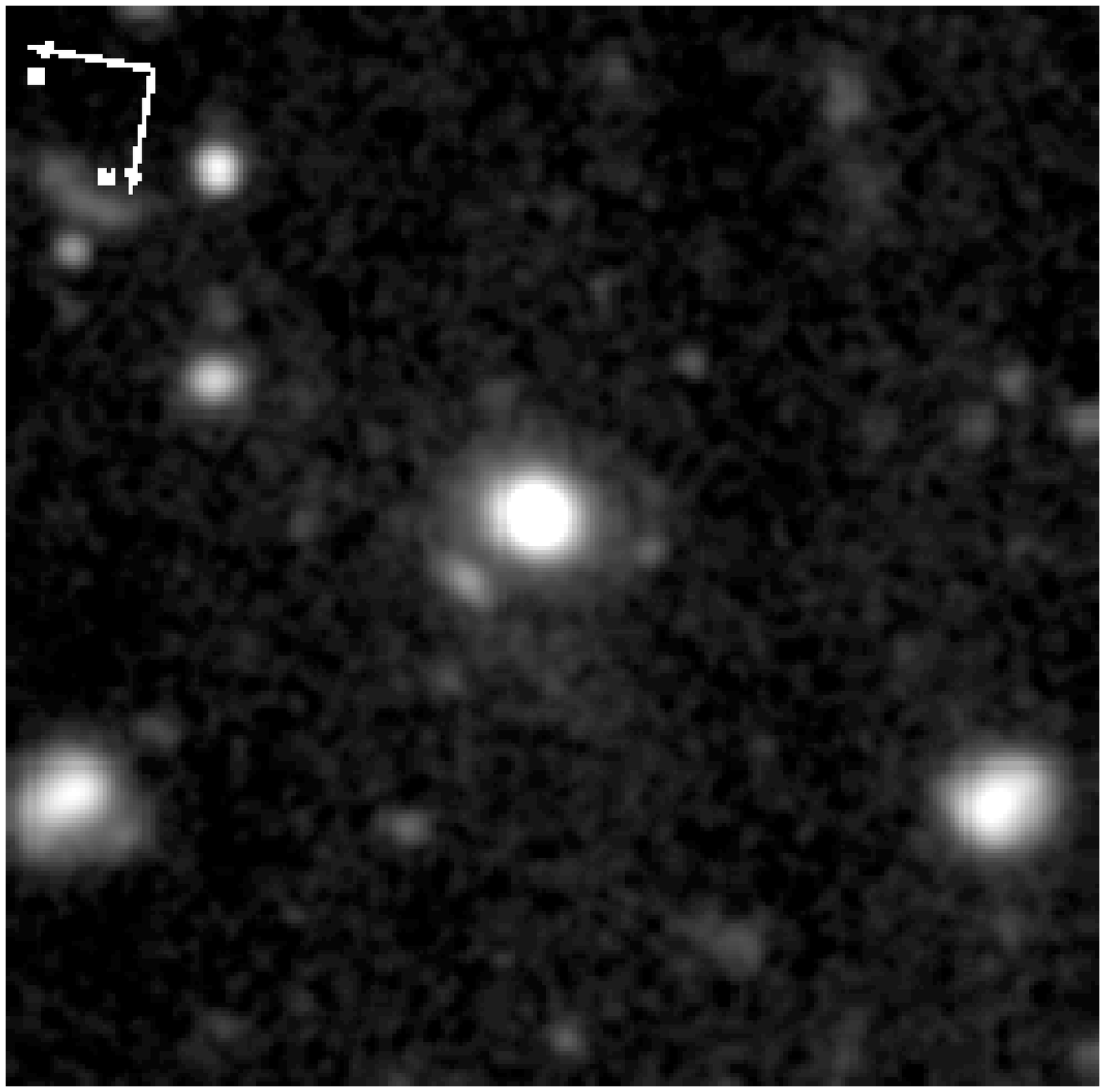}
\label{q0334}}
\subfigure[J0848+01, {O[L](U[L])}]{\includegraphics[width=7.5cm]{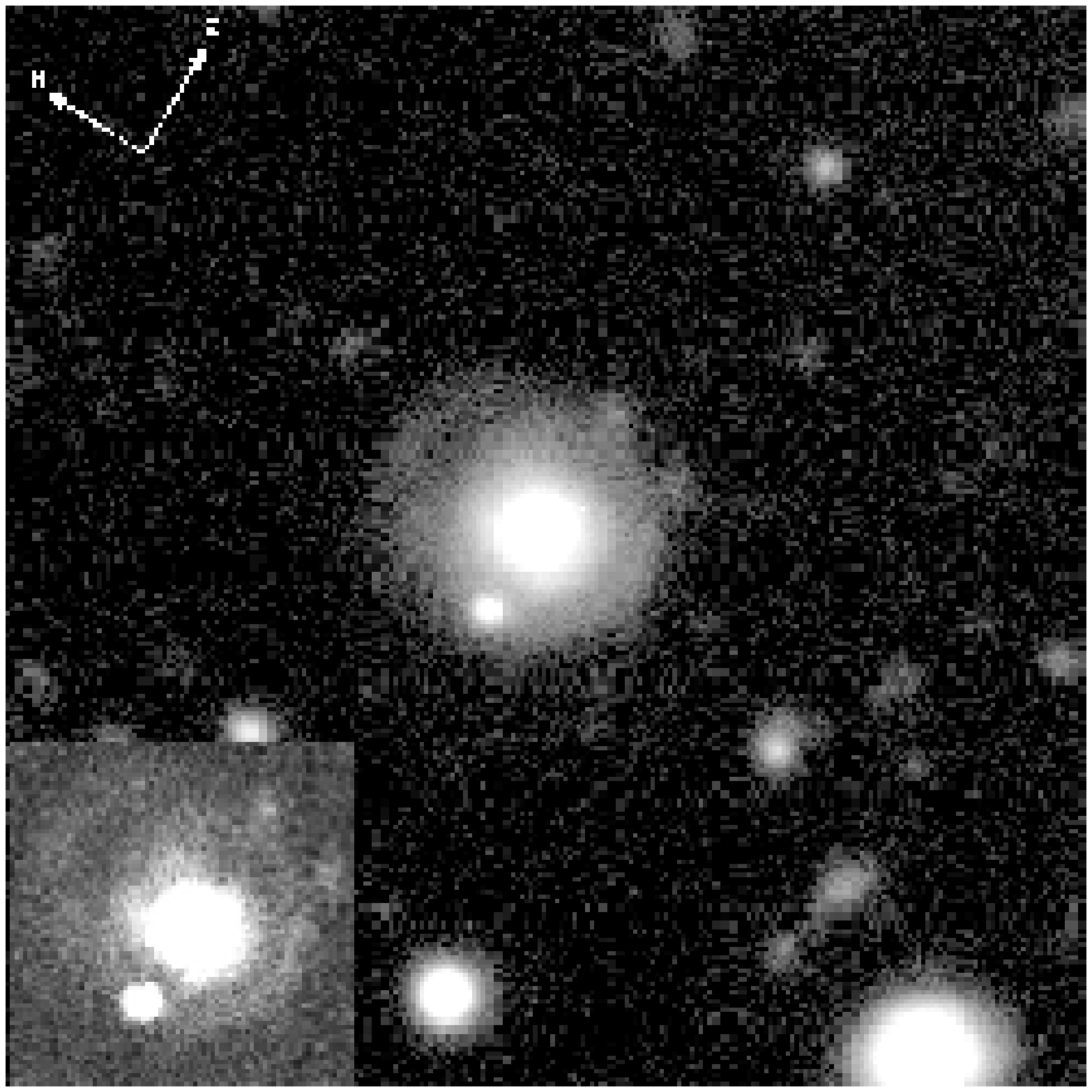}
\label{q0848}}
\subfigure[J0904-00, {O[L]}]{\includegraphics[width=7.5cm]{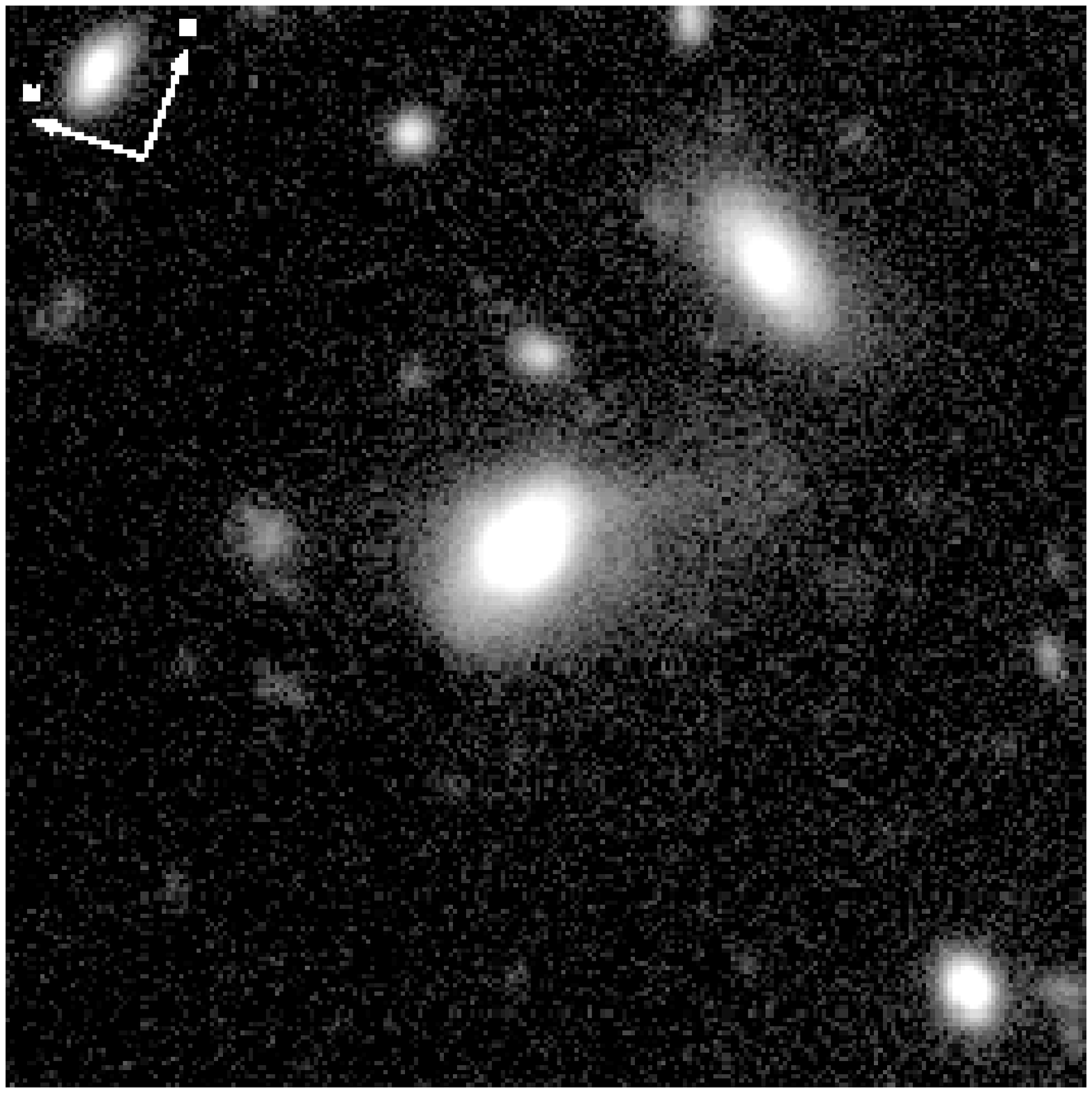}
\label{q0904}}
\subfigure[J0923+01, {M[L]}]{\includegraphics[width=7.5cm]{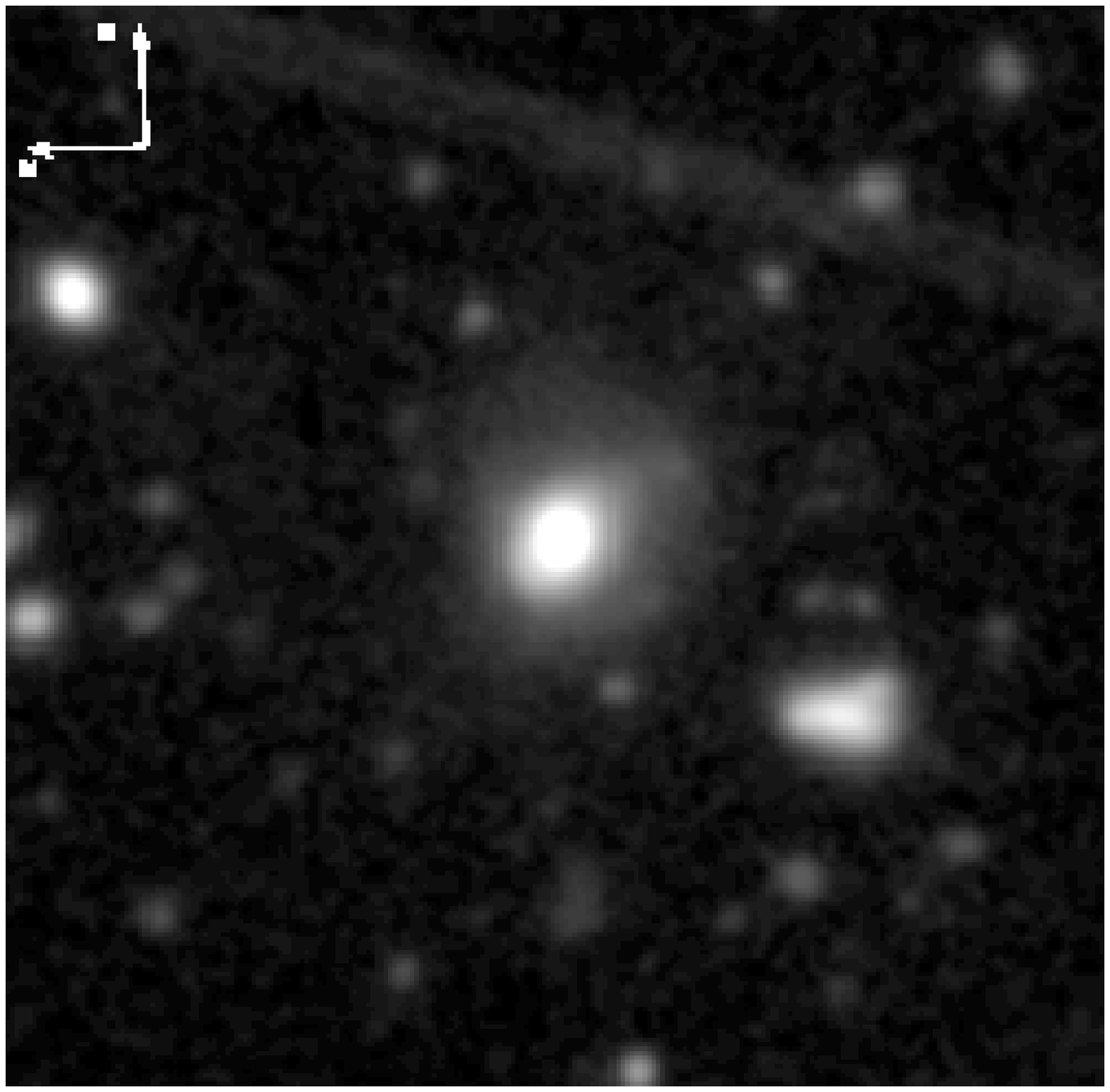}
\label{q0923}}
\subfigure[J0924+01, {M[L](S[D])}]{\includegraphics[width=7.5cm]{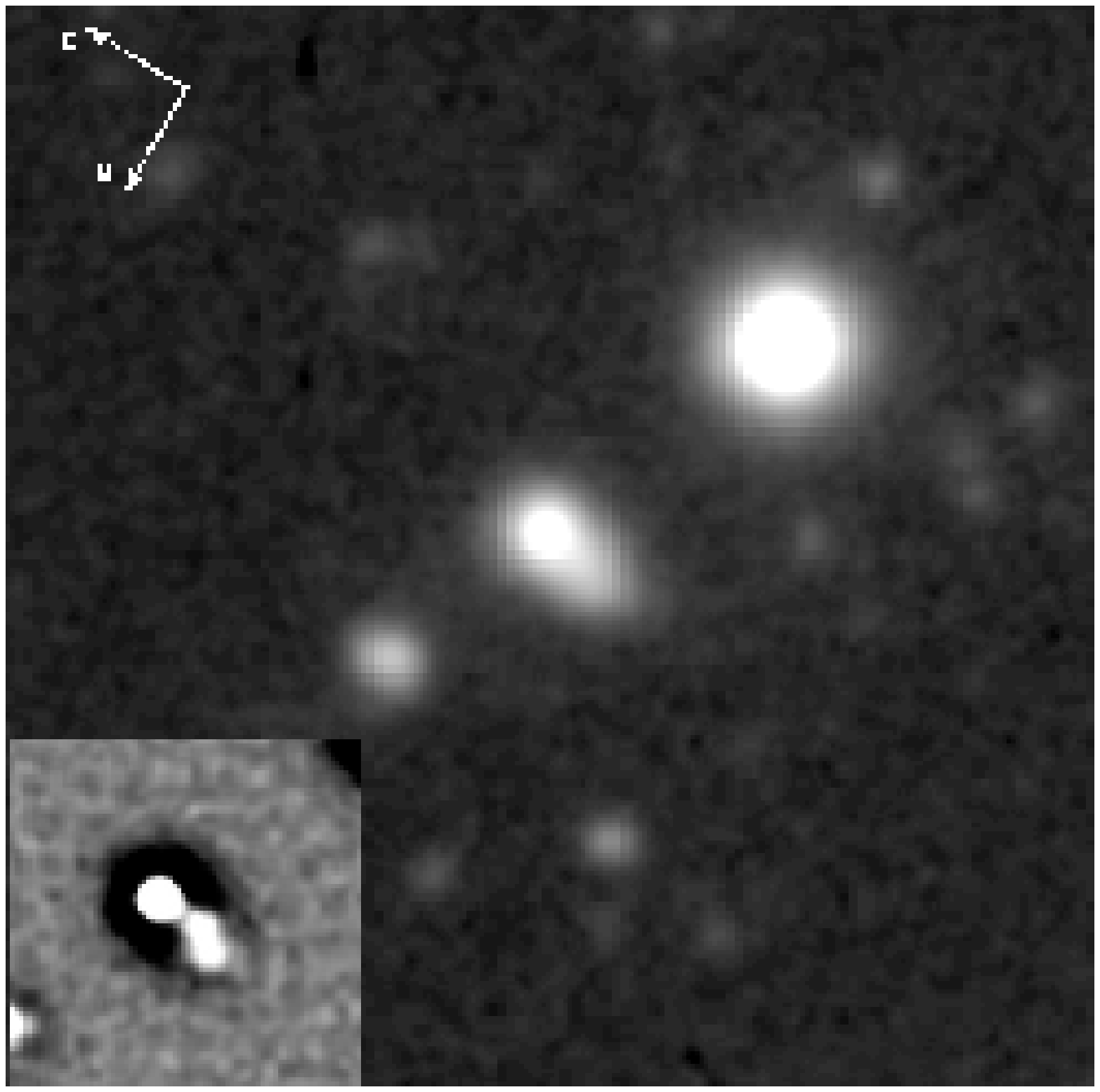}
\label{q0924}}
\end{figure*}

We used the fully reduced GMOS-S images for the detection of tidal features that provide evidence that the quasar host galaxy has been involved in an interaction or merger with another galaxy. The scheme used to classify the tidal features is based on that first used by \citet{heckman86} and is identical to that of RA11 and RA12. Summarising the different features, a tail (T) is a long, thin, usually radial structure (e.g. J2358-00 shown in Figure \ref{q2358}), whilst a fan (F) is shorter and broader than a tail (e.g. J0923+01 shown in Figure \ref{q0923}). A shell (S) is a curved feature that is tangential to the main body of the galaxy (e.g. J0848+01 shown in  Figure \ref{q0848}) and an amorphous halo (A) indicates an asymmetric halo with no definitive structure (e.g. J0123+00 and J0227+01 shown in Figures \ref{q0123} and \ref{q0227}). Bridges (B) extend between two galaxies linking them in tidal interactions and can extend over large distances (e.g. J2358-00 in Figure \ref{q2358}). Irregular (I) features do not fit into any of the above categories and can include features such as knots (e.g. J0217-00 in Figure \ref{q21700}). Double nuclei (2N) are those objects that have two brightness peaks within a projected separation of 10 kpc (e.g. J0332-00 in Figure \ref{q0332}).

Simulations have shown that the presence of all these features can be attributed to the host galaxy being involved in some kind of tidal encounter with another. This can take the form of mergers, both major and minor, or gravitational interactions that will not necessarily result in the coalescence of the two objects involved (\citealt*{quinn84, hernquist92, cattaneo05, lotz08, feldmann08}).

The detection of the features was made by eye, independently by three different people (PB, CT, CRA), and the results of each individual were then compared to confirm the detection of these features. This was necessary as, in the case of fainter features, detection and classification can be subjective. We found that in all cases, all three classifiers agreed on the presence or absence of tidal features. However, it was not always the case that all three agreed on the detailed classification of the feature (e.g. fan or shell). In such cases, PB determined the final classification.

One of the main aims of this study is to compare the results obtained for type II quasars with those of the control sample of intermediate redshift, early-type galaxies presented in RA12 and the PRG of RA11. Therefore, in order that our results be directly comparable, we have measured the surface brightnesses of the detected features of the type II quasar host galaxies in an identical manner to that applied in the latter study. We first calculated the average number of counts for each feature by taking several measurements using small apertures, the size of which was dependent on the extent of the feature. Using the same aperture, we then repeated exactly the same process in several regions local to the feature in order to subtract the contribution from the sky and the diffuse light from the galaxy. Because the results obtained from this technique are somewhat dependent on the exact placement of the apertures, we tested its reliability by CRA and PB making independent measurements of the same features for a number of the quiescent early types of RA12. In these cases, the difference in the surface brightness obtained was below 0.1 mag. We then corrected for Galactic extinction, $(1+z)^4$ surface brightness dimming and applied k-corrections. The derived values of surface brightness are given in Table \ref{surf_bright}. The purpose of measuring the surface brightness of the features is to determine the nature of the merger. For example, simulations show that morphological signatures produced in gas rich interactions are brighter than those produced in gas poor ones (\citealt*{naab06, bell06, mcintosh08}).

Once the initial classification was made, we applied three separate enhancement techniques to the images, using the Interactive Development Language (IDL), for the purposes of confirming the features detected in the original images, and detecting fainter features that were not evident in the unprocessed images. Note however, the detection and classification of the tidal features listed in Table \ref{surf_bright} is based purely on the unenhanced images. The three techniques employed are unsharp masking, image filtering and smoothed galaxy subtraction. Full details of how these techniques are applied can be found in RA11.

\section{Notes on Individual Objects \& Type II Quasar Images.}
\label{images}

\begin{figure*}
\centering
\subfigure[J0948+00, {O[L]}]{\includegraphics[width=7.5cm]{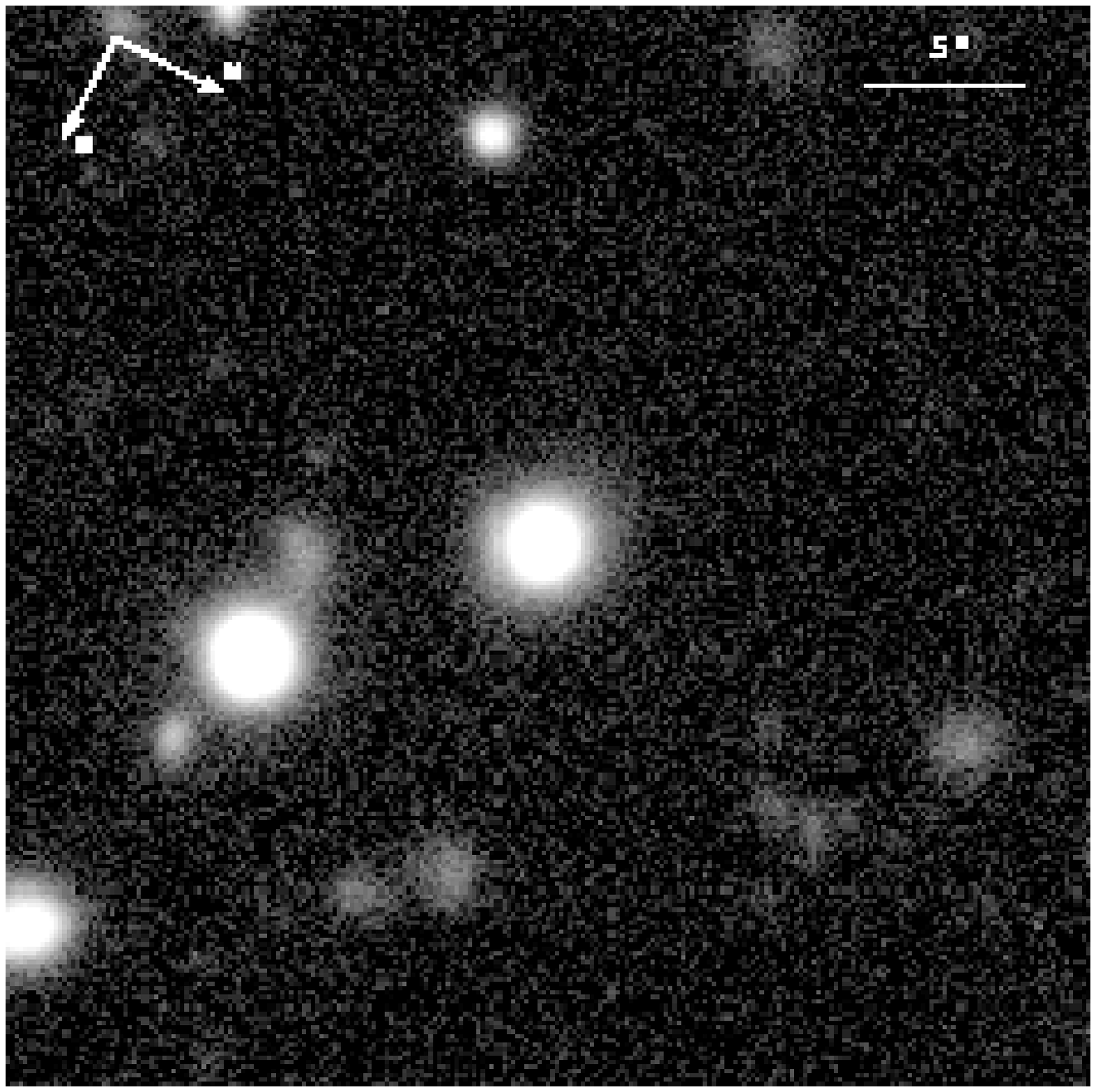}
\label{q0948}}
\subfigure[J2358-00, {M[D](U[D])}]{\includegraphics[width=7.5cm]{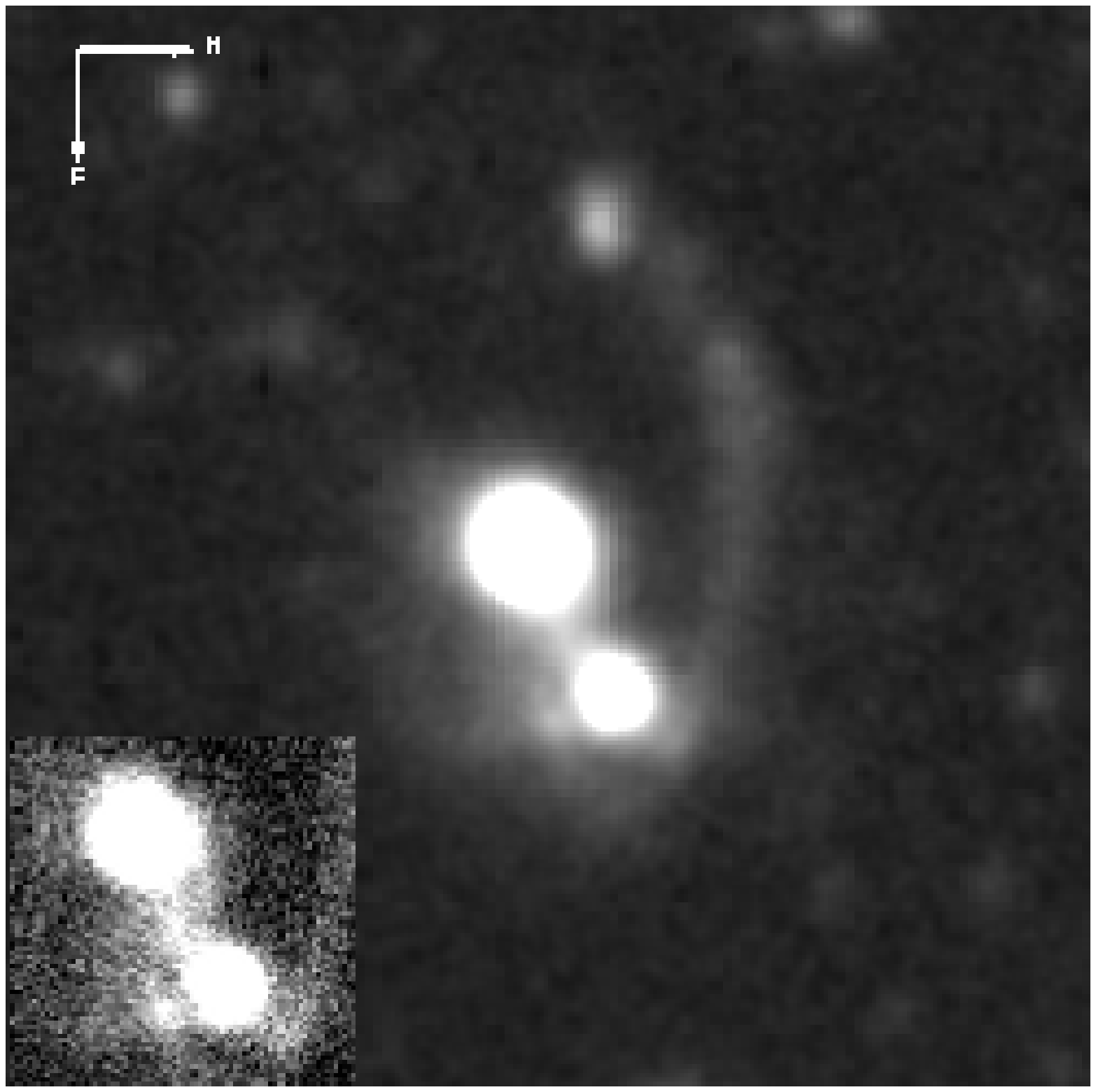}
\label{q2358}}
\caption{Gemini GMOS-S images of all 20 type II quasar host galaxies in this sample. All the images are $35\times35 $ arcsec, and the first image on each page shows the scale bar. All insert images are also to the same scale. The image processing technique used, if any, is given by the code beneath the image along with its abbreviated name. O: Original, unprocessed image, U: Unsharp mask, M: Median Filtered and S: Smoothed galaxy subtracted. For those images that include inserts, the code is given in brackets. The scale used for the image stretching is given in square brackets, L: log scale and D: default linear scale. We advise the reader that the images are best viewed on a computer screen.}
\label{all_images}
\end{figure*}

In order to improve the presentation of the images, we have applied all three image enhancement techniques mentioned in Section \ref{da} to our Gemini GMOS-S images. Figure \ref{all_images} shows the best resulting image for each object. Each main image is $35\times 35$ arcsec and the inserts are to the same scale. The captions for the sub-figures give details of the enhancement technique, if any, used in each case. The scale adopted for the image stretching is also given for each object, and the key is given in the main caption for the figure.             

\subsection{J0025-10}
In Figure \ref{q0025}, our Gemini GMOS-S image shows that this system is clearly interacting, since it has a double nucleus, which can be seen more clearly in the insert, with a projected separation of 4.5kpc. The north eastern nucleus is the quasar host. The galaxy also has two distinct tails, with the shorter, brighter tail pointing towards the north east and a longer tail pointing towards the south east. The VLT FORS2 long-slit spectroscopy of \citet{villar11} along PA0 and PA60 shows that the companion has a velocity shift of $0\pm70~km~s^{-1}$ relative to the quasar nucleus, and demonstrates that the south west nucleus and the two tidal tails are dominated by continuum rather than line emission.

\subsection{J0114+00}
Our deep Gemini GMOS-S image of this object, displayed in Figure \ref{q0114}, shows that this system has a second fainter nucleus to the north west at a projected separation of 8.0 kpc, which can be seen more clearly in the insert. There is also a shell extending to the south east with a surface brightness of $\mu_r^{corr} \simeq 23.1~mag~arcsec^{-2}$. 

\subsection{J0123+00}
Our deep image, displayed in Figure \ref{q0123}, clearly reveals that this quasar host galaxy is involved in an interaction. We can see a secondary nucleus at a projected separation of 7.6 kpc from the centre of the host galaxy. However, on the basis of the long-slit spectroscopy published in \citet{villar10}, it is likely that this feature is dominated by strong [OIII] line emission. There is also a tidal tail/bridge feature extending to the east with a surface brightness $\mu_r^{corr} \simeq 24.7~mag~arcsec^{-2}$ to a distance of 62 kpc from the centre of the host galaxy. \citet{villar10} also detect this extended feature in their narrow-band image centred on [OIII] $\lambda5007$, demonstrating that it leads to to a companion galaxy at the same redshift as J0123+00, and 100 kpc from its nucleus. In this object it is possible that all the extended tidal features are emission-line rather than continuum dominated.

\subsection{J0142+14}
Our deep Gemini GMOS-S image of this object, presented in Figure \ref{q0142}, shows that this quasar is hosted by a compact galaxy with no signs of morphological disturbance attributable to interactions or merger activity.

\subsection{J0159+14}
Figure \ref{q0159} shows our deep Gemini GMOS-S image of this object, which does not reveal any clear evidence of an interaction or merger event, suggesting that the quasar is hosted by a compact galaxy. However, the median filtered image does show some evidence for a low surface brightness tail/bridge feature linking J0159+14 to a faint companion galaxy 27 kpc to the north. Note however that as this feature was not detected in the unprocessed image, we do not include it in our classification.

\subsection{J0217-00}
This object is amongst the most spectacular in our sample, with clear signs of morphological disturbance indicating a merger or interaction (Figure \ref{q21700}). In the north west there is a tidal tail which appears to comprise of three knots, with a measured surface brightness of $\mu_r^{corr} \simeq 22.1~mag~arcsec^{-2}$. The long-slit VLT spectroscopy of \citet{villar11} along PA116 shows that this is a continuum emitting feature. There is also an irregular feature in the south-east, with a surface brightness of $\mu_r^{corr} \simeq 22.6~mag~arcsec^{-2}$, that also seems to take the form of a knot. \citet{villar11} confirm that this knot is at the same redshift as the galaxy and is quite probably a knot of star formation. The faintest feature is the shell on the north-east side which has a surface brightness of $\mu_r^{corr} \simeq 23.7~mag~arcsec^{-2}$.

\subsection{J0217-01}
It can be seen from the image presented in Figure \ref{q021701} that the quasar appears to reside in an undisturbed disc galaxy. We detect no signs of interactions or mergers, apart from what may be some faint companion galaxies at the western end of the disc. It appears that we are viewing this galaxy close to edge-on.

\subsection{J0218-00}
Figure \ref{q0218} clearly shows that this quasar host galaxy is one of the most tidally disturbed in the sample, with a shape reminiscent of a question mark. The quasar is hosted by the larger eastern galaxy of the pair, which has a projected separation of 12 kpc, and appears to be stretched and curled around itself. This arc-like tidal feature is the brightest in the sample with a mean surface brightness of $\mu_r^{corr} \simeq 20.6~mag~arcsec^{-2}$ There is also an irregular feature in the north west with a surface brightness $\mu_r^{corr} \simeq 22.6~mag~arcsec^{-2}$. Our Gemini long-slit spectrum of this object (PA171: \citealt{bessiere12}) passes through the eastern AGN and associated arc feature. We find to be continuum- rather than emission-line dominated.

\subsection{J0227+01}
This quasar host galaxy, shown in Figure \ref{q0227}, has a highly distorted appearance, with an amorphous halo and two overlapping shells, one in the east and the other in the south. The halo is the brightest feature and extends to 30 kpc from the centre of the galaxy, with surface brightness $\mu_r^{corr} \simeq 22.4~mag~arcsec^{-2}$. The southern shell has surface brightness $\mu_r^{corr} \simeq 24.1~mag~arcsec^{-2}$ and the eastern $\mu_r^{corr} \simeq 24.4~mag~arcsec^{-2}$. There is also a tail pointing towards the north to a projected distance of 55 kpc with a surface brightness $\mu_r^{corr} \simeq 24.6~mag~arcsec^{-2}$.

\subsection{J0234-07}
This quasar appears to be hosted by a compact galaxy. Our deep Gemini GMOS-S image, (Figure \ref{q0234}), shows no signs of morphological disturbance that could be attributed to an interaction or merger.

\subsection{J0249+00}
Our deep GMOS-S image (Figure \ref{q0249}) of this object shows a shell to the south of the host galaxy with surface brightness $\mu_r^{corr} \simeq 23.1~mag~arcsec^{-2}$, that extends in the direction of a companion object. The companion is also to the south, with a projected separation of 28 kpc, and the two galaxies are linked by a faint tidal bridge of $\mu_r^{corr} \simeq 24.0~mag~arcsec^{-2}$. Our Gemini long-slit spectrum (PA0: \citealt{bessiere12}) cuts through both the southern arc and the edge of the companion galaxy, but shows not strong [OIII] emission associated with either of these features, which are likely to be continuum-dominated.

\subsection{J0320+00}
This quasar host galaxy, shown in Figure \ref{q0320}, has an extended asymmetric halo in the north east/south west direction, with mean surface brightness $\mu_r^{corr} \simeq 23.5~mag~arcsec^{-2}$. It is not possible to define any clear structure within the halo itself, although there is a hint of a shell to the north east. This latter feature appears more prominent in our enhanced images, although we do not use it for the purposes of classification. There is no evidence that the brighter knot in the north west, which appears to be within the halo, is associated with the galaxy.

\subsection{J0332-00}
Figure \ref{q0332} clearly shows that this object has a double nucleus, with the quasar hosted by the western nucleus, and the second at a projected separation of 4 kpc to the east. There is a shell to the south west and also a fan in the north east direction that have surface brightnesses of $\mu_r^{corr} \simeq 22.3~mag~arcsec^{-2}$ and $\mu_r^{corr} \simeq 24.2~mag~arcsec^{-2}$ respectively, as well as what may be a diffuse tidal tail/bridge that extends in the direction of a disc galaxy in the south west, which is at a projected separation of 53 kpc. However, the latter galaxy does not appear to be tidally distorted, and we have no evidence that it is at the same redshift as J0332-00. Our Gemini long-slit spectrum (PA32: \citealt{bessiere12}), centred on the western nucleus, cuts through the shell to the south west as well as edge of the fainter nucleus to the east, but neither of these structures shows strong emission lines, suggesting that they are dominated by continuum emission.

\subsection{J0334+00}
Our deep image of this object (Figure \ref{q0334}), reveals a faint shell to the west of the host galaxy with $\mu_r^{corr} \simeq 24.1~mag~arcsec^{-2}$. There is also an object at a projected separation of 18 kpc to the north east, but because there is no obvious sign of co-aligned tidal features or a bridge, we do no consider it to be a companion.

\subsection{J0848+01}
Our Gemini GMOS-S image of this object, shown in Figure \ref{q0848}, reveals two clear shells, the brighter in the north with $\mu_r^{corr} \simeq 23.2~mag~arcsec^{-2}$ and the slightly dimmer in the south east with $\mu_r^{corr} \simeq 23.5~mag~arcsec^{-2}$. There is also a  smaller object to the west at a projected separation of 14 kpc, but we have no independent evidence that this is associated with the host galaxy. Our Gemini long-slit spectrum (PA174: \citealt{bessiere12}) shows no strong [OIII] emission associated with the northern shell, which is dominated by continuum emission.

\subsection{J0904-00}
This object, shown in Figure \ref{q0904}, has a tail in the south east direction, extending from the centre of the galaxy to a distance of 45 kpc. This tail becomes more diffuse along its length, and has a kink at its end pointing towards the east. The mean surface brightness taken along the entire length of the feature is $\mu_r^{corr}\simeq 23.2~mag~arcsec^{-2}$. There is also a  shell that appears to start at the  base of the tail and wraps $\sim$180 degrees around the galaxy. At its brightest, to the north west of the galaxy, this shell feature has a surface brightness of $\mu_r^{corr} \simeq 22.2~mag~arcsec^{-2}$. Our Gemini long-slit spectrum (PA280: \citealt{bessiere12}) shows no strong [OIII] emission specifically associated with the latter feature, which is likely to be continuum dominated.

\subsection{J0923+01}
Our Gemini GMOS-S image (Figure \ref{q0923}) shows that this is a morphologically disturbed host galaxy with a tail/fan feature to the south east that terminates in a shell ($\mu_r^{corr} \simeq 24.0~mag~arcsec^{-2}$) 27 kpc from the nucleus. There is also a shell feature to the south west of the nucleus with surface brightness $\mu_r^{corr}\simeq 23.4~mag~arcsec^{-2}$. Our Gemini long-slit spectrum (PA0: \citealt{bessiere12}) shows no strong [OIII] emission associated with the latter feature, which is likely to be continuum dominated.

\subsection{J0924+01}
The image of this host galaxy (Figure \ref{q0924}), has a prominent tidal tail pointing south to a projected distance of 17 kpc, with a surface brightness $\mu_r^{corr} \simeq 21.6~mag~arcsec^{-2}$.  Interestingly, the median filtered-image shown in Figure \ref{q0924}, reveals a faint bridge connecting J0924+01 to a companion galaxy 34 kpc to the north, which itself has a tidal tail that points away from the quasar host. However, since the latter features are not visible in the unenhanced image, we have not included them in the morphological classification presented in Table \ref{surf_bright}.

\subsection{J0948+00}
Figure \ref{q0948} shows that our deep Gemini GMOS-S image of this object does not revel any indication of morphological disturbance in the host galaxy of this type II quasar. It appears to be an undisturbed elliptical galaxy.

\subsection{J2358-00}
Our image of this system (Figure \ref{q2358}) reveals what is, without a doubt, the most spectacular quasar host galaxy in the sample. Here we have clear evidence of an interaction, possibly leading to a merger, with the two galaxies at a projected separation of 30 kpc. The quasar is hosted by the western galaxy in the pair and the two galaxies are linked by a bridge with a measured surface brightness of $\mu_r^{corr} \simeq 22.7~mag~arcsec^{-2}$, which can be seen more clearly in the insert. There is a long tidal tail extending from the companion to a projected distance of 80 kpc in the north west with a measured surface brightness of $\mu_r^{corr} \simeq 23.3~mag~arcsec^{-2}$, and also a fan in the south with surface brightness $\mu_r^{corr} \simeq 23.4~mag~arcsec^{-2}$. These features are also detected in HST images presented by \citet{zakamska06}, which show more detail within the main structures.

VLT long-slit spectroscopy by \cite{villar11} along PA60 and PA83, shows that the companion galaxy and tidal bridge are dominated by continuum emission. Moreover, our Gemini long-slit spectrum (PA331: \citealt{bessiere12}) shows no significant [OIII] emission where the slit intercepts the tidal tail to the north west of the quasar host galaxy.  \cite{villar11} also demonstrate that the companion has a velocity shift of $300\pm80~km~s^{-1}$ relative to the quasar, which is within the range expected for merging systems.  

\section{Results}
\label{results}

\begin{table*}
\centering
\begin{minipage}{140mm}
  
\caption{Full classification of the sample objects and the surface brightness of the detected morphological features: Column 1 gives the abbreviated SDSS identifier. Columns 2 gives the surface brightness dimming (mag arcsec$^{-2}$) taken from the NED database. Column 3 shows our morphological classification, with the features shown in square brackets not being considered as secure detections. These features are not considered in any of the analysis and are for information only. The morphological classification are 2N: double nuclei, T: tail, S: shell, A: amorphous halo, F: fan, B: bridge, I: irregular feature. Column 4 shows the surface brightness measured from the images and column 5 shows the surface brightness corrected for Galactic extinction, surface brightness dimming and k-corrected. Finally, column 6 indicates the morphological group that each object is classified into. The groups are defined as 1) The quasar host is part of a galaxy pair involved in a clear tidal interaction. 2) Other signs of morphological disturbance such as fans, shells and tails 3) A system with multiple nuclei, i.e. those within 10 kpc of each other. 4) An isolated galaxy with no signs of any morphological disturbance.}

\label{surf_bright}

\begin{tabular}{ l l l l l l  }
\hline
Abbreviated Name & Dimming & Morphology & $\mu_{AB}$ (mag arcsec$^{-2}$) & $\mu^{\emph{corr}}_{AB}$(mag arcsec$^{-2}$)& Group\\
\hline
J0025-10 & 1.15 & 2N, 2T		& 22.53, 24.91				& 20.90, 23.28					&	2,3\\
J0114+00 & 1.42 & 2N, S			& 25.2 						& 23.12							&	2,3\\
J0123+00 & 1.46 & 2N, B, [A]	& 26.8 						& 24.65							&	2,3\\
J0142+14 & --	& --			& --   						& --							&	4\\
J0159+14 & --	& [B]			& --						& --							&	4\\
J0217-00 & 1.28 & T, I, F		& 23.91,24.41, 25.56		& 22.06, 22.56, 23.71			&	2\\
J0217-01 & ---  & --			& --   						& --							&	4\\
J0218-00 & 1.37	&  I, A, [B]	& 22.65, 24.66				& 20.64, 22.65					&	1,2\\
J0227+01 & 1.34 & A, 2S, T		& 24.31,26.05, 26.33, 26.33	& 22.37, 24.11, 24.39, 24.63	&	2\\
J0234-07 & --	& --			&-- 						& --							&	4\\
J0249+00 & 1.48	& S, B			& 25.40, 26.22				& 23.13, 23.95					&	1,2\\
J0320+00 & 1.41	& I, [S]		& 25.74						& 23.49							&	2\\
J0332-00 & 1.17	& 2N, S, F, [B]	& 24.21, 26.15				& 22.34, 24.28					&	2,3\\
J0334+00 & 1.48	& S				& 26.59						& 24.16							&	2\\
J0848+01 & 1.31	& 2S			& 25.08, 25.41				& 23.18, 23.51					&	2\\
J0904-00 & 1.32	& T, S			& 25.08, 25.28				& 23.17, 23.37					&	2\\
J0923+01 & 1.42	& S, F, [T]		& 25.47, 26.06				& 23.42, 23.99					&	2\\
J0924+01 & 1.41	& T, [B]			& 23.70						& 21.63							&	2\\
J0948+00 & --	& --			& --						&--								&	4\\
J2358-00 & 1.46 & B, T, F		& 24.84, 25.45, 25.58		& 22.66, 23.27, 23.40			&	1,2\\
\hline
\end{tabular}
\end{minipage}  
\end{table*}

In this paper, we have followed the example of RA11 and references therein, in order to classify the detected features. Table \ref{surf_bright} gives the full classification of all the host galaxies.

The main result of this study is that, of the 20 type II quasar host galaxies in this sample, fifteen $(75\pm20\%)$ show clear evidence of morphological disturbance at relatively high levels of surface brightness, $\tilde{\mu}_r^{corr} =  23.37~mag~arcsec^{-2}$ with a range $\Delta \mu_r^{corr}\simeq[20.64,~24.65]~mag~arcsec^{-2}$.

Following the example of RA11, we have grouped the galaxies by morphology (Tables \ref{surf_bright} \& \ref{disturb}), though we have changed the scheme slightly. RA11 used five groups while we find four sufficient, since none of the quasar host galaxies in this sample have detected dust features. The groupings are as follows.

\begin{enumerate}
\renewcommand{\theenumi}{(\arabic{enumi})}
\item Galaxies that are involved in a tidal encounter with another that will not necessarily lead to a merger (B).
\item Galaxies that show evidence of morphological disturbance due to interactions or mergers such as shells, fans and amorphous halos (A,I,S,F).
\item Galaxies that have two nuclei within 10 kpc (2N). 
\item Isolated galaxies that are not disturbed.
\end{enumerate}

The classification of each of the host galaxies into the morphological groups is given in Table \ref{surf_bright}. We point out that the above categories are not exclusive, and a host galaxy may be classified into more than one group. We find that for the type II sample, 7 ($35\pm13\%$) are in the pre-coalescence phase of a galaxy interaction and are classified in either group 1 or 3. Another 8 ($40\pm14\%$) are in the coalescence or post coalescence phase of a merger and are classified only into group 2. The remaining 5 ($25\pm11\%$) show no signs of interaction and are classified into group 4. 

These findings indicate that, if an interaction/merger event is responsible for the triggering of the nuclear activity, there is no preferred phase in which it occurs. AGN activity is seen to occur both before, during and after the coalescence of the black holes, although we must bear in mind that there may be more than one phase of quasar activity. This is in accordance with the results of RA11 (Table \ref{disturb}) for the SLRG, where they find that the radio activity is seen to occur at all phases of the interaction/merger, with roughly equal proportions categorised as pre- and post-coalescence. \citet{tadhunter11}, also present similar findings based on studies of the stellar populations in the subset of radio galaxies that show significant star formation. In the latter work, spectral synthesis modelling has been carried out in order to date the starburst event, which simulations suggest occurs close to the time of coalescence of the SMBH \citep{hopkins08}. They find that radio-loud AGN activity may be concurrent with the starburst, and therefore the merging of the SMBH, or there may be a significant delay between the peak of starburst activity and the triggering of AGN activity (see also \citealt{tadhunter05}). However, they also find that the more luminous radio-loud AGN are triggered close to the point of coalescence, which may indicate that, for the most luminous AGN such as those studied here, there is a limited period around the peak of the merger over which AGN activity occurs.

\begin{figure}
\includegraphics[trim = 11mm 2mm 4mm 5mm, clip, width=9cm ]{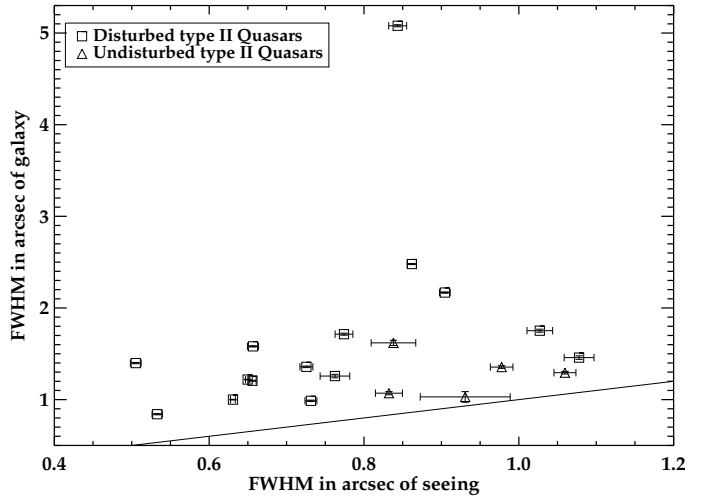}
\caption{Shows the FWHM of the type II quasar host galaxies broken down into morphologically disturbed (open squares) and undisturbed (open triangles), measured assuming a Gaussian core,  against the FWHM of the stars measured from the same images. The line shows the seeing of the image.}
\label{compactness}
\end{figure}

Note that the 25\% of objects that appear undisturbed are potentially interesting in their own right. Of these five objects, one (J0217-01, Figure \ref{q021701}) is an apparently undisturbed edge on disc, while the other four (J0142+14 Figure \ref{q0142}, J0159+14 Figure \ref{q0159}, J0234-07 Figure \ref{q0234} and J0948+00 Figure \ref{q0948}) appear to be spatially compact objects. In order to determine if there is any real difference between the host galaxies of the disturbed majority and these four undisturbed quasars, other than the lack of tidal features, we have used 2-D Gaussian fits to determine the FWHM of the cores of the type II quasar images, and compared them with values for foreground stars in the same image. Figure \ref{compactness} shows the results of the analysis, with the black line marking the seeing and the type II sample broken down into disturbed and undisturbed host galaxies. As can be seen from Figure \ref{compactness}, the four undisturbed galaxies have a FWHM close to the seeing of the images, demonstrating that they are barely resolved, although these are the images with some of the worst seeing, which may mean that it is not possible to detect faint tidal features on larger scales that are present. It will be interesting in the future to use high resolution HST images to compare the surface brightness profiles of the disturbed and undisturbed type II quasar objects in our sample.

\begin{table*}
\centering
  \begin{minipage}{140mm}
  \caption{A morphological classification of the galaxies in the type II quasar sample, the PRG at $z> 0.2$ of RA11, and quiescent galaxies of the $0.3 < z < 0.41$ EGS comparison sample, the comparison sample including discs and the full EGS sample of RA12 in the same absolute magnitude range as the type II host galaxies(104 galaxies). We group the galaxies by morphology with groups 1 \& 3 being pre-coalescence systems, group 2 being coalescing or post-coalescence systems and group 4 being non interacting systems. Column 1 \& 2 give the morphology and the group respectively. Columns 3, 4 \& 5 give the percentage of the sample of type II quasars, SLRG and quiescent galaxies that are classified in each group. Please note that if a galaxy is classified in group 2 but is also in groups 1 or 3, we consider it a pre-coalescence system.}
  
  \label{disturb}
  \begin{tabular}{l c c c c c c}
  \hline
  Morphology 						& Group			&	Type II 	&	SLRG	&	Comparison	& Comparison &	EGS Sample	\\
  									&				&	Quasars		&			&				& + discs	 &	\\						
  \hline
  Pre-coalescence					&	1, 3		&	35\%		&	50\%	&	28\%		&	20\%	&	21\%\\
  Coalescence or post-coalescence 	&	2			&	40\%		&	45\%	&	39\%		&	29\%	&	33\%\\
  No sign of interaction			&	4			&	25\%		&	5\%		&	33\%		&	51\%	&	46\%\\
  \hline
  Any sign of disturbance		&	1,2,3		&	$75 \pm 20\%$	&	$95^{~+5}_{-21}\%$	&	$67\pm14\%$	&	$49\pm10\%$	&	$54\pm7\%$\\
  \hline
  \end{tabular}
  \end{minipage}       
\end{table*}

\section{Discussion}
\label{discussion}
\begin{figure}
\centering
\subfigure[]{
\includegraphics[trim = 8mm 0mm 0mm 0mm, clip,width=8.5cm]{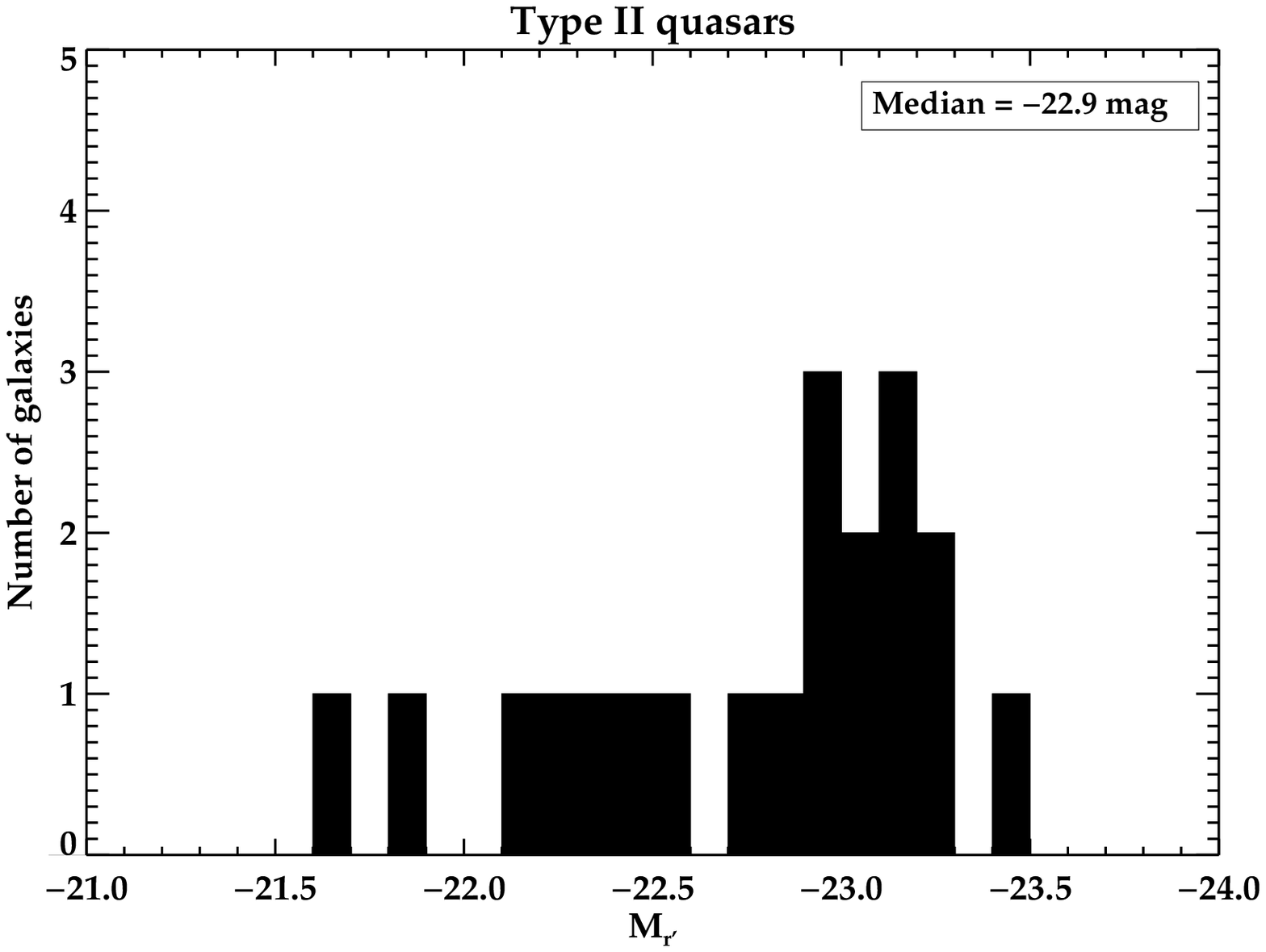}
\label{type2_absmag}}
\subfigure[]{
\includegraphics[trim = 8mm 0mm 0mm 0mm, clip,width=8.5cm]{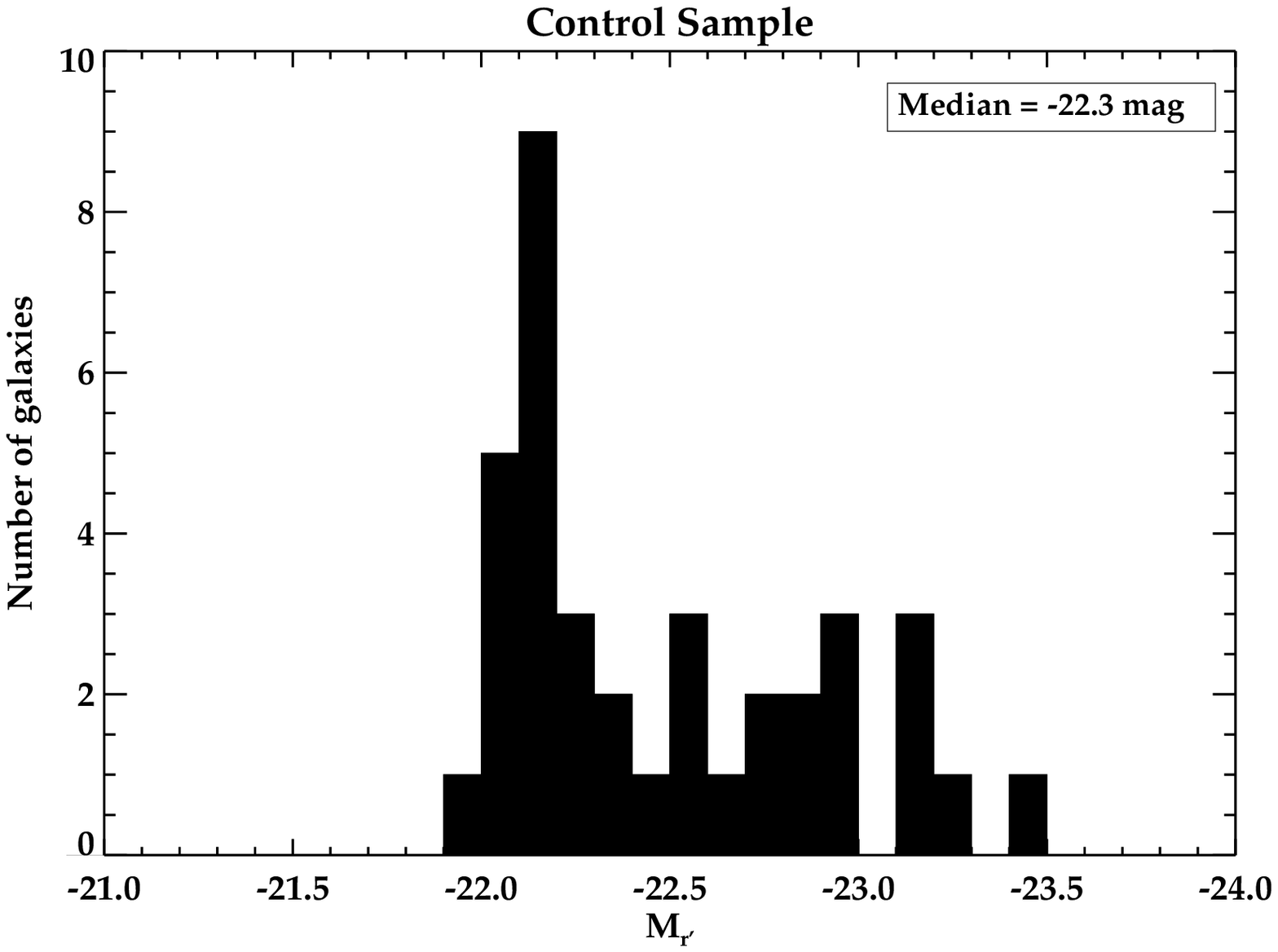}
\label{control_absmag}}
\subfigure[]{
\includegraphics[trim = 8mm 0mm 0mm 0mm, clip,width = 8.5cm]{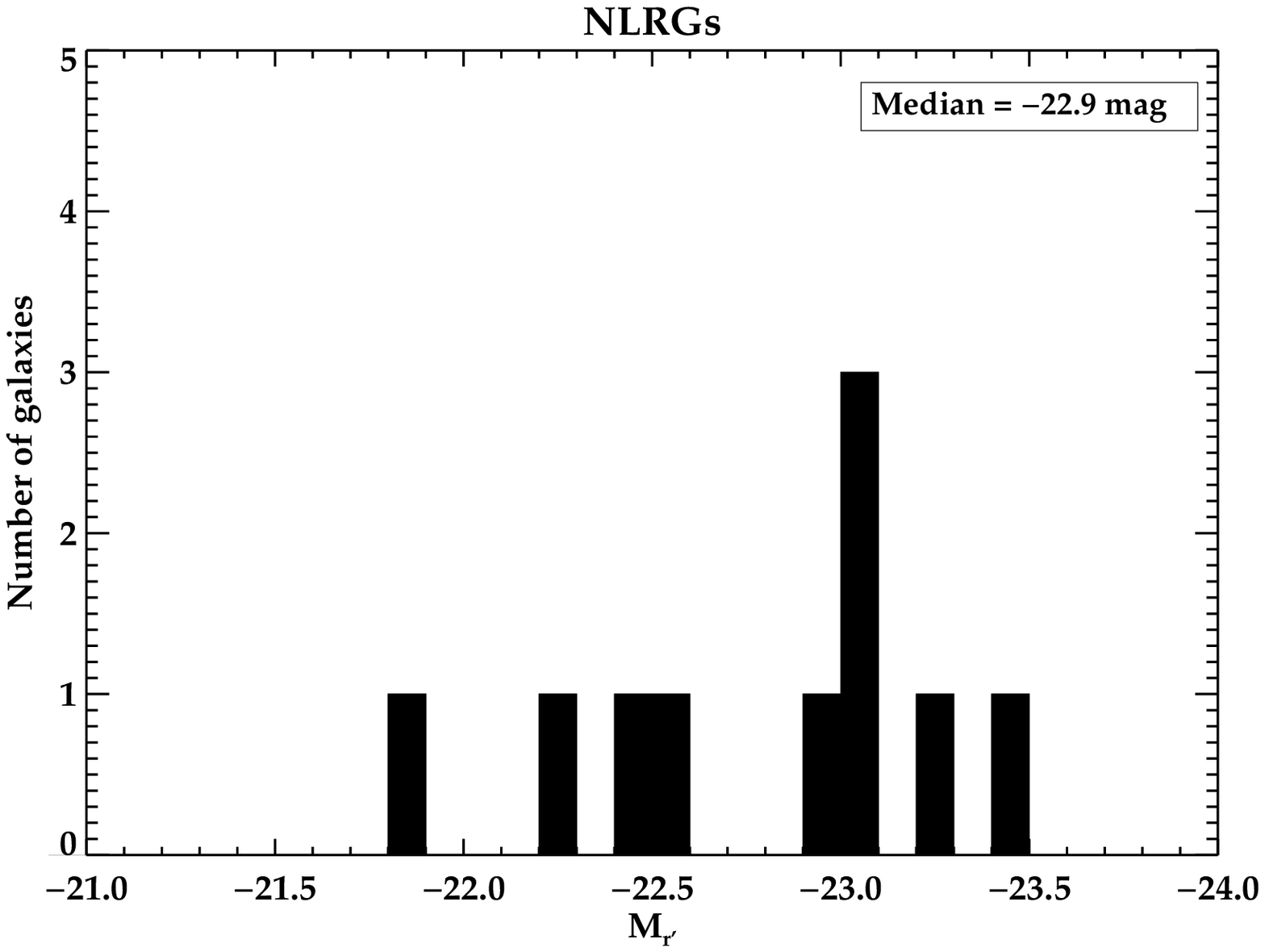}
\label{2Jy_absmag}}
\caption{The distribution of the absolute magnitudes of (a) the type II quasar host galaxies, (b) the $0.3< z < 0.41$ part of the EGS comparison sample and (c) the $0.2 < z < 0.7$ 2Jy NLRG. The median value of M$_{r^\prime}$ is given for the portion of sample used for the comparison in each panel.}
\label{comp_mag}
\end{figure}

\subsection{Comparison with quiescent early-type sample}
\label{comp_control_sample}
In this section we compare our results to those obtained for a comparison sample of quiescent galaxies taken from deep Subaru images of the Extended Groth Strip (EGS). The full criteria used to select objects as suitable for the full control sample are given in RA12, who made use of the \emph{Rainbow Cosmological Surveys database}\footnote{$https://rainbowx.fis.ucm.es/Rainbow_Database$}, which includes a compilation of photometric and spectroscopic data, along with value-added products such a photometric redshifts, stellar masses, star formation rates, and synthetic rest-frame magnitudes for several deep cosmological fields (\citealt{perez08,barro09,barro11}). To summarise, galaxies were first selected by colour to isolate those that are on the `red sequence' using the colour selection criterion (M$_u$-M$_g)>1.5$ \citep{blanton06}. They were then further selected to be early-types, which involved visual inspection of the images to determine whether the galaxies were discs or ellipticals. The galaxies that were considered to be definite discs were disregarded, and those that were classified as possible discs were maintained. The reason for making this distinction is that the vast majority of PRG, which the full control sample was selected to match, are associated with elliptical galaxies. The full EGS control sample was also selected to have the same host luminosity and redshift range as the $0.2 < z < 0.7$ portion of the 2Jy sample. The Subaru images were considered to be suitable because they have similar filter, depth and resolution to the Gemini GMOS-S images of RA11. 

\begin{figure}
\centering
\subfigure[]{
\includegraphics[trim = 8mm 0mm 0mm 0mm, clip,width=8.5cm]{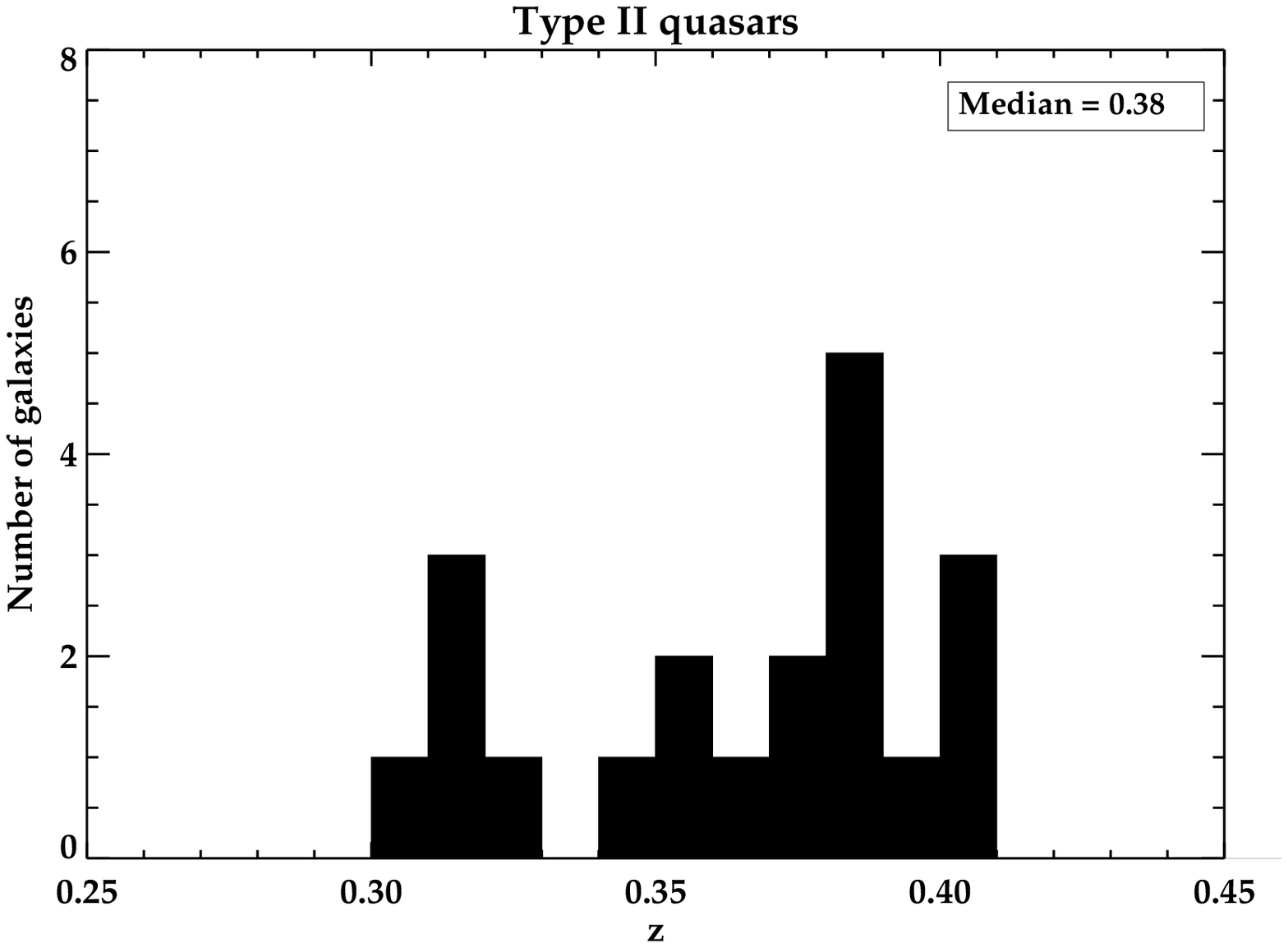}
\label{type2_zrange}}
\subfigure[]{
\includegraphics[trim = 8mm 0mm 0mm 0mm, clip,width=8.5cm]{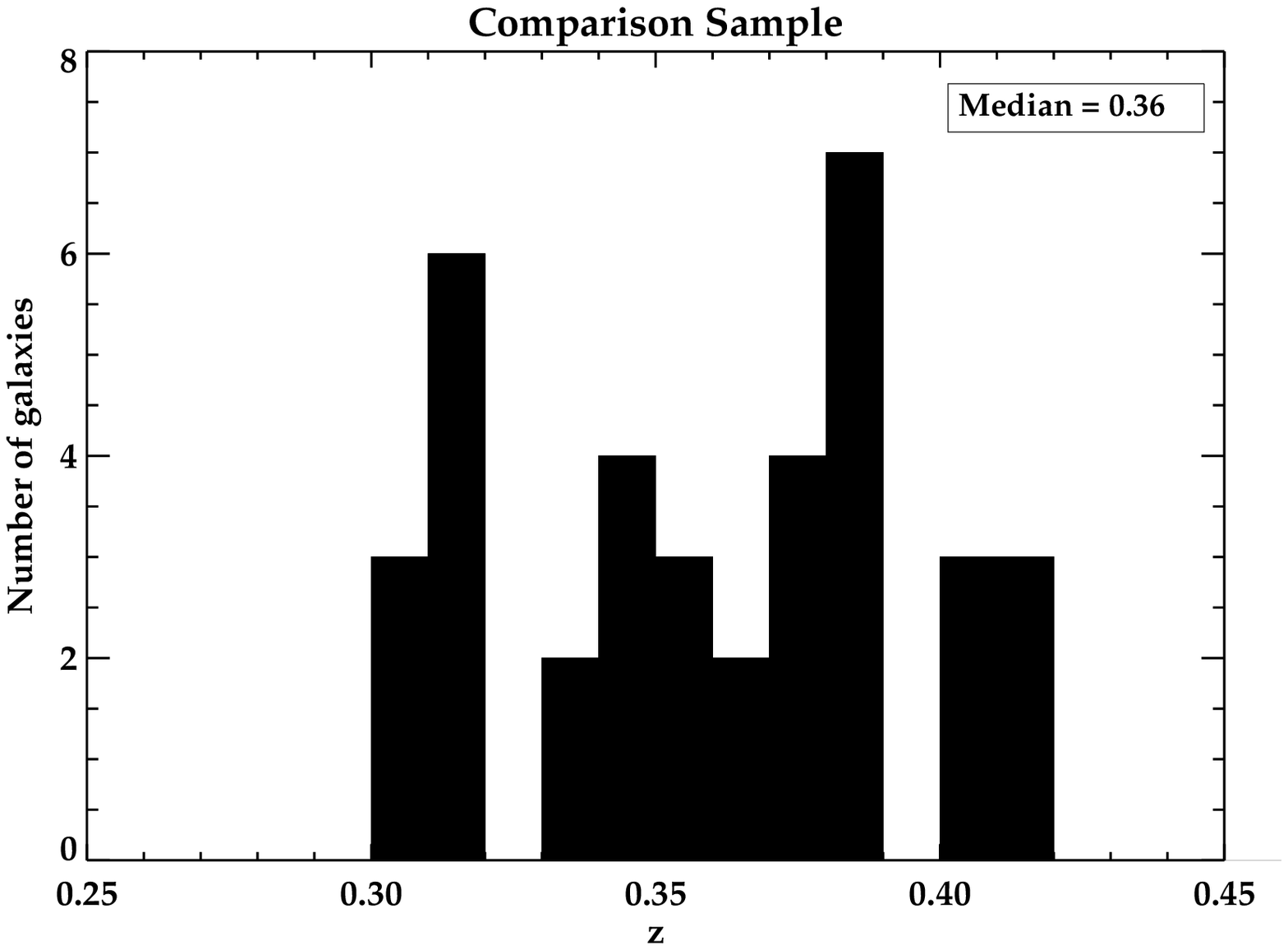}
\label{control_zrange}}
\caption{A comparison of the redshift distribution of (a) the type II quasars and (b) our comparison sample. The median redshift for each sample is shown in the appropriate panel. A K-S test returns the probability P=0.643 that the redshift distribution is drawn from the same parent population.}
\label{comp_z}
\end{figure}

The 2Jy sample extends over a greater redshift range than the type II quasars studied here, and this is reflected in the redshift range of the EGS quiescent galaxies. Therefore, for comparison with the type II quasar host galaxies, we have elected to use only the galaxies from the parent EGS sample that are in the redshift range $0.3 < z < 0.41$, to reflect the distribution of the sample studied here. This leaves us with a comparison sample of 36 quiescent galaxies, classified as either ellipticals or possible discs, within the same absolute magnitude (Figures \ref{type2_absmag} \& \ref{control_absmag}) and redshift range (Figure \ref{comp_z}) as the type II quasars.

The results of our morphological analysis of the $0.3 < z < $ EGS comparison sample are presented in Table \ref{disturb}. These show that there is no statistically significant difference in the rate of disturbance detected in the type II quasars and the comparison sample, with $75\pm20\%$ of the type II quasars and $67\pm14\%$ of the comparison sample showing evidence of tidal features. If we only consider features in the comparison sample that fall within the same range of surface brightness as the type II quasars, the proportion of tidally disturbed early-type galaxies in the comparison sample falls to $50\pm6\%$. When considering the classification of the galaxies into the different morphological groups, we also find no significant difference between the type II quasars and comparison sample (see Table \ref{disturb}).

\begin{figure}
\centering
\subfigure[]{
\includegraphics[trim = 11mm 2mm 3mm 0mm, clip,width=8.5cm]{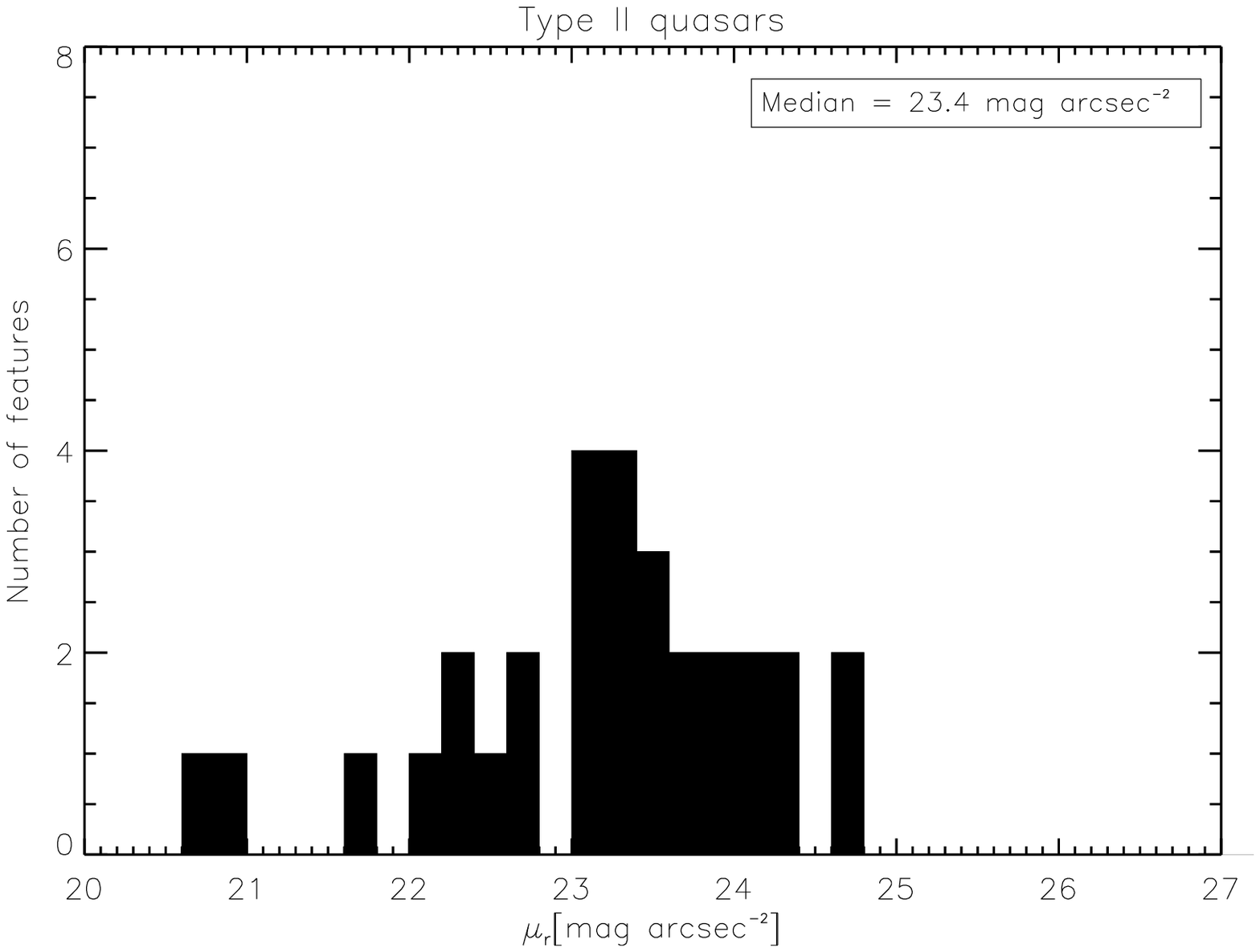}
\label{type2_surf_bright}}
\subfigure[]{
\includegraphics[trim = 11mm 2mm 3mm 0mm, clip,width=8.5cm]{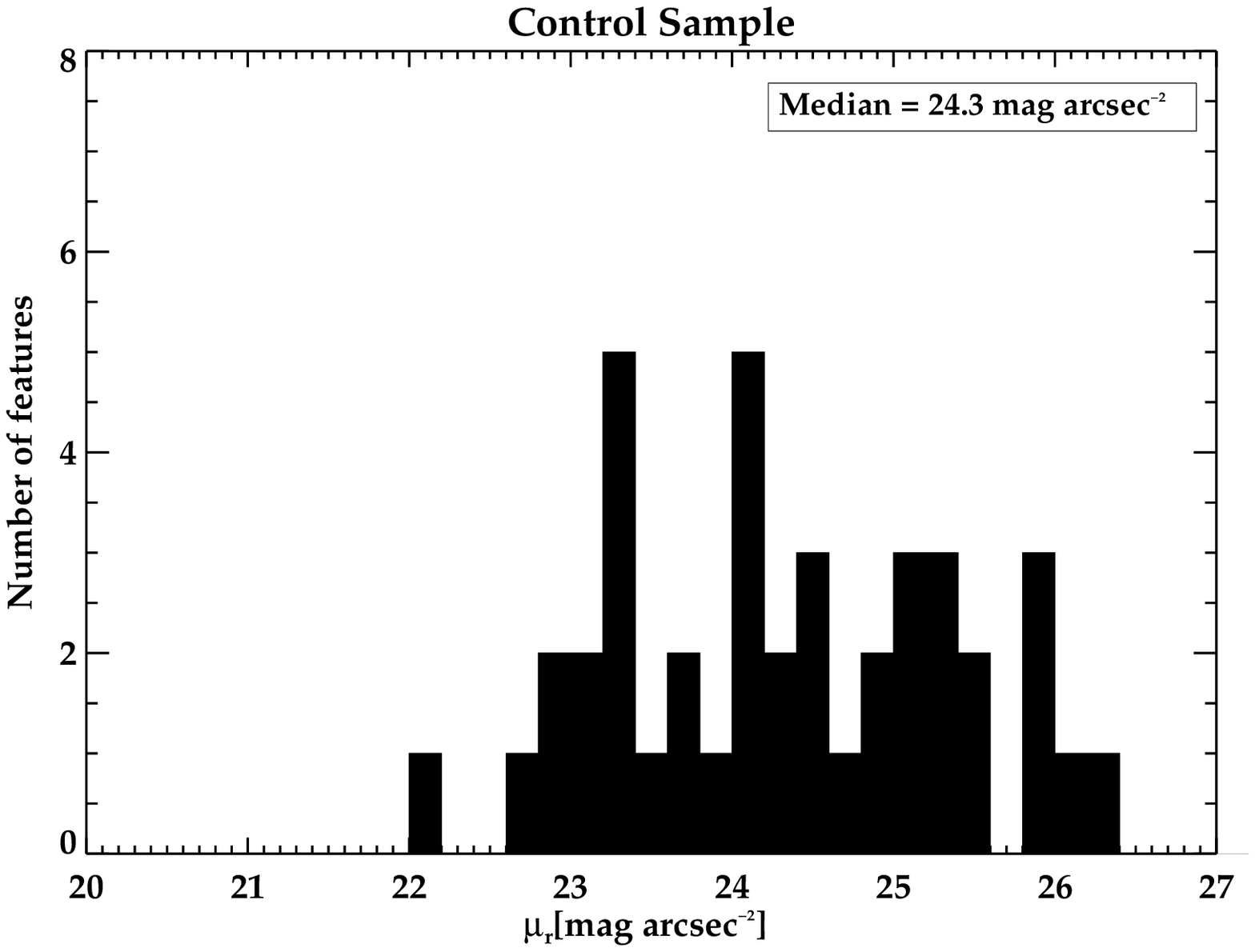}
\label{control_surf_bright}}
\subfigure[]{
\includegraphics[trim = 11mm 2mm 3mm 0mm, clip,width=8.5cm]{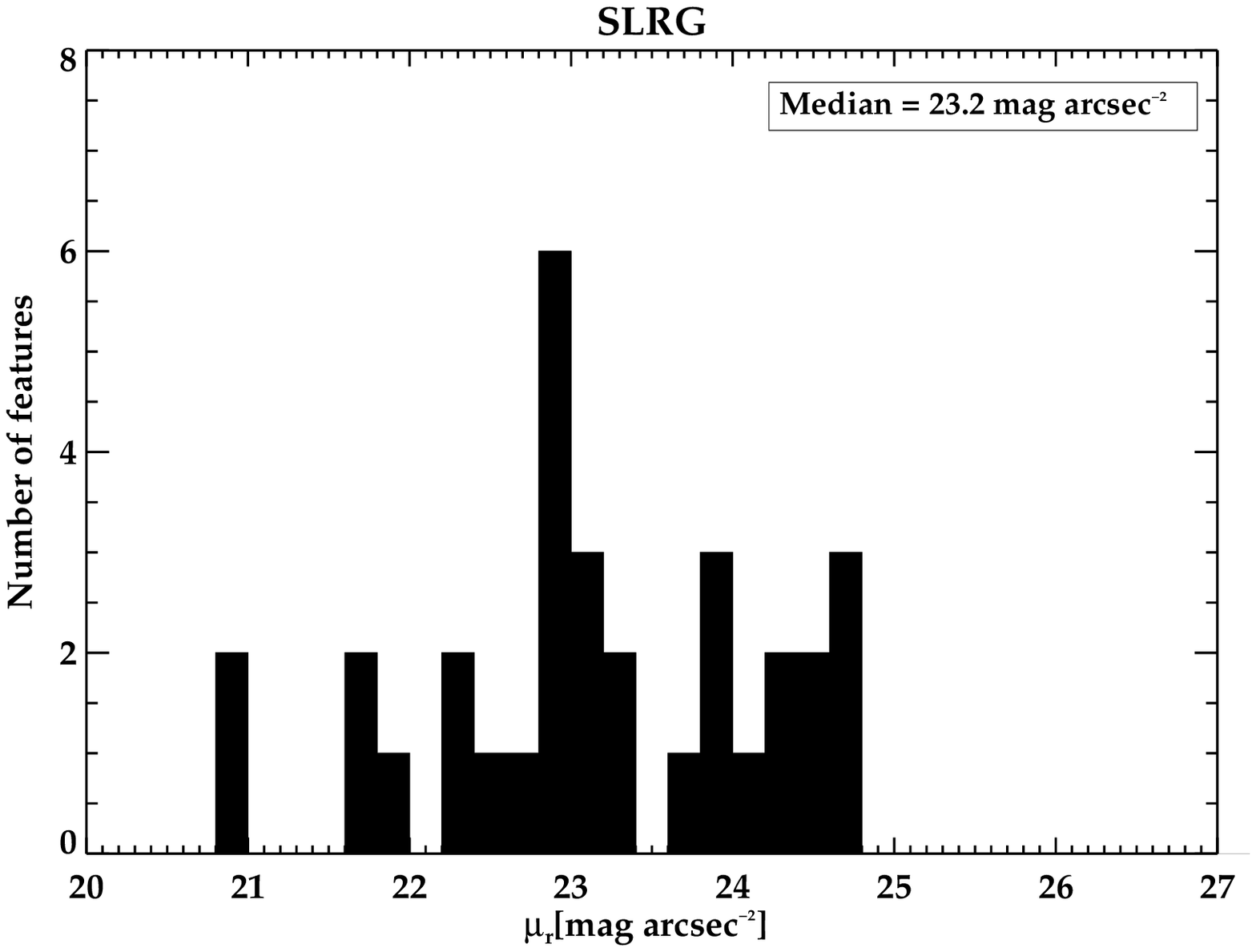}
\label{2jy_surf_bright}}
\caption{A comparison of the distribution of the surface brightness of the detected features in the $r^\prime$ band of (a) the type II quasar sample, (b) the comparison sample of quiescent early type galaxies, and (c) the $0.2 < z < 0.7$ 2Jy SLRG. The median value of $\mu_{r^\prime}$ ($mag~arcsec^{-2}$) is given in each panel for that sample.}
\label{sb_features}
\end{figure}

The values for the surface brightnesses of the detected features of the comparison sample are taken from RA12 and are summarised in Table \ref{surf_bright_comp} and Figure \ref{control_surf_bright}, converted to $r^\prime$ magnitudes using colours for elliptical galaxies taken from \citet{fukugita95}. We see that, the median value of surface brightness for the features in the comparison sample is almost a magnitude fainter than the type II sample ($\tilde{\mu}_r^{corr} = 24.3~mag~arcsec^{-2}$ compared with $\tilde{\mu}_r^{corr} = 23.4~mag~arcsec^{-2}$  for the type II quasars), the range of surface brightness  also extends to fainter values than the type II quasars, with the comparison sample having a range $\Delta \mu_r \simeq [22.11, 26.06]~mag~arcsec^{-2}$. Therefore, the features detected in the type II quasars are up to 2 magnitudes brighter than those detected in the comparison sample. We can compare the distribution of the surface brightnesses in these two samples by performing a Kolmogorov-Smirnov (K-S) test. This returns a probability P=0.002 that the two distributions are drawn from the same parent population, which is significant at the 99.8 \% ($> 3\sigma$) level, strongly indicating that the two samples are drawn from different populations. Figure \ref{sb_features} clearly demonstrates this difference between distributions, with Figure \ref{control_surf_bright} shifted towards fainter magnitudes in relation to Figures \ref{type2_surf_bright} and \ref{2jy_surf_bright}.

\subsubsection{Including the discs}
A possible weakness of the comparison made in the previous section is that we have only included elliptical and possible discs in our EGS comparison sample. While such a comparison sample selection is appropriate for PRG, which are almost invariable associated with giant elliptical galaxies, it is not clear whether it is so appropriate for luminous radio-quiet AGN, because some (albeit a low fraction) of the latter are known to be hosted by disc galaxies (\citealt{dunlop03, greene09}). Indeed one of the type II quasar objects discussed in this paper (J0217-01) is a clear disc galaxy. Therefore, we repeat the comparison of the previous section, but reinstate the quiescent galaxies that were rejected from the EGS control sample of RA12 because they were visually classified by the authors as discs. This leads to the inclusion of an extra 15 quiescent galaxies with $0.3< z < 0.41$ and absolute magnitudes in the same range as the type II quasar host galaxies, bringing the total sample to 51 quiescent galaxies. Visual inspection suggests that the vast majority of the discs are early-type (S0) and therefore all magnitudes have been converted to $r^\prime$ using colours for S0 galaxies from \citet{fukugita95}.

We find that, of these 15 galaxies, only one displays any sign of morphological disturbance, which is in the form of a fan. We therefore classify this galaxy into group 2 and the remaining 14 into group 4. When including the discs, we find that $49\pm10\%$ of the full sample of 51 show evidence for morphological disturbance. Again, if we only include those galaxies with at least one feature as bright as the dimmest type II feature (and double nuclei), this figure falls to $36\pm9\%$. Moreover, inclusion of the measured surface brightness of the one feature associated with the disc galaxies does not change the statistics of the distribution of the surface brightnesses for the comparison sample.

Table \ref{disturb} shows clearly that only including quiescent galaxies that have been visually classified as early-types (Section \ref{comp_control_sample}) leads to a conservative comparison between the rates of disturbance found in the type II quasars and those found in the quiescent red sequence galaxies.

This is also true  if we compare the detected rates of disturbance in the type II quasars with those for the 104 early-type galaxies from the complete EGS control sample that are in the same absolute magnitude range. In this case, we find that $54\pm7\%$ show signs of interaction (Table \ref{disturb}), falling to $45\pm7\%$ if we only consider those with a feature in the same surface brightness range as those detected in the type II quasars. The median and range of surface brightnesses remains very similar (Table \ref{surf_bright_comp}) to that for the comparison sample. This comparison emphasises the fact that, by considering only early-type galaxies in the same redshift range as the type II quasars, the comparison between the rates of disturbance may be a conservative one.

\subsubsection{Surface brightnesses}

The fact that, in all the quiescent samples considered here, the surface brightness of the detected features is up to 2 magnitudes fainter than in the type II quasar host galaxies could indicate that the interactions experienced by these quiescent early-type galaxies are different in nature to those being experienced by both the type II quasars and PRG. Examples of these differences may be that they are gas poor interactions rather than gas rich ones, or minor mergers instead of major ones. In such cases, we would not expect the interaction to lead to luminous AGN activity if major, gas-rich mergers are required to trigger and sustain such activity.  

Alternatively, the difference in the surface brightness of the features between the two groups could be due to the fact that, for the comparison sample, we are seeing these interaction events at later stages than in the 2Jy and type II quasar groups. Simulations (e.g. \citealt{springel05, dimatteo05, hopkins08})  suggest that, if major mergers and interactions between galaxies were the triggering event for luminous AGN activity, then this would preferentially occur close to the peak of the process, around the point of coalescence of the SMBH. However, considering the separation of the two nuclei in the pre-coalescence systems, such as J2358-00 shown if Figure \ref{q2358}, we may be observing some of the type II objects in this sample 100-200 Myr before coalescence (though it could be even earlier for J0123+00), it is not clear how long after the coalescence they might be observed. Although there is evidence for substantial time delays in lower luminosity PRG and AGN \citep*{wild10,tadhunter11}, it is not clear that this applies to high luminosity AGN, and it remains uncertain how long after the merger it is possible for quasar-like AGN activity to be triggered. If we assume that there is a period after the coalescence of the SMBH in which the AGN activity can be triggered ($t_{tr}$), then this means that there is a limited window of $\sim 200+t_{tr}$ Myr in which the activity can occur. 

\begin{figure}
\includegraphics[width=8.5cm, height= 5cm]{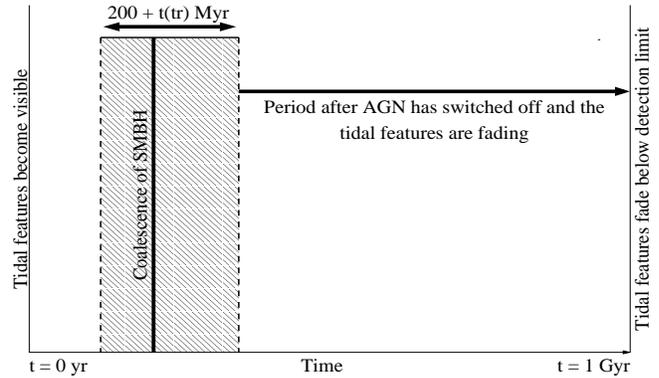}
\caption{A representation of the time scales over which the morphological features will be visible in comparison to the period in which the quasar will be active. The shaded area is the period around the coalescence of the SMBH which simulations suggest is when the AGN is most likely to be triggered. The AGN is then switched off fairly early in the process of the merger and then the features gradually fade until they fall below our detection limit.}
\label{time}
\end{figure}

Assuming that the tidal features will gradually fade over the lifetime of the interaction until they fall below our detection limit, the morphological features associated with galaxies that are hosting a quasar should appear brighter at the time that the quasar is active, fading as time progresses until the AGN switches off. Figure \ref{time} shows a visual representation of this scenario, with the shaded area indicating the time around the coalescence of the SMBH in which we would expect the AGN to be triggered. It also shows that, for a significant proportion of the time over which the interaction signatures are visible, we would not expect the system to be capable of hosting a quasar. We must bear in mind that an AGN phase is a snapshot in time in the history of a galaxy and it may be that, in the comparison sample, we are viewing these interaction events at a time before AGN activity has commenced or after it has waned, leaving the fading tidal debris of a historic merger induced AGN. 

\subsection{Comparison with the 2Jy sample}
\label{comp_2jy_samp}

As previously stated, one of the aims of this work is to compare the rates of disturbance for type II quasars with those found in the 2Jy sample of PRG studied by RA11 which comprises a sample of the PRG selected by \citet{tadhunter93,tadhunter98}. To summarise the selection criteria, they required that all objects had $S_{2.7GHz} > 2.0$ Jy, steep radio spectra indices $\alpha_{2.7}^{4.8} >0.5 ~(F_{\nu} \propto \nu^{-\alpha})$, declinations $\delta < +10^{\circ}$ and redshifts $0.05<z<0.7$. For further details on the criteria used for selection and completeness, refer to \citet{tadhunter93,tadhunter98} and RA11. By comparing our results with those found by RA11, we wish to identify any common factors between the two samples which may isolate a single triggering mechanism. 

To allow a direct comparison between the two, we have used only the $0.2 < z < 0.7$ portion of the 2Jy sample. This gives a sub-sample of 22 PRG comprising 21 SLRG and 1 weak line radio galaxy (WLRG).  We  discount the latter object from consideration because it has an emission line luminosity that is almost two orders of magnitude lower than the 2 Jy radio galaxies of similar redshift and radio power, and is therefore not comparable to the powerful quasars studied here. This leaves a final sub-sample of 21 SLRG at $0.2 < z < 0.7$. Note that the lower redshift limit corresponds to the redshift at which, due to the correlations between radio power, emission line luminosity and redshift, the radio galaxies have a similar lower limit to their emission line luminosity as the type II quasars in our sample. 

As mentioned in Section \ref{sample_obs}, our original sample selection was not based on the radio properties of the type II quasars. However, because we are making a comparison here between the morphological properties  of the type II quasar and PRG host galaxies, we also compare their radio properties and AGN luminosities. In order to do so, we utilise the February 2012 version of the Faint Images of the Radio Sky at Twenty-centimeters (FIRST)\footnote{http://sundog.stsci.edu/} catalogue. This survey is intended to cover 10,000 degrees$^{2}$ of the North and South Galactic caps, to a $5\sigma$ detection threshold of 1.0 mJy \citep*{becker95}. The maps produced have 1.8 arcsec pixels, a typical rms of 0.15 mJy and a resolution of 5 arcsec.

Following the example of \citet{zakamska04}, we searched the catalogue for any matches within 3 arcsec of each object, producing a total of 11 matches. This method allows for the matching of any core radio component associated with the source but does not include the flux associated with any extended lobes or jets. In order that all emission associated with an object is included, we extended the search radius to 90 arcsec, which at the typical redshift of this sample equates to $\sim$0.5 Mpc, resulting in 6 further matches. We then visually inspected these matches to determine if they are associated with the sources. They were considered to be associated if they had a double lobed morphology centred on the core position, even if no core component is detected, or if they have an asymmetric radio component and a core component \citep{zakamska04}. Imposing this criteria left a total of 4 out of the 6 matches. We then summed the integrated flux at 1.4 GHz for each component associated with the source to obtain the total flux.

Two objects in our type II quasar sample (J0142+14 and J0159+14) are not in the footprint of the FIRST survey. In order to ensure we had radio data for all our sources, we utilised the NVSS survey \citep{condon98} also conducted at 1.4 GHz, but with a $5 \sigma$ detection threshold of $\sim$2.5 mJy. Neither of these objects had matches with 90 arcsec.

\begin{figure}
\includegraphics[trim = 9mm 2mm 0mm 4mm, clip, width=9cm ]{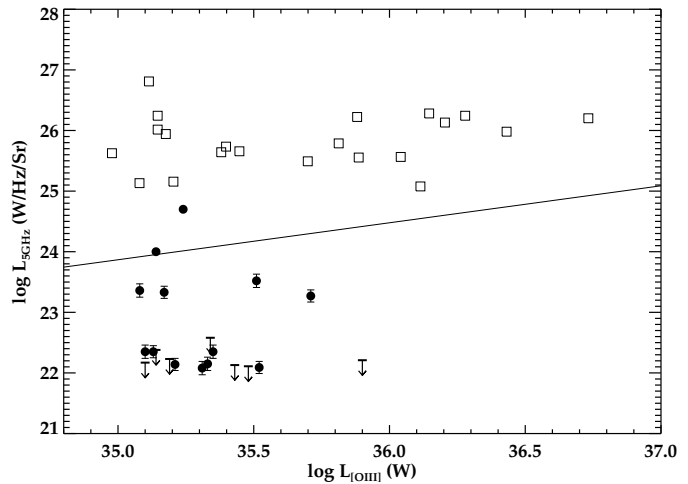}
\caption{The values of $L_{5GHz}$ (W Hz$^{-1}$ Sr$^{-1})$  plotted against $L_{[OIII]}$ (W) for the full sample of type II quasars (solid symbols) and PRG from RA11 (open squares). For those type II quasars that were detected by the FIRST survey (circles), $L_{5GHz}$ was calculated from the total integrated flux at 1.4 GHz. If the spectral index is known, we used that value for making the calculations, and no error bars are shown. In all other cases a value of $\alpha = -0.75$ was used and the error bars were derived by calculating $L_{5GHz}$ using $\alpha = -0.5$ and $\alpha = -1$. The horizontal lines represent those objects that were undetected by either FIRST or NVSS. In these cases, the detection threshold of the survey (1.0 mJy for FIRST and 2.3 mJy for NVSS) was used as an upper limit. The squares represent the 21 SLRG used as a comparison to the type II quasars. The line represents the boundary between radio-loud and radio-quiet AGN.}
\label{radio_power}
\end{figure}

We then converted the total integrated fluxes to luminosities at 5 GHz ($L_{5GHz}$) using spectral indices taken from \citet{zakamska04}. However, in most cases, the spectral index is unknown so, in these cases, we used an assumed value $\alpha = -0.75$ because, this is a typical value for radio-loud AGN \citep*{donnelly87}. The values of both $L_{5GHz}$ and $L_{[OIII]}$ for the PRG are taken from \citet{dicken09} and references therein. Figure \ref{radio_power} shows the comparison of radio power for both the PRG (open squares) and type II quasars (solid symbols). Using the correlation found by \citet*{xu99} between radio-power and emission line luminosity, the line represents the boundary between their definition of radio-loud and radio-quiet AGN. As can be seen from Figure \ref{radio_power}, the majority of the type II sample can be considered to be radio-quiet, with only one object falling squarely in radio-loud territory (J0114+00) and one on the boarder line between the two groups (J0848-01). This is in accordance with the \citet{zakamska04} findings that $\sim$10\% of type II quasars are radio-loud.

Figure \ref{radio_power} also demonstrates that the [OIII] luminosities of the SLRG, which are indicative of the AGN luminosities, extend to higher values than those of the type II quasars. However, in this paper, we are considering the link between \emph{quasar} activity and the rates of interactions/mergers, so in that respect, our real concern is that the lower limit of $L_{[OIII]}$ of the two samples should be consistent with each other and with all the objects having quasar-like luminosities. Although there are no hard and fast rules regarding the boundary between what constitutes a Seyfert and a quasar, we have made the distinction at $L_{[OIII]} > 10^{8.5} L_\odot$ W. Figure \ref{radio_power} demonstrates that all the objects studied here fulfil this criteria, except for 1 SLRG (PKS 0859-25) which has $L_{[OIII]} = 10^{8.4} L_\odot$. We maintain this object for the purpose of comparison, however, as it is on the borderline of our original selection criteria.

In terms of the aims of this work, which is to study the rates of interaction in \emph{powerful quasars}, the SLRG provide a good comparison in terms of optical luminosity, as they fulfil our criteria for the lower limit to the AGN power.

\begin{figure}
\includegraphics[trim =9mm 2mm 4mm 5mm, clip, width=9cm]{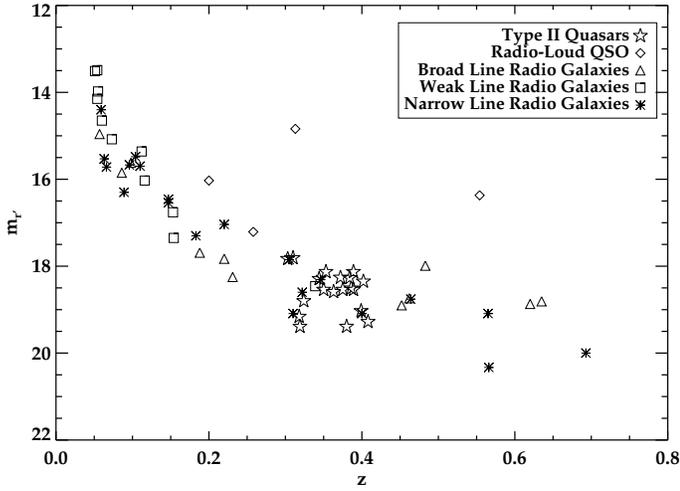}
\caption{A comparison of the 30 kpc magnitude of this sample of type II quasars compared to the 30 kpc magnitude of the 2Jy radio galaxy sample of RA11. The type II quasars are shown as open stars with the PRG sample broken down into their optical classifications of radio-loud quasars (open diamonds), broad-line radio galaxies (open triangles), weak-line radio galaxies (open squares) and narrow-line radio galaxies (asterisks). Both samples have been corrected for Galactic extinction and k-corrected.}
\label{compare_app_mags}
\end{figure}

In order to determine whether the host galaxies of the two types of AGN are of similar luminosity, we have compared the apparent $r^\prime$ band magnitudes of the samples against redshift. Figure \ref{compare_app_mags} shows both the type II quasars (open squares) and the entire 2Jy sample split into the various optical classifications given in RA11. This was done to aid the comparison between the two, since broad line radio galaxies are likely to be inherently more luminous due to the unobscured AGN contribution to their global brightness. As can be seen from Figure \ref{compare_app_mags}, if we discount those PRG in which the central engine is directly visible (e.g. broad line radio galaxies), the type II quasars have similar apparent magnitudes to PRG in the same redshift range. It can clearly be seen that the inclusion of these unobscured PRG would artificially push the entire sample to brighter magnitudes, and therefore we do not include them in further comparison of the luminosities of the host galaxies.

\begin{table}
\caption{A comparison of the surface brightnesses of the tidal features detected in the type II quasars, the EGS comparison sample, the EGS comparison sample including discs, the complete EGS sample of RA12 in the same absolute magnitude range as the type II quasar host galaxies, and the $0.2 < z < 0.7$ portion of the 2 Jy SLRG sample of RA11. Column 1 gives the sample, column 2 lists the median of the surface brightness of the features and column 3 corresponds to the range of surface brightness of the detected features. All values of surface brightness are given in the $r^prime$ band in mag arcsec$^{-2}$, and have been corrected for Galactic extinction, surface brightness dimming and k-corrected. }
  \label{surf_bright_comp}
  \begin{tabular}{l c c c}
  \hline
  Sample					& Redshift			&	Median 	&	Range\\  
  							& Range				&	Value	&		\\
  \hline	
  Type II					& $0.3,~ 0.41$	&	23.37	&	20.90, 24.65\\  
  EGS Comparison			& $0.3,~ 0.41$	&	24.30	&	22.11, 26.06\\
  EGS Comparison incl. discs& $0.3,~ 0.41$	&	24.17	&	22.11, 26.06\\
  EGS Complete				& $0.2,~ 0.7$	&	23.95	&	22.03, 26.06\\
  2Jy PRG					& $0.2,~ 0.7$	&	23.18	&	20.94, 24.78\\
  \hline
  \end{tabular}
  \end{table}

Figure \ref{type2_absmag} and \ref{2Jy_absmag} shows the distribution of M$_{r^\prime}$ for the type II quasars and narrow line radio galaxies (NLRG) of the 2Jy sample. We only compare with the 10 NLRG because, in this way, we are likely to be comparing the intrinsic luminosities of the host galaxies, without needing to account for the direct quasar light as mentioned above. Carrying out a K-S test on these two populations returns a probability of P=0.275, implying that we cannot reject the null hypothesis that the two distributions are drawn from the same parent population. The two groups have the same median M$_{r^\prime}$ magnitude (-22.9 mag), while the range of magnitudes is $\Delta M_{r^\prime} \simeq [-23.4, -21.6]$ mag for the type II quasars and $\Delta$M$_{r^\prime} \simeq[-23.4, -21.8]$ mag for the NLRG. This demonstrates that we detect no significant difference between the absolute magnitudes of the host galaxies of radio-loud and radio-quiet quasars, which is contrary to the findings of some previous studies such as \citet{kirhakos99}, \citet{dunlop03} and \citet{hyvonen07} who detected differences in host galaxy absolute magnitude of up to $\sim$1 mag.

We find that, considering the $0.2 < z < 0.7$ 2Jy sample of RA11, the median surface brightness of the detected features in the $r^\prime$ band is $\tilde{\mu}_r^{corr} = 23.18~mag~arcsec^{-2}$ with a range $\Delta \mu_r^{corr} \simeq [20.94, 24.78]~mag~arcsec^{-2}$ (Table \ref{surf_bright_comp}). These values of surface brightness are in the same range as those found for the type II quasar sample, with a very similar median (23.4 mag arcsec$^{-2}$ for the type II quasars and 23.2 mag arcsec$^{-2}$ for the 2Jy PRG). The probability returned by the K-S test when comparing the surface brightness of all the detected features of the 2Jy PRG sub-sample and the type II quasars is P = 0.517, indicating that the distribution of surface brightnesses of the detected features may be drawn from the same parent distribution. This comparison is shown in Figures \ref{type2_surf_bright} and \ref{2jy_surf_bright}, and the similarity in both the median value and range of surface brightness of the detected features support the hypothesis that the two types of AGN may be triggered by the same type of merger events, since the surface brightness and type of features present will be dependent on the characteristics of the progenitor galaxies and the parameters of the merger.

We also compare the proportions of the two samples that are classified into the different morphological groups described above. It can  be seen from Table \ref{disturb} that we find similar rates of disturbance in the two samples, with a higher proportion of the 2Jy sample of RA11 showing signs of morphological disturbance, although we must still consider the small sample sizes when making this comparison. Moreover, in both cases there is no single phase of the interaction in which the nuclear activity has been triggered. The similarities between the rates of disturbance and the range of features observed in the two samples are consistent with the idea that the two types are triggered in a similar manner.

\subsection{The volume density of type II quasars, disturbed early-type galaxies and 2Jy PRG}
\label{vd}

\subsubsection{Type II quasars and early-type galaxies}

\begin{table*}
\centering
	\begin{minipage}{140mm}	
    \caption{The volume density of the type II quasar sample. We also include the volume density of both the PRG and comparison samples as derived in this work and that of RA12. Columns 1 and 2 list the sample we are considering and the work in which the volume density was derived. Column 3 corresponds to the lower luminosity limit used to derive the volume density. Column 4 gives the assumed redshift at which the volume density is derived and Column 5 gives the volume density when considering these lower luminosity limits and assumed redshifts. }
    \label{lum_lim}
	\begin{tabular}{ l l l c r  }
	\hline
	Sample	&	Work	&	Lower Limit	&	Assumed redshift	&	Volume Density (Mpc$^{-3}$)\\ 
	\hline 
	Type II			&	This				&	L$_{[OIII]}\simeq 1.2\times 10^{35}$ W 						&	0.3	&	$3\times 10^{-7}$\\	
	PRG			 	&	RA12				&	P$_{151MHz}\simeq 1.3\times 10^{25}$ W Hz$^{-1}$ sr$^{-1}$	&	0.0	&	$2\times 10^{-7}$ \\
	PRG			 	&	This				&	P$_{151MHz}\simeq 2.0\times 10^{26}$ W Hz$^{-1}$ sr$^{-1}$	&	0.3	&	$3\times 10^{-9}$\\
	PRG				&	RA12				&	P$_{151MHz}\simeq 1.3\times 10^{25}$ W Hz$^{-1}$ sr$^{-1}$	&	0.5	&	$1\times 10^{-6}$ \\
	Control sample	& RA12					&	M$_B=-20.3$ mag												&	0.1	&	$4\times 10^{-4}$\\
	Comparison Sample 	& This					&	M$_B=-20.4$ mag												&	0.3	&	$9\times 10^{-4}$\\	
	Control sample 	& RA12					&	M$_B=-20.3$ mag												&	0.5	&	$8\times 10^{-4}$\\ 	
	\hline	  
	\end{tabular}
	\end{minipage}       
\end{table*}

We can test the idea that all interacting early-type galaxies will, at some point in the encounter, go through a quasar phase by following the methodology of RA12. In this case, the volume density of quasars ($\rho_{QSO}$) is related to the volume density of morphologically disturbed ellipticals ($\rho_{DE}$) by the equation:
\begin{equation}
\frac{\rho_{QSO}}{\rho_{DE}}=0.01\left ( \frac{t_{QSO}}{10Myr} \right )\left ( \frac{t_{DE}}{1 Gyr} \right )^{-1},
\end{equation}
where $t_{QSO}$ is the typical lifetime of a quasar, and $t_{DE}$ is the time-scale on which interaction signatures will remain visible. If we assume a lifetime for the quasar of $\sim10^7-10^8$ years \citep{martini01} and that the features will remain visible above the surface brightness detection limit of the images for $\sim$$10^9$ years \citep{lotz08}, then we would expect quasars to make up a fraction of ~0.01-0.1 of the total population of disturbed ellipticals.

We can determine a value for the volume density of type II quasars by integrating the $L_{[OIII]}$ luminosity function of \citet{bongiorno10} above the lower $L_{[OIII]}$ limit of our sample, ($log_{10}(L_{[OIII]}/L_{\sun}) = 8.50$) and assuming a redshift of $z= 0.3$\footnote{We assume this z rather than the median of the sample because, in order to determine the volume density of disturbed ellipticals, we make use of work by \citet{faber07}, who provide figures for z=0.3. Therefore, to be directly comparable, we assume this redshift for all the samples we analyse here.}. This yields a result for the volume density of type II quasars of $N_{QSO2} =2.7 \times 10^{-7}~ Mpc^{-3}$. Multiplying this by the fraction of type II quasar host galaxies in our sample that show evidence of tidal disruption ($f_{DQSO2} = 0.75$), and applying the results of \citet{reyes08}, who find a ratio of $\sim$1.2:1 of type II to type I quasars in this luminosity and redshift range, the total volume density of morphologically disturbed quasars in this redshift range, including both type I and type II quasars is $\sim 4.5\times 10^{-7}~ Mpc^{-3}$.

If we then integrate the luminosity function for red early-type galaxies of \cite{faber07} above the lowest B band luminosity ($M_B=-20.4~mag$) of the type II quasars at $z = 0.3$, we find a volume density of $8.6 \times 10^{-4}~ Mpc^{-3}$. Multiplying this figure by the proportion of galaxies from the comparison sample that have morphological features above the minimum surface brightness of the type II quasar morphological features ($f_D\sim0.62$) gives a volume density of disturbed early-type galaxies of $\rho_{DE} \sim 5.3 \times 10^{-4}~ Mpc^{-3}$. Then, comparing the latter volume density with that of quasars at similar redshifts, we obtain $\frac{\rho_{QSO}}{\rho_{DE}}=8.5 \times 10^{-4}$, which is at least a factor of 10 less than expected under the simple assumption that all merging galaxies go through a quasar phase. One explanation for this result is that not all disturbed early-type galaxies are capable of hosting a quasar. As mentioned before, apart from merger phase, other factors are likely to have a strong influence on the outcome of the merger and the gas flows into the nuclear region. These include whether the galaxies are gas rich or gas poor, the mass ratio of the two galaxies, the orbital parameters of the interaction and the masses of the black holes.

\subsubsection{Type II quasars and 2Jy PRG}
\label{IIPRG}
As shown in Section \ref{comp_2jy_samp}, the similarities in the rates of morphological disturbance, the mix of tidal features of the PRG and quasar II objects, and the host galaxy magnitudes  support the idea that both classes are  triggered in a similar way. We can also test the idea that quasars cycle through radio-loud and radio-quiet phases during a particular triggering event (e.g. \citealt{nipoti05}), by making use of the correlation between radio power and emission line luminosity that has been found to exist for PRG (\citealt{tadhunter98, rawlings90, baum89}). This correlation can be used to ensure that we are only considering PRG of comparable optical luminosity to the type II quasars, thereby allowing a comparison of the volume densities of the two classes of AGN, and thus testing the possibility that all quasars go through at least one radio-loud phase. 

To determine the volume density of PRG ($\rho_{PRG}$), we integrate the radio luminosity function of \citet{willott01} above the radio luminosity limit of the 2Jy radio sample at $z>0.2$, which is $P_{151MHz}\simeq 2.0\times 10^{26}~ W~ Hz^{-1}~ sr^{-1}$. This corresponds approximately to the same emission line luminosity as the lowest $L_{[OIII]}$ value in the type II sample\footnote{The lowest [OIII] emission line luminosity measured for the SLRG in the $z\geq0.2$ 2Jy sample is $L_{[OIII]} = 9.4\times10^{34}~W$}, allowing us to compare quasars of the same intrinsic emission line luminosity. We find that the volume density of powerful radio galaxies at $z=0.3$  is $3.1\times10^{-9}~ Mpc^{-3}$. 

The integration of this luminosity function was also performed in RA12, however, they used a lower limit of $P_{151MHz}\simeq 1.3\times 10^{25}~ W~ Hz^{-1}~ sr^{-1}$ at $z=0.5$ and obtained a volume density of $1\times10^{-6}~ Mpc^{-3}$. Table \ref{lum_lim} clearly demonstrates that the volume density of radio galaxies is highly sensitive to the lower radio power limit used. This is due to the fact that the radio luminosity function falls very steeply at the highest luminosities: an increase of a factor of 10 in the lower radio power limit leads to a decrease in the volume density by a factor of 100. Table \ref{lum_lim} demonstrates this clearly by showing the volume densities derived for the different samples considered, with the different lower limits used.

Using the results derived above for the volume densities of type II quasars and PRG of similar emission line luminosity, we find that $\frac{\rho_{RG}}{\rho_{QSO}}=6.8\times10^{-3}$ or $\sim$0.7\%. Utilising this result, we can also apply time scale arguments to test whether radio-loud AGN and quasars are the same objects seen at different phases in their evolution. If we assume that the typical time-scale of a radio-loud phase is $\sim$$10^7-10^8$ years (\citealt*{blundell99,shabala08}), and that $\sim$0.7\% of quasars of comparable luminosity are currently in a radio-loud phase, this would imply that the lifetime of the radio-quiet phase is $\sim$$10^9-10^{10}$ years. However, this is much larger than the currently favoured typical lifetime of radio-quiet quasars which is thought to be $\sim$$10^7-10^8$ years (\citealt{norman88,kauffmann00,martini01,kelly10}), similar to that of PRG. Therefore, based on their relative volume densities and estimated duty cycles, it is unlikely that most radio-quiet quasars go through a radio-loud phase during a particular quasar triggering event. This result may instead indicate that other factors, apart from accretion history, are important in determining whether an AGN is radio-loud, for example the spin of the SMBH (e.g. \citealt*{baum95, sikora07}).

\section{Conclusions}
\label{conclusions}

In this work we present the results of a morphological study of a complete sample of 20 type II quasars at intermediate redshifts ($0.3 < z < 0.41$), selected from the catalogue of candidate objects of \citet{zakamska03}. We also compare our results with those obtained for a comparison sample of quiescent red galaxies (RA12) and a sample of PRG (RA11). We interpret these results in the context of triggering quasar activity, and understanding the relationship between radio-quiet and radio-loud quasars. The main results are:

\begin{itemize}
\item{$75\pm20\%$ of the complete sample of type II quasars analysed in this study show evidence of morphological disturbance at relatively high levels of surface brightness. The median value of surface brightness of the features is $\tilde{\mu}_r^{corr} =  23.4~mag~arcsec ^{-2}$ and the range is $\Delta \mu_r^{corr} \simeq [20.6, 24.7]~mag~arcsec ^{-2}$.}
\item{$35\pm13\%$ of the host galaxies are in the pre-coalescence phase of a merger or are involved in a gravitational interaction that will not necessarily lead to a merger. $40\pm14\%$ of the host galaxies are in the coalescence or post-coalescence phase of a merger. The remaining $25\pm11\%$ of the host galaxies show no morphological signs of mergers or interaction. These results indicate that, if a merger is responsible for triggering the AGN activity, this can happen before, during or after the merging of the SMBH}
\item{When comparing the rates of morphological disturbance found in the type II quasars ($75\pm20\%$) and a sample of quiescent red galaxies ($68\pm14\%$), we find similar rates of interaction in both samples. However, the morphological features detected in the comparison sample are up to 2 magnitudes fainter than those found in the type II quasars, which may highlight a fundamental difference in the types of mergers that the two groups are undergoing.}
\item{When comparing to the SLRG of RA11, we find that the detected morphological features have a very similar range and median value of surface brightness, indicating that in both radio-loud and radio-quiet AGN, the host galaxies may be undergoing the same types of mergers.}
\item{It is unlikely that a single triggering event will lead to a quasar that cycles through both radio-loud and radio-quiet phases. This is because, from the values derived in Section \ref{IIPRG}}, only $\sim$0.7\% of quasars of comparable emission line luminosity are radio-loud, suggesting a lifetime for the radio-quiet phase of $\sim$$10^9-10^{10}$ years --- much larger than current observationally based estimates.
\end{itemize}

These results are consistent with the idea that powerful quasars are triggered in galaxy mergers/interactions, although it is evident that not all morphologically disturbed early-type galaxies are capable of hosting a powerful quasar-like AGN. Clearly the detailed nature of the interaction (e.g. major/minor, gas-rich/gas-poor) is likely to have an influence on the outcome. It is also evident that mechanisms other than major, gas-rich mergers must also be capable of triggering quasar activity, as we find that $25\pm11\%$ of the quasar host galaxies show no signs of tidal interactions.

 In order to further characterise the host galaxies of the type II quasars in our sample, and determine at which point in the merger process the AGN is triggered, spectral synthesis modelling, is required to date the young stellar populations and determine the star formation histories of the host galaxies (e.g. \citealt{tadhunter11, tadhunter02, greene09,liu09,holt07}). 

\section{Acknowledgements}
PB acknowledges support in the form of an STFC Ph.D. studentship. CRA acknowledges the Spanish Ministry of Science and Innovation (MICINN) through project Consolider-Ingenio 2010 Program grant CSD2006-00070: First Science with the GTC (http://www.iac.es/consolider-ingenio-gtc/) and the ESTALLIDOS project PN AYA2010-21887-C04.04. This work is based on observations obtained at the Gemini Observatory, which is operated by the Association of Universities for Research in Astronomy, Inc., under a cooperative agreement with the NSF on behalf of the Gemini partnership: the National Science Foundation (United States), the Science and Technology Facilities Council (United Kingdom), the National Research Council (Canada),CONICYT (Chile), the Australian Research Council (Australia), Minist\'{e}rio da Ci\^{e}ncia, Tecnologia e Inova\c{c}\H{a}o (Brazil) and Ministerio de Ciencia, Tecnolog\'{i}a e Innovaci\'{o}n Productiva Argentina). The Gemini programs under which the data were obtained are GS-2009B-Q-87 and GS-2011B-Q-42.This research has made use of the NASA/ IPAC Infrared Science Archive, which is operated by the Jet Propulsion Laboratory, California Institute of Technology, under contract with the National Aeronautics and Space Administration. Finally, we would like to acknowledge the valuable constructive feedback and suggestions of the anonymous referee.

\bsp
\label{lastpage}

\begin{thebibliography}{99}
\bibitem[\protect\citeauthoryear{Bahcall et al.}{1997}]{bahcall97} Bahcall J.N., Kirhakos S., Saxe D.H., Schneider D.P., 1997, ApJ, 479, 642
\bibitem[\protect\citeauthoryear{Barro et al.}{2009}]{barro09} Barro G., et al. 2009. A\&A, 494, 63
\bibitem[\protect\citeauthoryear{Barro et al.}{2011}]{barro11} Barro G., et al. 2011, ApJS, 193, 13
\bibitem[\protect\citeauthoryear{Baum \& Heckman}{1989}]{baum89} Baum S.A., Heckman T., 1989, ApJ, 336, 702 
\bibitem[\protect\citeauthoryear{Baum, Zirbel \& O'Dea}{Baum et al.}{1995}]{baum95} Baum S.A., Zirbel E.L., O'Dea C.P., 1995, ApJ, 451, 88
\bibitem[\protect\citeauthoryear{Bell et al.}{2006}]{bell06} Bell E.F., et al., 2006, ApJ, 640, 241
\bibitem[\protect\citeauthoryear{Becker, White \& Helfand}{Becker et al.}{1995}]{becker95} Becker R., White R., Helfand D., 1995, ApJ, 450, 559
\bibitem[\protect\citeauthoryear{Bennert et al.}{2008}]{bennert08} Bennert N., Canalizo G., Jungwiert B., Stockton A., Schweizer F., Peng C.Y., Lacy M., 2008, ApJ, 677, 846
\bibitem[\protect\citeauthoryear{Bessiere et al.}{2012}]{bessiere12} Bessiere P.S., Tadhunter C.N. et al., 2012, in preparation
\bibitem[\protect\citeauthoryear{Blanton}{2006}]{blanton06} Blanton M.R., 2006, ApJ, 648, 268
\bibitem[\protect\citeauthoryear{Blundell, Rawlings \& Willott}{Blundell et al.}{1999}]{blundell99} Blundell K.M., Rawlings S., Willott C.J., 1999, AJ, 117, 677
\bibitem[\protect\citeauthoryear{Bongiorno et al.}{2010}]{bongiorno10} Bongiorno A., et al., 2010, A\&A, 510, 56
\bibitem[\protect\citeauthoryear{Cardelli, Clayton \& Mathis}{1989}]{cardelli89} Cardelli J.A., Clayton G.C., Mathis J.S., 1989, ApJ, 345, 245
\bibitem[\protect\citeauthoryear{Cattaneo et al.}{2005}]{cattaneo05} Cattaneo A., Combes F., Colombi S., Bertin E., Melchior A.L., 2005, MNRAS, 359, 1237
\bibitem[\protect\citeauthoryear{Cisternas et al.}{2011}]{cisternas11} Cisternas M., 2011, ApJ, 726, 57
\bibitem[\protect\citeauthoryear{Combes}{2001}]{combes01} Combes F., 2001, in I.Aretxaga, D.Kunth, \& R.M{\'u}jica ed.,, Advanced Lectures on the Starburst-AGN, Fueling the AGN. p.223
\bibitem[\protect\citeauthoryear{Condon et al.}{1998}]{condon98} Condon J., Cotton W., Greisen E., Yin Q., Perley R., Taylor G., Broderick J., 1998, AJ, 115, 1693
\bibitem[\protect\citeauthoryear{Dicken et al.}{2009}]{dicken09} Dicken D., Tadhunter C., Axon D., Morganti R., Inskip K., Holt J., Gonz{\'a}lez Delgado R., Groves B., 2009, ApJ, 694, 268
\bibitem[\protect\citeauthoryear{Dickson et al.}{1995}]{dickson95} Dickson R., Tadhunter C., Shaw M., Clark N., Morganti R., 1995, MNRAS, 273, 29
\bibitem[\protect\citeauthoryear{di Matteo, Springel \& Hernquist}{di Matteo et al.}{2005}]{dimatteo05} di Matteo T., Springel V., Hernquist L., 2005, Nature, 433, 604
\bibitem[\protect\citeauthoryear{Donnelly, Partridge \& Windhorst}{Donnelly et al.}{1987}]{donnelly87} Donnelly R., Partridge R., Windhorst R., 1987, ApJ, 321, 94
\bibitem[\protect\citeauthoryear{Dunlop et al.}{2003}]{dunlop03} Dunlop J.S., McLure R.J., Kukula M.J., 	Baum S.A., O'Dea C.P., Hughes D.H., 2003, MNRAS, 340, 1095
\bibitem[\protect\citeauthoryear{Ellison et al.}{2011}]{ellison11} Ellison S.L., Patton D.R., Mendel J.T., Scudder J.M., 2011, MNRAS, 418, 2043
\bibitem[\protect\citeauthoryear{Faber et al.}{2007}]{faber07} Faber S.M., et al., 2007, ApJ, 665, 265
\bibitem[\protect\citeauthoryear{Feldmann, Mayer \& Carollo}{Feldmann et al.}{2008}]{feldmann08} Feldmann R., Mayer L., Carollo C.M., 2008, ApJ, 684, 1062 
\bibitem[\protect\citeauthoryear{Ferrarese \& Merritt}{2000}]{ferrarese00} Ferrarese L., Merritt D., 2000, ApJL, 539, L9
\bibitem[\protect\citeauthoryear{Frei \& Gunn}{1994}]{frei94} Frei Z., Gunn J.E., 1994, AJ, 108, 1476
\bibitem[\protect\citeauthoryear{Fukugita, Shimasaku \& Ichikawa}{Fukugita et al.}{1995}]{fukugita95} Fukugita M., Shimasaku K., Ichikawa T., 1995, PASP, 107, 945
\bibitem[\protect\citeauthoryear{Gebhardt et al.}{2000}]{gebhardt00} Gebhardt K., et al., 2000, ApJL, 539, L13 
\bibitem[\protect\citeauthoryear{Greene et al.}{2009}]{greene09} Greene J,.E., Zakamska N.L., Liu X., Barth A.J., Ho L.C., 2009, ApJ, 702, 441
\bibitem[\protect\citeauthoryear{Grogin et al.}{2003}]{grogin03} Grogin N.A. et al., 2003, ApJ, 595, 685
\bibitem[\protect\citeauthoryear{Heckman et al.}{1986}]{heckman86} Heckman T.M., Smith E.P., Baum S.A., van Breugel W.J.M., Miley G.K., Illingworth G.D., Bothun G.D., Balick B., 1986, ApJ, 311, 526
\bibitem[\protect\citeauthoryear{Hernquist \& Spergel}{1992}]{hernquist92} Hernquist L., Spergel D.N., 1992, ApJL, 399, L117
\bibitem[\protect\citeauthoryear{Holt et al.}{2007}]{holt07} Holt J., Tadhunter C., Gonz{\'a}lez Delgado R., Inskip K., Rodriguez Zaurin J., Emonts B., Morganti R., Wills K., 2007, MNRAS, 381, 611 
\bibitem[\protect\citeauthoryear{Hook et al.}{2004}]{hook04} Hook I.M., J{\o}rgensen I., Allington-Smith J.R., Davies R.L., Metcalfe N., Murowinski R.G., Crampton D., 2004, PASP, 116, 425
\bibitem[\protect\citeauthoryear{Hopkins et al.}{2006}]{hopkins06} Hopkins P.F., Hernquist L., Cox T.J., Di Matteo T., Robertson B., Springel V., 2006, ApJS, 163,, 1
\bibitem[\protect\citeauthoryear{Hopkins et al.}{2008}]{hopkins08} Hopkins P.F., Hernquist L., Cox T.J., Kere{\v s} D., 2008, ApJS, 175, 356 
\bibitem[\protect\citeauthoryear{Hyv{\"o}nen et al.}{2007}]{hyvonen07} Hyv{\"o}nen T., Kotilainen J.K., {\"O}rndahl E., Falomo R., Uslenghi M., 2007, AAP, 462, 525
\bibitem[\protect\citeauthoryear{Jogee}{2006}]{jogee06} Jogee S., 2006, in Alloin D., ed, Physics of Active Galactic Nuclei at all Scales. Springer Verlag, Berlin, p. 143
\bibitem[\protect\citeauthoryear{Kauffmann \& Haehnelt}{2000}]{kauffmann00} Kauffmann G., Haehnelt M., 2000, MNRAS, 311, 576
\bibitem[\protect\citeauthoryear{Kelly et al.}{2010}]{kelly10} Kelly B.C., Vestergaard M., Fan X., Hopkins P., Hernquist L., Siemiginowska A., 2010, ApJ, 719, 1315
\bibitem[\protect\citeauthoryear{Kirhakos et al.}{1999}]{kirhakos99} Kirhakos S., Bahcall J.N., Schneider D.P., Kristian J., 1999, ApJ, 520, 67
\bibitem[\protect\citeauthoryear{Kormendy \& Richstone}{1995}]{kormendy95} Kormendy J., Richstone D., 1995, ARA\&A, 33, 581
\bibitem[\protect\citeauthoryear{Kormendy, Bender \& Cornell}{Kormendy et al.}{2011}]{kormendy11} Kormendy J., Bender R., Cornell M.E., 2011, Nature, 469, 374
\bibitem[\protect\citeauthoryear{Liu et al.}{2009}]{liu09}Liu X., Zakamska N.L., Greene J.E., Strauss M.A., Krolik J.H., Heckman T.M., 2009, ApJ, 702, 1098
\bibitem[\protect\citeauthoryear{Lotz et al.}{2008}]{lotz08} Lotz J. M., Jonsson P., Cox T. J.,  Primack J. R. 2008, MNRAS, 391, 1137
\bibitem[\protect\citeauthoryear{Madau et al.}{1996}]{madau96} Madau P., Ferguson H. C., Dickinson M. E., Giavalisco M., Steidel C. C., Fruchter A. 1996, MNRAS, 283, 1388
\bibitem[\protect\citeauthoryear{Magorrian et al.}{1998}]{magorrian88} Magorrian J. et al., 1998, AJ, 115, 2285
\bibitem[\protect\citeauthoryear{Martini \& Weinberg}{2001}]{martini01} Martini P., Weinberg D.H., 2001, 547, 12
\bibitem[\protect\citeauthoryear{McIntosh et al.}{2008}]{mcintosh08} McIntosh D. H., Guo Y., Hertzberg J., Katz N., Mo H. J., van den Bosch F. C., Yang X. 2008, MNRAS, 388, 1537
\bibitem[\protect\citeauthoryear{McLure et al.}{1999}]{mclure99} McLure R.J., Kukula M.J., Dunlop J.S., Baum S.A., O'Dea C.P., Hughes D.H., 1999, MNRAS, 308, 377
\bibitem[\protect\citeauthoryear{Naab, Khochfar \& Burkert}{Naab et al.}{2006}]{naab06} Naab T., Khochfar S.,  Burkert A. 2006, ApJL, 636, L81
\bibitem[\protect\citeauthoryear{Nipoti, Blundell \& Binney}{Nipoti et al.}{2005}]{nipoti05} Nipoti C., Blundell K.M., Binney J., 2005, MNRAS, 361, 633
\bibitem[\protect\citeauthoryear{Norman \& Scoville}{1988}]{norman88} Norman C., Scoville N., 1988, ApJ, 332, 124
\bibitem[\protect\citeauthoryear{P{\'e}rez-Gonz{\'a}lez et al.}{2008}]{perez08} P{\'e}rez-Gonz{\'a}lez P.G., Trujillo I., Barro G., Gallego J., Zamorano J., Conselice C.J., 2008, ApJ, 687, 50
\bibitem[\protect\citeauthoryear{Quinn}{1984}]{quinn84} Quinn P. J. 1984, ApJ, 279, 596
\bibitem[\protect\citeauthoryear{Ramos Almeida et al.}{2011}]{ramos11} Ramos Almeida C., Tadhunter C.N., Inskip K.J., Morganti R., Holt J., Dicken D., 2011, MNRAS, 410, 1550
\bibitem[\protect\citeauthoryear{Ramos Almeida et al.}{2012}]{ramos12} Ramos Almeida C., Bessiere P.S., Tadhunter C.N., P{\'e}rez-Gonz{\'a}lez P.G., Barro G., Inskip K.J., Morganti R., Holt J., Dicken D., 2012, MNRAS, 419, 687
\bibitem[\protect\citeauthoryear{Rawlings et al.}{1990}]{rawlings90} Rawlings S., Saunders R., Miller P., Jones M.E., Eales S.A., 1990, MNRAS, 246, 21
\bibitem[\protect\citeauthoryear{Reyes et al.}{2008}]{reyes08} Reyes R. et al., 2008, AJ, 136, 2373
\bibitem[\protect\citeauthoryear{Serber et al.}{2006}]{serber06} Serber W., Bahcall N., M{\'e}nard B., Richards G., 2006, ApJ, 643, 68
\bibitem[\protect\citeauthoryear{Shabala et al.}{2008}]{shabala08} Shabala S.S., Ash S., Alexander P., Riley J.M., 2008, MNRAS, 388, 625
\bibitem[\protect\citeauthoryear{Sikora, Stawarz \& Lasota}{Sikora et al.}{2007}]{sikora07} Sikora M., Stawarz {\L}., Lasota J.P., 2007, ApJ, 658, 815
\bibitem[\protect\citeauthoryear{Smith \& Heckman}{1989}]{smith89} Smith E. P., Heckman T. M. 1989, ApJ, 341, 658
\bibitem[\protect\citeauthoryear{Springel, Di Matteo \& Hernquist}{Springel et al.}{2005}]{springel05} Springel V., Di Matteo T.,  Hernquist L. 2005, MNRAS, 361, 776
\bibitem[\protect\citeauthoryear{Tadhunter et al.}{1992}]{tadhunter92} Tadhunter C., Scarrott S., Draper P., Rolph C., 1992, MNRAS, 256, 53
\bibitem[\protect\citeauthoryear{Tadhunter et al.}{1993}]{tadhunter93} Tadhunter C., Morganti R., di Serego-Alighieri S., Fosbury R.A., Danziger I.J., 1993, MNRAS, 263, 999
\bibitem[\protect\citeauthoryear{Tadhunter et al.}{1998}]{tadhunter98} Tadhunter C., Morganti R., Robinson A., Dickson R., Villar-Martin M., Fosbury R.A.E., 1998, MNRAS, 298, 1035
\bibitem[\protect\citeauthoryear{Tadhunter et al.}{2002}]{tadhunter02} Tadhunter C., Dickson R., Morganti R., Robinson T., Wills K., Villar-Martin M., Hughes M., 2002, MNRAS, 330,977
\bibitem[\protect\citeauthoryear{Tadhunter et al.}{2005}]{tadhunter05} Tadhunter C., Robinson T.G., Gonz{\'a}lez Delgado R.M., Wills K., Morganti R., 2005, MNRAS, 356, 480
\bibitem[\protect\citeauthoryear{Tadhunter et al.}{2011}]{tadhunter11} Tadhunter C. et al., 2011, MNRAS, 412, 960
\bibitem[\protect\citeauthoryear{Toomre \& Toomre}{1972}]{toomre72} Toomre A., Toomre J., 1972, ApJ, 178, 623
\bibitem[\protect\citeauthoryear{Ueda et al.}{2003}]{ueda03} Ueda Y., Akiyama M., Ohta K., Miyaji T., 2003, ApJ, 598, 886
\bibitem[\protect\citeauthoryear{Villar-Mart{\'{\i}}n et al.}{2010}]{villar10} Villar-Mart{\'{\i}}n M., Tadhunter C., P{\'e}rez E., Humphrey A., Mart{\'{\i}}nez-Sansigre A., Delgado R.G., P{\'e}rez-Torres M., 2010, MNRAS, 407, L6
\bibitem[\protect\citeauthoryear{Villar-Mart{\'{\i}}n et al.}{2011}]{villar11} Villar-Mart{\'{\i}}n M., Tadhunter C., Humphrey A., Encina R.F., Delgado R.G., Torres M.P., Mart{\'{\i}}nez-Sansigre A., 2011, MNRAS, 416, 262
\bibitem[\protect\citeauthoryear{Villar-Mart{\'{\i}}n et al.}{2012}]{villar12} Villar-Mart{\'{\i}}n M., Cabrera-Lavers A., Bessiere P., Tadhunter C., Rose M., de Breuck C., 2012, accepted MNRAS, arxiv 1202.1729
\bibitem[\protect\citeauthoryear{Wild, Heckman \& Charlot}{Wild et al.}{2010}]{wild10} Wild V., Heckman T., Charlot S., 2010, MNRAS, 405, 933
\bibitem[\protect\citeauthoryear{Willott et al.}{2001}]{willott01} Willott C.J., Rawlings S., Blundell K.M., Lacy M., Eales S.A., 2001, MNRAS, 322, 536
\bibitem[\protect\citeauthoryear{York et al.}{2000}]{york00} York D.G. et al., 2000, AJ, 120, 1579
\bibitem[\protect\citeauthoryear{Xu, Livio \& Baum}{Xu et al.}{1999}]{xu99} Xu C., Livio M., Baum S., 1999, AJ, 118, 1169
\bibitem[\protect\citeauthoryear{Zakamska et al.}{2003}]{zakamska03} Zakamska N.L. et al., 2003, AJ, 126,, 2125
\bibitem[\protect\citeauthoryear{Zakamska et al.}{2004}]{zakamska04} Zakamska N.L., Strauss M.~A., Heckman T.~M., Ivezi{\'c} {\v Z}., Krolik J.~H., 2004, AJ, 128,, 1002
\bibitem[\protect\citeauthoryear{Zakamska et al.}{2005}]{zakamska05} Zakamska N.L. et al., AJ, 132, 1496
\bibitem[\protect\citeauthoryear{Zakamska et al.}{2006}]{zakamska06} Zakamska N.L. et al., 2006, AJ, 132, 1496

\end{thebibliography}
\end{document}